# Loop Corrections in Non-Linear Cosmological Perturbation Theory [*]

Román Scoccimaro

*Department of Physics and Enrico Fermi Institute, University of Chicago, Chicago, IL 60637 and*
*NASA/Fermilab Astrophysics Center, Fermi National Accelerator Laboratory, P.O.Box 500, Batavia, IL 60510*

Joshua Frieman

*NASA/Fermilab Astrophysics Center, Fermi National Accelerator Laboratory, P.O.Box 500, Batavia, IL 60510*
*and Department of Astronomy and Astrophysics, University of Chicago, Chicago, IL 60637*

(September 7, 1995)

Using a diagrammatic approach to Eulerian perturbation theory, we analytically calculate the variance and skewness of the density and velocity divergence induced by gravitational evolution from Gaussian initial conditions, including corrections *beyond* leading order. Except for the power spectrum, previous calculations in cosmological perturbation theory have been confined to leading order (tree level)—we extend these to include loop corrections. For scale-free initial power spectra, $P(k) \sim k^n$ with $-2 \leq n \leq 2$, the one-loop variance $\sigma^2 \equiv \langle \delta^2 \rangle = \sigma_\ell^2 + 1.82\sigma_\ell^4$ and the skewness $S_3 = \langle \delta^3 \rangle / \sigma^4 = 34/7 + 9.8\sigma_\ell^2$, where $\sigma_\ell$ is the rms fluctuation of the density field to linear order. (These results depend weakly on the spectral index $n$, due to the non-locality of the non-linear solutions to the equations of motion.) Thus, loop corrections for the (unsmoothed) density field begin to dominate over tree-level contributions (and perturbation theory presumably begins to break down) when $\sigma_\ell^2 \simeq 1/2$. For the divergence of the velocity field, loop dominance does not occur until $\sigma_\ell^2 \approx 1$. We also compute loop corrections to the variance, skewness, and kurtosis for several non-linear approximation schemes, where the calculation can be easily generalized to 1-point cumulants of higher order and arbitrary number of loops. We find that the Zel'dovich approximation gives the best approximation to the loop corrections of exact perturbation theory, followed by the Linear Potential approximation (LPA) and the Frozen Flow approximation (FFA), in qualitative agreement with the relative behavior of tree-level results. In LPA and FFA, loop corrections are infrared divergent for spectral indices $n \leq -1$; this is related to the breaking of Galilean invariance in these schemes.

## I. INTRODUCTION

There is growing evidence that the large-scale structure of the Universe grew via gravitational instability from small primordial fluctuations in the matter density. At early epochs, the growth of density perturbations can be described by linear perturbation theory, provided that the linear power spectrum $P(k)$ falls off less steeply than $k^4$ for small $k$ [1–3]. In the linear regime, perturbation Fourier modes evolve independently of one another, conserving the statistical properties of the primordial fluctuations. In particular, if the primordial fluctuations are Gaussian random fields (as expected from the simplest inflationary models), they remain Gaussian in linear theory. In this case, the statistical properties of the density and velocity fields are completely determined by the two-point correlation function or the power spectrum. When the fluctuations become non-linear, coupling between different Fourier modes becomes important, inducing non-trivial correlations that modify the statistical properties of the cosmological fields. For Gaussian initial conditions, this causes the appearance of higher-order reduced correlations ($p$-point cumulants), which constitute independent statistics that can be measured in observational data and numerical simulations, even when the departure from the linear regime is small.

Non-linear cosmological perturbation theory (NLCPT) provides a framework for analytic calculations of these effects in this weakly non-linear regime. The assumption is that even when the density and velocity fields are highly non-linear on the smallest scales, one can accurately describe the evolution of the large-scale properties of these fields during the first stages of non-linear evolution using a systematic perturbative approach. The validity of this ansatz has not yet been proven theoretically, but its predictions have been checked against numerical simulations. In fact, comparison of one-point statistics with N-body simulations shows that *leading order* NLCPT provides an adequate

---

description even on scales where *next-to-leading order* and higher order perturbative contributions would be expected to become important. The surprising agreement of leading order calculations with numerical simulations raises interesting questions: does it imply that the next-to-leading order corrections are generically small, or are they merely 'accidentally' small for the cases that have been numerically studied? Or does it instead imply that the whole series of contributions beyond leading order conspire to nearly cancel? The answer to this puzzle should tell us something interesting about gravitational dynamics and the transition to the non-linear regime. These questions clearly motivate the study of corrections beyond leading order (loop corrections) in NLCPT. In any case, the calculation of next-to-leading order corrections should help us to better understand the successes and limitations of the perturbative approach to gravitational instability.

To study the dynamics of gravitational instability, one must choose between an Eulerian or Lagrangian description; each has its advantages and drawbacks. In the Eulerian approach, one considers the motion of matter relative to some fixed coordinate system, whereas in the Lagrangian formalism one follows the trajectories of individual fluid elements. Perturbation theory is formulated quite differently in the two approaches: in the Eulerian framework, the small expansion parameter is the linear rms density fluctuation; in the Lagrangian formulation, the expansion is done about 'inertial' fluid element trajectories. As a result, the Lagrangian description is more successful in describing large density contrasts that can arise before multi-streaming (orbit crossing) occurs; for an account of Lagrangian perturbation theory, see [4] and references therein. On the other hand, the Eulerian approach is conceptually straightforward and has the advantage that it works directly with the density and velocity fields whose statistics we want to quantify. In this paper, we use the Eulerian description, which can be solved to arbitrary order in perturbation theory for an Einstein-de Sitter Universe [5].

Leading order calculations in non-linear theory began with Peebles [6], who used second order perturbation theory to obtain the skewness for the unsmoothed density field (assuming Gaussian initial conditions), $S_3 = \langle\delta^3\rangle/\langle\delta^2\rangle^2 = 34/7$. Fry [7] used second and third order perturbation theory to calculate the leading order three and four-point functions, respectively. He also showed how one can associate tree diagrams with each perturbative contribution to leading order. The predictions for the three-point function were subsequently found to agree with numerical simulations [8]. Goroff et al. [5] developed a systematic approach to NLCPT, obtaining recursion relations for the density and velocity fields to arbitrary order in an Einstein-de Sitter universe. They showed how the perturbation series could be represented in terms of Feynman diagrams, with leading order contributions corresponding to tree diagrams, next-to-leading order to one-loop diagrams, and so on. They also introduced the normalized one-point cumulants $S_p$ ($p \geq 3$), which are measures of the skewness ($p = 3$), kurtosis ($p = 4$), and higher central moments of the density field. They obtained the leading-order $S_p$ for $p = 3, 4, 5$ for the Gaussian-smoothed density field by Monte Carlo integration for $\Omega = 1$ cold dark matter (CDM), but they did not do any explicit loop calculations. These calculations were extended to non-flat Universes ($\Omega \neq 1$) using Lagrangian perturbation theory [9,10], with the result that the skewness of the unsmoothed density field depends very weakly on $\Omega$. Recently, equivalent results were obtained in Eulerian perturbation theory [11].

To compare the results of NLCPT with numerical simulations and with observations, the one-point cumulants must be calculated for the density field smoothed over a finite volume. Analytic leading-order results for $S_3$ were obtained in [12] for scale-free initial power spectra, $P(k) \sim k^n$, for Gaussian ($n \geq -1$) and top-hat smoothing ($-3 \leq n \leq 0$); with smoothing, $S_3$ decreases with increasing $n$. Comparison of these results with N-body simulations showed a very good agreement at $\sigma^2 \approx 0.5$, while Monte Carlo integration of the smoothed skewness for standard CDM agreed with N-body results up to scales where $\sigma^2 \approx 1$. These results were later generalized by Bernardeau [13], who presented analytic calculations of $S_3, S_4, T_3$ and $T_4$ ($T_p$ being the normalized cumulants for the divergence of the velocity field) for top-hat smoothing using Eulerian perturbation theory for arbitrary $\Omega$, $\Lambda$, and $P(k)$. The dependence on $\Lambda$ was found to be extremely small, with only $T_3$ and $T_4$ being sensitive to $\Omega$ (see also [14]). For Gaussian smoothing, $S_4$ was obtained by Monte Carlo integration in [15], and semianalytically in [16], where similar results are presented for $T_4$ and also analytic computations of $S_3$ and $T_3$.

An important generalization of all these results is the calculation of the whole series (i.e., for *all p*) of one-point cumulants at tree-level. This was done for unsmoothed fields by Bernardeau [17,18] and generalized to include top-hat smoothing for $S_p$ and $T_p$ [19], who showed that the tree-level cumulant generating function obeys the equations of motion of the spherical collapse model. Therefore, all the information concerning one-point cumulants at tree-level is encoded in the spherical collapse dynamics. In particular, the tree-level smoothed one-point cumulants are independent of the variance $\sigma(R)$ and depend on the smoothing scale $R$ only through derivatives of the spectral index $n$ with respect to $R$; thus, for scale-free linear power spectra, the leading-order $S_p(R)$ are constants. From the series of leading-order cumulants, the one-point probability function $P(\delta)d\delta$ was reconstructed and shown to agree with CDM N-body simulations even when $\sigma \approx 1.5$ [20]. A detailed comparison between leading-order NLCPT predictions and N-body simulations for the $S_p$ parameters (up to $p = 10$) was carried out in [21,22], showing very good agreement up to scales where $\sigma \approx 1$.

One-loop corrections in NLCPT have been studied in the literature for the 2-point cumulant in real space (the two-point correlation function) and Fourier space (the power spectrum) [23–30]. These calculations show how power is transferred between large and small scales; the agreement with N-body simulations is good for $\sigma^2 \leq 1$ (when long wavelength modes dominate the non-linear contribution) and gradually breaks down when $\sigma$ increases above unity [29,30]. Higher order loop corrections to the power spectrum were considered in [31], including the full contributions up to 2 loops and the most important terms for high $k$ in 3- and 4-loop contributions. Very recently, one-loop corrections to the variance were studied numerically in [32] for gaussian smoothing.

Comparison of NLCPT predictions with galaxy surveys is more intricate. Due to the possibility of a non-trivial bias mechanism during the process of structure formation, the statistical properties of the galaxy distribution are not necessarily equivalent to those of the underlying matter distribution predicted by theory. Moreover, when considering higher-order correlations, the relation between the galaxy and matter density correlations is complicated by the fact that the bias between galaxies and mass is likely to be non-linear [33]. However, *assuming* the validity of NLCPT (as demonstrated by comparison with numerical simulations), one can use the observed galaxy higher-order correlations [34–37] to put constraints on the non-linear bias [38,39] and test models of large-scale structure formation [40,41]. Another possibility is to test perturbation theory by including some form of bias in numerical simulations [42].

Calculation of loop corrections to one-point cumulants constitutes a natural development of the subject and should provide further insight into the weakly non-linear regime. Numerical simulations have shown that the $S_p$ parameters depart from the tree-level perturbation theory predictions in the non-linear regime [43–45]. This is to be expected, since the fluid equations of motion which are the starting point for the perturbative expansion become invalid when multistreaming and vorticity develop. However, it is important to know whether NLCPT can provide an understanding of this departure in the first stages of non-linear evolution or whether this behavior is instead due to non-perturbative and/or physical effects not included in the analytic treatments so far.

In this paper, we develop the diagrammatic approach to Eulerian perturbation theory, providing the calculational rules and technical machinery to perform loop calculations. As applications, we calculate analytically the one-loop corrections to the variance and skewness for the unsmoothed density and divergence of the velocity fields, assuming Gaussian initial conditions. Since we deal with unsmoothed fields, these results cannot yet be checked against numerical simulations. In subsequent papers, we will present results for the one-loop bispectrum (the three-point function in Fourier space) and the variance and skewness for the smoothed density field.

When fluctuations become strongly non-linear, perturbation theory breaks down and one must rely on N-body simulations to follow the dynamical evolution. Alternatively, one can formulate analytic approximation schemes that provide some insight into the physics of the non-linear regime. This was first suggested by Zel'dovich [46], who initiated the Lagrangian approach to gravitational instability by extrapolating the linear solution in Lagrangian space into the non-linear regime. The Zel'dovich approximation gives an analytic understanding of the formation of walls and filaments in gravitational clustering. Recently, alternative approximations have been suggested, e.g., to extrapolate the linear velocity field into the non-linear regime [47] (the frozen flow approximation), and to extrapolate the linear gravitational potential [48,49] (the linear potential approximation). This set of non-linear approximations have in common that they are all consistent in linear theory with the exact dynamics, their differences appearing at second order in a perturbative expansion.

We also calculate loop corrections to one-point cumulants in these non-linear approximation schemes; these calculations are much easier to perform than for the exact dynamics and can be carried out to higher order and number of loops. Although these approximations do not provide a very accurate estimation of the exact one-point cumulants (as was first noticed for the Zel'dovich approximation [50]), they nevertheless give order of magnitude estimates which are useful when corresponding results for the exact dynamics are unknown. A comparison of the tree-level predictions in these approximations for one-point cumulants has been given in [51–53]. A comparative study of non-linear approximations in the strongly non-linear regime was carried out in [54]. One-loop corrections to $S_3$ and $S_4$ for the Zel'dovich approximation in the unsmoothed case where estimated in [20] by taking moments of the (regularized) one-point probability distribution. Here we provide a perturbative calculation and extend the results to the divergence of the velocity field and to two-loop corrections for $S_3$ and $T_3$. We also calculate up to 3-loop corrections to the variance of the density and divergence of the velocity fields.

The paper is organized as follows. In Sec. II we discuss solutions (to arbitrary order in perturbation theory) to the density and divergence of the velocity fields in the exact dynamics and non-linear approximations for the Einstein-de Sitter universe. Section III is devoted to the statistical description of cosmological fields and the perturbative expansion for 1-point cumulants. The diagrammatic approach to perturbation theory is the subject of Sec. IV, where we discuss diagrammatic expansions for the variance, skewness, and kurtosis. In Sec. V and Sec. VI we apply the formalism developed in the previous sections to the calculation of loop corrections for density and divergence of the velocity fields, respectively. The connection of Galilean invariance to the cancellation of infrared divergences is considered in Section VII. Section VIII contains our conclusions. Technical material regarding the integrations needed to carry out the calculations is given in the Appendices.

## II. EULERIAN PERTURBATION THEORY

### A. The Equations of Motion

The N-body problem in which particles interact only through gravity is quite difficult to treat analytically, so some simplifications are necessary in order to make progress. The exact evolution equations are equivalent to a hierarchy of $N$ coupled integro-differential equations for the $n$-particle distribution functions in phase space ($n = 1, ..., N$) known as the BBGKY hierarchy (see e.g. [6]). By assuming that the particles move without collisions (an excellent approximation for weakly interacting dark matter particles) under the influence of the smooth gravitational potential of the density field, one can decouple this hierarchy into a mean field equation, the collisionless Boltzmann equation, which describes the effective evolution of the 1-particle distribution function. Since we are interested in the study of the weakly non-linear regime, we can simplify this further by taking velocity moments of the Boltzmann equation and assuming that particle trajectories do not cross (single stream approximation). We also assume the particles have negligible thermal motions and move non-relativistically, i.e., that the universe is dominated by cold dark matter (pressureless dust). This leaves us with the first two moments (density and velocity fields) as a complete description of the system.

The relevant equations therefore correspond to conservation of mass and momentum and the Poisson equation for a self-gravitating perfect fluid with zero pressure in a homogeneous and isotropic universe [6]:

$$\frac{\partial \delta(\mathbf{x}, \tau)}{\partial \tau} + \nabla \cdot \{[1 + \delta(\mathbf{x}, \tau)]\mathbf{v}(\mathbf{x}, \tau)\} = 0, \tag{2.1a}$$

$$\frac{\partial \mathbf{v}(\mathbf{x}, \tau)}{\partial \tau} + \mathcal{H}(\tau)\,\mathbf{v}(\mathbf{x}, \tau) + [\mathbf{v}(\mathbf{x}, \tau) \cdot \nabla]\mathbf{v}(\mathbf{x}, \tau) = -\nabla \Phi(\mathbf{x}, \tau), \tag{2.1b}$$

$$\nabla^2 \Phi(\mathbf{x}, \tau) = \frac{3}{2}\Omega \mathcal{H}^2(\tau)\delta(\mathbf{x}, \tau)\ . \tag{2.1c}$$

Here, $\mathbf{x}$ denotes comoving spatial coordinates, the density contrast $\delta(\mathbf{x}, \tau) \equiv \rho(\mathbf{x}, \tau)/\bar{\rho} - 1$, with $\bar{\rho}(\tau)$ the mean density of matter, $\mathbf{v} \equiv d\mathbf{x}/d\tau$ represents the velocity field fluctuations about the Hubble flow, $\mathcal{H} \equiv d\ln a/d\tau = Ha$ is the conformal expansion rate, $a(\tau)$ is the cosmic scale factor, $\tau = \int dt/a$ is the conformal time, $\Phi$ is the gravitational potential due to the density fluctuations, and the density parameter $\Omega = \bar{\rho}/\rho_c = 8\pi G\bar{\rho}a^2/3\mathcal{H}^2$. Note that we have implicitly assumed the Newtonian approximation to general relativity, valid for non-relativistic matter on scales less than the Hubble length $a\mathcal{H}^{-1}$. We take the velocity field to be irrotational, so it can be completely described by its divergence $\theta \equiv \nabla \cdot \mathbf{v}$. By Kelvin's circulation theorem, which follows from taking the curl of Eq. (2.1b), vorticity cannot be generated during the non-linear evolution, and any initial vorticity decays as $a^{-2}$. This conclusion holds true, however, only as long as Eqs. (2.1) are valid; in particular, multi-streaming and shocks can generate vorticity. We will refer to Eqs. (2.1) as the "exact dynamics" (ED) (although for the reasons stated above this is a slight abuse in terminology), to make a distinction with the modified dynamics introduced by the non-linear approximations to be discussed later. Equations (2.1) are valid in an arbitrary homogeneous and isotropic background Universe, which evolves according to the Friedmann equations:

$$\frac{\partial \mathcal{H}(\tau)}{\partial \tau} = -\frac{\Omega}{2}\mathcal{H}^2(\tau) + \frac{\Lambda}{3}a^2(\tau) \tag{2.2a}$$

$$(\Omega - 1)\mathcal{H}^2(\tau) = k - \frac{\Lambda}{3}a^2(\tau), \tag{2.2b}$$

where $\Lambda$ is the cosmological constant, and the spatial curvature constant $k = -1, 0, 1$ for $\Omega_{tot} < 1$, $\Omega_{tot} = 1$ and $\Omega_{tot} > 1$ respectively (where $\Omega_{tot} \equiv \Omega + \Lambda a^2/(3\mathcal{H}^2)$).

When the non-linear terms in Eqs. (2.1) are neglected, different Fourier modes evolve independently. Therefore, it is natural to Fourier transform the perturbation equations and work in momentum space. Our convention for the Fourier transform of a field $A(\mathbf{x}, \tau)$ is:

$$\tilde{A}(\mathbf{k}, \tau) = \int \frac{d^3x}{(2\pi)^3} \exp(-i\mathbf{k} \cdot \mathbf{x})\ A(\mathbf{x}, \tau). \tag{2.3}$$

When the non-linear terms in Eqs. (2.1) are taken into account, the equations of motion display the coupling between different Fourier modes [5] characteristic of non-linear theories. Taking the divergence of Equation (2.1b) and Fourier transforming the resulting equations of motion we get:

$$\frac{\partial \tilde{\delta}(\mathbf{k},\tau)}{\partial \tau} + \tilde{\theta}(\mathbf{k},\tau) = -\int d^3k_1 \int d^3k_2 \delta_D(\mathbf{k}-\mathbf{k}_1-\mathbf{k}_2)\alpha(\mathbf{k},\mathbf{k}_1)\tilde{\theta}(\mathbf{k}_1,\tau)\tilde{\delta}(\mathbf{k}_2,\tau), \tag{2.4a}$$

$$\frac{\partial \tilde{\theta}(\mathbf{k},\tau)}{\partial \tau} + \mathcal{H}(\tau)\tilde{\theta}(\mathbf{k},\tau) + \frac{3}{2}\Omega\mathcal{H}^2(\tau)\tilde{\delta}(\mathbf{k},\tau) = -\int d^3k_1 \int d^3k_2 \delta_D(\mathbf{k}-\mathbf{k}_1-\mathbf{k}_2)\beta(\mathbf{k},\mathbf{k}_1,\mathbf{k}_2)\tilde{\theta}(\mathbf{k}_1,\tau)\tilde{\theta}(\mathbf{k}_2,\tau), \tag{2.4b}$$

($\delta_D$ denotes the three-dimensional Dirac delta distribution) where the functions

$$\alpha(\mathbf{k},\mathbf{k}_1) \equiv \frac{\mathbf{k}\cdot\mathbf{k}_1}{k_1^2}, \qquad \beta(\mathbf{k},\mathbf{k}_1,\mathbf{k}_2) \equiv \frac{k^2(\mathbf{k}_1\cdot\mathbf{k}_2)}{2k_1^2 k_2^2} \tag{2.5}$$

encode the non-linearity of the evolution (mode coupling) and come from the non-linear terms in the continuity equation (2.1a) and the Euler equation (2.1b) respectively. From equations (2.4) we see that the evolution of $\tilde{\delta}(\mathbf{k},\tau)$ and $\tilde{\theta}(\mathbf{k},\tau)$ is determined by the mode coupling of the fields at all pairs of wave-vectors $\mathbf{k}_1$ and $\mathbf{k}_2$ whose sum is $\mathbf{k}$, as required by translation invariance in a spatially homogeneous Universe [29].

### B. Perturbation Theory Solutions for $\Omega = 1$ and $\Lambda = 0$

Equations (2.4) are very difficult to solve in general, since they are coupled integro-differential equations with no small dimensionless parameter. The perturbative approach to the problem is to temporarily introduce a small parameter $\lambda$ on the RHS of (2.4) to organize a perturbative expansion and then let $\lambda \to 1$ at the end of the calculation [55]. This can be done by taking the fields to be $\mathcal{O}(\lambda)$ and expanding the general solution for $\delta$ and $\theta$ in powers of $\lambda$. The coefficients of these series can be determined at each order recursively as a function of the lower order ones. Ideally one would then like to sum the whole series to recover the solution to the original problem. That is of course not possible in practice, so by keeping the first few terms one hopes to get a good approximation to the answer. Since the small parameter $\lambda$ is related to the rms density fluctuations, the whole approach is expected to break down when these fluctuations become of order one. In this regime, all the terms in the perturbative series are of the same order and the perturbation expansion does not make sense without a resummation.

We consider a matter-dominated Einstein-de Sitter Universe, for which $\Omega = 1$ and $\Lambda = 0$. From Eq. (2.2a) we obtain $a \propto \tau^2$ and $3\Omega\mathcal{H}^2/2 = 6/\tau^2$. Therefore, Eqs. (2.4) become homogeneous in $\tau$ and we can formally solve them with the following perturbative expansion [5,28,29]:

$$\tilde{\delta}(\mathbf{k},\tau) = \sum_{n=1}^{\infty} a^n(\tau)\delta_n(\mathbf{k}), \qquad \tilde{\theta}(\mathbf{k},\tau) = \mathcal{H}(\tau)\sum_{n=1}^{\infty} a^n(\tau)\theta_n(\mathbf{k}), \tag{2.6}$$

where only the fastest growing mode is taken into account. At small $a$, the series are dominated by their first terms, and since $\theta_1(\mathbf{k}) = -\delta_1(\mathbf{k})$ from the continuity equation, $\delta_1(\mathbf{k})$ completely characterizes the linear fluctuations. The equations of motion (2.4) determine $\delta_n(\mathbf{k})$ and $\theta_n(\mathbf{k})$ in terms of the linear fluctuations,

$$\delta_n(\mathbf{k}) = \int d^3q_1 \ldots \int d^3q_n \delta_D(\mathbf{k}-\mathbf{q}_1-\ldots-\mathbf{q}_n) F_n^{(s)}(\mathbf{q}_1,\ldots,\mathbf{q}_n)\delta_1(\mathbf{q}_1)\ldots\delta_1(\mathbf{q}_n), \tag{2.7a}$$

$$\theta_n(\mathbf{k}) = -\int d^3q_1 \ldots \int d^3q_n \delta_D(\mathbf{k}-\mathbf{q}_1-\ldots-\mathbf{q}_n) G_n^{(s)}(\mathbf{q}_1,\ldots,\mathbf{q}_n)\delta_1(\mathbf{q}_1)\ldots\delta_1(\mathbf{q}_n), \tag{2.7b}$$

where the kernels $F_n^{(s)}$ and $G_n^{(s)}$ are symmetric homogeneous functions of the wave vectors $\{\mathbf{q}_1,\ldots,\mathbf{q}_n\}$ with degree zero. They are constructed from the fundamental mode coupling functions $\alpha(\mathbf{k},\mathbf{k}_1)$ and $\beta(\mathbf{k},\mathbf{k}_1,\mathbf{k}_2)$ according to the recursion relations ($n \geq 2$, see [5,29] for a derivation):

$$F_n(\mathbf{q}_1,\ldots,\mathbf{q}_n) = \sum_{m=1}^{n-1} \frac{G_m(\mathbf{q}_1,\ldots,\mathbf{q}_m)}{(2n+3)(n-1)}\left[(2n+1)\alpha(\mathbf{k},\mathbf{k}_1)F_{n-m}(\mathbf{q}_{m+1},\ldots,\mathbf{q}_n) + 2\beta(\mathbf{k},\mathbf{k}_1,\mathbf{k}_2)G_{n-m}(\mathbf{q}_{m+1},\ldots,\mathbf{q}_n)\right],$$

(2.8a)

$$G_n(\mathbf{q}_1, \ldots, \mathbf{q}_n) = \sum_{m=1}^{n-1} \frac{G_m(\mathbf{q}_1, \ldots, \mathbf{q}_m)}{(2n+3)(n-1)} \Big[ 3\alpha(\mathbf{k}, \mathbf{k}_1) F_{n-m}(\mathbf{q}_{m+1}, \ldots, \mathbf{q}_n) + 2n\beta(\mathbf{k}, \mathbf{k}_1, \mathbf{k}_2) G_{n-m}(\mathbf{q}_{m+1}, \ldots, \mathbf{q}_n) \Big],$$

(2.8b)

(where $\mathbf{k}_1 \equiv \mathbf{q}_1 + \ldots + \mathbf{q}_m$, $\mathbf{k}_2 \equiv \mathbf{q}_{m+1} + \ldots + \mathbf{q}_n$, $\mathbf{k} \equiv \mathbf{k}_1 + \mathbf{k}_2$, and $F_1 = G_1 \equiv 1$) and the symmetrization procedure:

$$F_n^{(s)}(\mathbf{q}_1, \ldots, \mathbf{q}_n) = \frac{1}{n!} \sum_\pi F_n(\mathbf{q}_{\pi(1)}, \ldots, \mathbf{q}_{\pi(n)}),$$

(2.9a)

$$G_n^{(s)}(\mathbf{q}_1, \ldots, \mathbf{q}_n) = \frac{1}{n!} \sum_\pi G_n(\mathbf{q}_{\pi(1)}, \ldots, \mathbf{q}_{\pi(n)}),$$

(2.9b)

where the sum is taken over all the permutations $\pi$ of the set $\{1, \ldots, n\}$. For example, for $n=2$ we have:

$$F_2^{(s)}(\mathbf{q}_1, \mathbf{q}_2) = \frac{5}{7} + \frac{1}{2} \frac{\mathbf{q}_1 \cdot \mathbf{q}_2}{q_1 q_2} \left( \frac{q_1}{q_2} + \frac{q_2}{q_1} \right) + \frac{2}{7} \frac{(\mathbf{q}_1 \cdot \mathbf{q}_2)^2}{q_1^2 q_2^2},$$

(2.10a)

$$G_2^{(s)}(\mathbf{q}_1, \mathbf{q}_2) = \frac{3}{7} + \frac{1}{2} \frac{\mathbf{q}_1 \cdot \mathbf{q}_2}{q_1 q_2} \left( \frac{q_1}{q_2} + \frac{q_2}{q_1} \right) + \frac{4}{7} \frac{(\mathbf{q}_1 \cdot \mathbf{q}_2)^2}{q_1^2 q_2^2}.$$

(2.10b)

The complexity of the symmetrized kernels increases rapidly with $n$. For example, the number of terms in $F_n^{(s)}$ and $G_n^{(s)}$ is 134 each for $n=3$ and 8523 for $n=4$. Explicit expressions for the unsymmetrized kernels $F_3$ and $F_4$ are given in [5]. The perturbation theory kernels have the following properties [5,56]:

1. As $\mathbf{k} = \mathbf{q}_1 + \ldots + \mathbf{q}_n$ goes to zero, but the individual $\mathbf{q}_i$ do not, $F_n^{(s)} \propto k^2$. This is a consequence of momentum conservation in center of mass coordinates.

2. As some of the arguments of $F_n^{(s)}$ or $G_n^{(s)}$ get large but the total sum $\mathbf{k} = \mathbf{q}_1 + \ldots + \mathbf{q}_n$ stays fixed, the kernels vanish in inverse square law. That is, for $p \gg q_i$, we have:

$$F_n^{(s)}(\mathbf{q}_1, \ldots, \mathbf{q}_{n-2}, \mathbf{p}, -\mathbf{p}) \approx G_n^{(s)}(\mathbf{q}_1, \ldots, \mathbf{q}_{n-2}, \mathbf{p}, -\mathbf{p}) \propto k^2/p^2.$$

(2.11)

3. If one of the arguments $\mathbf{q}_i$ of $F_n^{(s)}$ or $G_n^{(s)}$ goes to zero, there is an infrared divergence of the form $\mathbf{q}_i/q_i^2$. This comes from the infrared behavior of the mode coupling functions $\alpha(\mathbf{k}, \mathbf{k}_1)$ and $\beta(\mathbf{k}, \mathbf{k}_1, \mathbf{k}_2)$. There are no infrared divergences as partial sums of several wavevectors go to zero.

### C. Perturbation Theory Kernels for Non-Linear Approximations

When the fluctuations become strongly non-linear, perturbation theory breaks down, and one has to resort to N-body simulations to study the subsequent evolution. On the other hand, several non-linear approximation schemes have been proposed in the literature which allow analytic calculations beyond the domain of linear perturbation theory. These are approximations in the sense that they replace the Poisson equation by a given ansatz which is true only in linear theory for the exact dynamics [52], and therefore they are neither exact nor asymptotic to the exact solution beyond linear order (except for special cases). In this section we derive the perturbation theory kernels in Fourier space for the Einstein-de Sitter model for these approximations. Similar analysis has been done in real space for second order perturbation theory in [52] and up to third order in [51].

### 1. Zel'dovich Approximation (ZA)

In this approximation [46,57], the motion of each particle is given by its initial Lagrangian displacement: the dynamics of fluid elements is therefore governed purely by "inertia". In Eulerian space, this is equivalent to replacing the Poisson equation by the ansatz [52]:

$$\mathbf{v}(\mathbf{x}, \tau) = -\frac{2}{3\mathcal{H}(\tau)} \nabla \Phi(\mathbf{x}, \tau), \quad (2.12)$$

which is the relation between velocity and gravitational potential valid in linear theory. The important point about ZA is that a small perturbation in Lagrangian fluid element paths carries a large amount of non-linear information about the the corresponding Eulerian quantities, since the Lagrangian picture is intrinsically non-linear in the density field. This leads to non-zero Eulerian perturbation theory kernels at every order. ZA works reasonably well as long as streamlines of flows do not cross each other. However, multistreaming develops at the location of pancakes leading to the breakdown of ZA [57]. The equations of motion in Fourier space are:

$$\frac{\partial \tilde{\delta}(\mathbf{k}, \tau)}{\partial \tau} + \tilde{\theta}(\mathbf{k}, \tau) = -\int d^3k_1 \int d^3k_2 \delta_D(\mathbf{k} - \mathbf{k}_1 - \mathbf{k}_2) \alpha(\mathbf{k}, \mathbf{k}_1) \tilde{\theta}(\mathbf{k}_1, \tau) \tilde{\delta}(\mathbf{k}_2, \tau), \quad (2.13a)$$

$$\frac{\partial \tilde{\theta}(\mathbf{k}, \tau)}{\partial \tau} - \frac{\mathcal{H}(\tau)}{2} \tilde{\theta}(\mathbf{k}, \tau) = -\int d^3k_1 \int d^3k_2 \delta_D(\mathbf{k} - \mathbf{k}_1 - \mathbf{k}_2) \beta(\mathbf{k}, \mathbf{k}_1, \mathbf{k}_2) \tilde{\theta}(\mathbf{k}_1, \tau) \tilde{\theta}(\mathbf{k}_2, \tau) \quad . \quad (2.13b)$$

These equations, together with the perturbative expansion (2.6), lead to the recursion relations ($n \geq 2$):

$$F_n^Z(\mathbf{q}_1, \ldots, \mathbf{q}_n) = \sum_{m=1}^{n-1} G_m^Z(\mathbf{q}_1, \ldots, \mathbf{q}_m) \left[ \frac{\alpha(\mathbf{k}, \mathbf{k}_1)}{n} F_{n-m}^Z(\mathbf{q}_{m+1}, \ldots, \mathbf{q}_n) + \frac{\beta(\mathbf{k}, \mathbf{k}_1, \mathbf{k}_2)}{n(n-1)} G_{n-m}^Z(\mathbf{q}_{m+1}, \ldots, \mathbf{q}_n) \right], \quad (2.14a)$$

$$G_n^Z(\mathbf{q}_1, \ldots, \mathbf{q}_n) = \sum_{m=1}^{n-1} G_m^Z(\mathbf{q}_1, \ldots, \mathbf{q}_m) \frac{\beta(\mathbf{k}, \mathbf{k}_1, \mathbf{k}_2)}{(n-1)} G_{n-m}^Z(\mathbf{q}_{m+1}, \ldots, \mathbf{q}_n), \quad (2.14b)$$

In this case, we can factor out the wave vector dependence in the denominator by defining:

$$F_n^Z(\mathbf{q}_1, \ldots, \mathbf{q}_n) \equiv \frac{1}{q_1^2 \ldots q_n^2} \mathcal{F}_n^Z(\mathbf{q}_1, \ldots, \mathbf{q}_n), \quad (2.15a)$$

$$G_n^Z(\mathbf{q}_1, \ldots, \mathbf{q}_n) \equiv \frac{k^2}{q_1^2 \ldots q_n^2} \mathcal{G}_n^Z(\mathbf{q}_1, \ldots, \mathbf{q}_n), \quad (2.15b)$$

where, again, $\mathbf{k} = \mathbf{q}_1 + \ldots \mathbf{q}_n$. The recursion relations become:

$$n \mathcal{F}_n^Z(\mathbf{q}_1, \ldots, \mathbf{q}_n) = \sum_{m=1}^{n-1} \mathcal{G}_m^Z(\mathbf{q}_1, \ldots, \mathbf{q}_m)(\mathbf{k} \cdot \mathbf{k}_1) \mathcal{F}_{n-m}^Z(\mathbf{q}_{m+1}, \ldots, \mathbf{q}_n) + k^2 \mathcal{G}_n^Z(\mathbf{q}_1, \ldots, \mathbf{q}_n), \quad (2.16a)$$

$$2(n-1) \mathcal{G}_n^Z(\mathbf{q}_1, \ldots, \mathbf{q}_n) = \sum_{m=1}^{n-1} \mathcal{G}_m^Z(\mathbf{q}_1, \ldots, \mathbf{q}_m)(\mathbf{k}_1 \cdot \mathbf{k}_2) \mathcal{G}_{n-m}^Z(\mathbf{q}_{m+1}, \ldots, \mathbf{q}_n), \quad (2.16b)$$

where $\mathcal{G}_1^Z \equiv 1$ and $\mathcal{F}_1^Z(\mathbf{q}) \equiv q^2$. From these equations it is clear that all the wave-vector dependence in the denominator of the kernels has been factored out in (2.15) and is not present in $\mathcal{G}$ or $\mathcal{F}$. In fact, after symmetrization, the density field kernel takes the simple form [50]:

$$F_n^{Z(s)}(\mathbf{q}_1, \ldots, \mathbf{q}_n) = \frac{1}{n!} \frac{(\mathbf{k} \cdot \mathbf{q}_1)}{q_1^2} \ldots \frac{(\mathbf{k} \cdot \mathbf{q}_n)}{q_n^2}. \quad (2.17)$$

Note that this expression means that in ZA, property (1) for $F_n^{(s)}$ given in the previous section is changed to $F_n^{Z(s)} \propto k^n$. So, in ZA, higher order kernels are underestimated for large scales compared to ED. For one-dimensional perturbations, the ZA becomes exact, and therefore its symmetrized kernels (where dot products are replaced by ordinary multiplication) agree with the symmetrized kernels of the exact dynamics given in Section II B. In 3 dimensions, for $n = 2$ we have (compare (2.10)):

$$F_2^{Z(s)}(\mathbf{q}_1, \mathbf{q}_2) = \frac{1}{2} + \frac{1}{2}\frac{\mathbf{q}_1 \cdot \mathbf{q}_2}{q_1 q_2}\left(\frac{q_1}{q_2} + \frac{q_2}{q_1}\right) + \frac{1}{2}\frac{(\mathbf{q}_1 \cdot \mathbf{q}_2)^2}{q_1^2 q_2^2}, \tag{2.18a}$$

$$G_2^{Z(s)}(\mathbf{q}_1, \mathbf{q}_2) = \frac{1}{2}\frac{\mathbf{q}_1 \cdot \mathbf{q}_2}{q_1 q_2}\left(\frac{q_1}{q_2} + \frac{q_2}{q_1}\right) + \frac{1}{2}\frac{(\mathbf{q}_1 \cdot \mathbf{q}_2)^2}{q_1^2 q_2^2}. \tag{2.18b}$$

In this case, the number of terms in $F_n^{Z(s)}$ and $G_n^{Z(s)}$ increases more slowly than in the exact dynamics kernels. There are 26 (16) terms in $F_3^{Z(s)}$ ($G_3^{Z(s)}$) and 237 (127) in $F_4^{Z(s)}$ ($G_4^{Z(s)}$).

### 2. Linear Potential Approximation (LPA)

In this approximation [48,49], the gravitational potential is substituted by its linear value. Therefore, the full Poisson equation (2.1c) is replaced by:

$$\nabla^2 \Phi(\mathbf{x}, \tau) = \frac{3}{2}\mathcal{H}^2(\tau)\delta_1(\mathbf{x}, \tau), \tag{2.19}$$

where $\delta_1(\mathbf{x}, \tau) = a(\tau)\delta_1(\mathbf{x})$ is the solution to the linearized equations of motion. The idea behind this approximation is that since $\Phi \propto \delta/k^2$, the gravitational potential is dominated by long-wavelength modes more than the density field, and therefore ought to obey linear perturbation theory longer. In fact, N-body simulations show that $\Phi$ evolves much more slowly than the density field [48]. The equations of motion in Fourier space are:

$$\frac{\partial \tilde{\delta}(\mathbf{k}, \tau)}{\partial \tau} + \tilde{\theta}(\mathbf{k}, \tau) = -\int d^3 k_1 \int d^3 k_2 \delta_D(\mathbf{k} - \mathbf{k}_1 - \mathbf{k}_2)\alpha(\mathbf{k}, \mathbf{k}_1)\tilde{\theta}(\mathbf{k}_1, \tau)\tilde{\delta}(\mathbf{k}_2, \tau), \tag{2.20a}$$

$$\frac{\partial \tilde{\theta}(\mathbf{k}, \tau)}{\partial \tau} + \mathcal{H}(\tau)\tilde{\theta}(\mathbf{k}, \tau) + \frac{3}{2}\mathcal{H}^2(\tau)\tilde{\delta}_1(\mathbf{k}, \tau) = -\int d^3 k_1 \int d^3 k_2 \delta_D(\mathbf{k} - \mathbf{k}_1 - \mathbf{k}_2)\beta(\mathbf{k}, \mathbf{k}_1, \mathbf{k}_2)\tilde{\theta}(\mathbf{k}_1, \tau)\tilde{\theta}(\mathbf{k}_2, \tau), \tag{2.20b}$$

leading to the following recursion relations ($n \geq 2$):

$$F_n^{LP}(\mathbf{q}_1, \ldots, \mathbf{q}_n) = \sum_{m=1}^{n-1} G_m^{LP}(\mathbf{q}_1, \ldots, \mathbf{q}_m)\left[\frac{\alpha(\mathbf{k}, \mathbf{k}_1)}{n}F_{n-m}^{LP}(\mathbf{q}_{m+1}, \ldots, \mathbf{q}_n) + \frac{2\beta(\mathbf{k}, \mathbf{k}_1, \mathbf{k}_2)}{n(2n+1)}G_{n-m}^{LP}(\mathbf{q}_{m+1}, \ldots, \mathbf{q}_n)\right], \tag{2.21a}$$

$$G_n^{LP}(\mathbf{q}_1, \ldots, \mathbf{q}_n) = \sum_{m=1}^{n-1} G_m^{LP}(\mathbf{q}_1, \ldots, \mathbf{q}_m)\frac{2\beta(\mathbf{k}, \mathbf{k}_1, \mathbf{k}_2)}{(2n+1)}G_{n-m}^{LP}(\mathbf{q}_{m+1}, \ldots, \mathbf{q}_n). \tag{2.21b}$$

Making the analogous definitions to (2.15), we find:

$$n\mathcal{F}_n^{LP}(\mathbf{q}_1, \ldots, \mathbf{q}_n) = \sum_{m=1}^{n-1} \mathcal{G}_m^{LP}(\mathbf{q}_1, \ldots, \mathbf{q}_m)(\mathbf{k} \cdot \mathbf{k}_1)\mathcal{F}_{n-m}^{LP}(\mathbf{q}_{m+1}, \ldots, \mathbf{q}_n) + k^2 \mathcal{G}_n^{LP}(\mathbf{q}_1, \ldots, \mathbf{q}_n), \tag{2.22a}$$

$$(2n+1)\mathcal{G}_n^{LP}(\mathbf{q}_1, \ldots, \mathbf{q}_n) = \sum_{m=1}^{n-1} \mathcal{G}_m^{LP}(\mathbf{q}_1, \ldots, \mathbf{q}_m)(\mathbf{k}_1 \cdot \mathbf{k}_2)\mathcal{G}_{n-m}^{LP}(\mathbf{q}_{m+1}, \ldots, \mathbf{q}_n). \tag{2.22b}$$

Note that (2.22a) is exactly equivalent to its Zel'dovich approximation counterpart (2.15a), whereas (2.22b) has a different numerical coefficient from (2.16b). For $n = 2$ we have:

$$F_2^{LP(s)}(\mathbf{q}_1, \mathbf{q}_2) = \frac{1}{2} + \frac{7}{20} \frac{\mathbf{q}_1 \cdot \mathbf{q}_2}{q_1 q_2} \left( \frac{q_1}{q_2} + \frac{q_2}{q_1} \right) + \frac{1}{5} \frac{(\mathbf{q}_1 \cdot \mathbf{q}_2)^2}{q_1^2 q_2^2}, \tag{2.23a}$$

$$G_2^{LP(s)}(\mathbf{q}_1, \mathbf{q}_2) = \frac{1}{5} \frac{\mathbf{q}_1 \cdot \mathbf{q}_2}{q_1 q_2} \left( \frac{q_1}{q_2} + \frac{q_2}{q_1} \right) + \frac{2}{5} \frac{(\mathbf{q}_1 \cdot \mathbf{q}_2)^2}{q_1^2 q_2^2}. \tag{2.23b}$$

The number of terms in the symmetrized kernels for $n > 2$ is equal to the ZA case.

### 3. Frozen Flow Approximation (FFA)

In this approximation [47] the velocity field is assumed to remain linear. Therefore, we have the relation:

$$\theta(\mathbf{x}, \tau) = \theta_1(\mathbf{x}, \tau) = -\mathcal{H}(\tau) \delta_1(\mathbf{x}, \tau). \tag{2.24}$$

Streamlines are kept frozen to their initial configuration, therefore the dynamics remains forever in the single-stream regime. The equation of motion for $\tilde{\delta}(\mathbf{k}, \tau)$ is:

$$\frac{\partial \tilde{\delta}(\mathbf{k}, \tau)}{\partial \tau} - \mathcal{H}(\tau) \delta_1(\mathbf{k}, \tau) = \mathcal{H}(\tau) \int d^3 k_1 \int d^3 k_2 \delta_D(\mathbf{k} - \mathbf{k}_1 - \mathbf{k}_2) \alpha(\mathbf{k}, \mathbf{k}_1) \tilde{\delta}_1(\mathbf{k}_1, \tau) \tilde{\delta}(\mathbf{k}_2, \tau), \tag{2.25}$$

where $\delta_1(\mathbf{k}, \tau) = a(\tau) \delta_1(\mathbf{k})$. This leads to the perturbation theory kernels ($n \geq 2$):

$$F_n^{FF}(\mathbf{q}_1, \ldots, \mathbf{q}_n) = \frac{1}{n!} \frac{(\mathbf{k} \cdot \mathbf{q}_1)}{q_1^2} \frac{[(\mathbf{k} - \mathbf{q}_1) \cdot \mathbf{q}_2]}{q_2^2} \cdots \frac{(\mathbf{q}_n \cdot \mathbf{q}_n)}{q_n^2}, \tag{2.26a}$$

$$G_n^{FF}(\mathbf{q}_1, \ldots, \mathbf{q}_n) = 0. \tag{2.26b}$$

For $n = 2$, we have:

$$F_2^{FF(s)}(\mathbf{q}_1, \mathbf{q}_2) = \frac{1}{2} + \frac{1}{4} \frac{\mathbf{q}_1 \cdot \mathbf{q}_2}{q_1 q_2} \left( \frac{q_1}{q_2} + \frac{q_2}{q_1} \right). \tag{2.27}$$

In this case, the number of terms in $F_n^{FF(s)}$ is 16 and 125 for $n = 3$ and $n = 4$, respectively.

It is worth noticing that all these non-linear approximations involve neglecting "$\alpha$ coupling" (for ZA and LPA and also "$\beta$ coupling" for FFA) in the velocity divergence kernels, which comes from the fact that none of these approximations solves Poisson equation. This leads to the result that both kernels in these approximations do not involve scalar products of wave-vectors in their denominators. This fact will significantly simplify the calculation of 1-point cumulants.

### III. STATISTICAL DESCRIPTION OF FLUCTUATIONS

The starting point for a statistical description of fluctuations in cosmology is the "Fair Sample Hypothesis" [6,58]. This asserts that fluctuations can be described by homogeneous and isotropic random fields (so that our Universe is a random realization from a statistical ensemble) and that within the accessible part of the Universe there are many independent samples that can be considered to approximate a statistical ensemble, so that spatial averages are equivalent to ensemble averages ("ergodicity").

A full description of a set of random fields is provided by their joint probability distribution functional. However, except for special cases, the determination of this functional cannot be carried out exactly [20], and in practice one considers only the simplest statistical tools which describe some particular statistical property of the fluctuations. A very important set of functions are the moments of the probability functional. The (equal-time) $p^{\text{th}}$-order, $p$-point moment for the density field is defined by the ensemble average (denoted by angle brackets) of the product of $p$ fields at $p$ different points (in the following, we give formulas in real space, but analogous expressions hold in Fourier space; see, e.g., [59] for a general discussion):

$$\mu_p(\mathbf{x}_1, \ldots, \mathbf{x}_p) \equiv \langle \delta(\mathbf{x}_1) \ldots \delta(\mathbf{x}_p) \rangle, \tag{3.1}$$

where $\mu_p(\mathbf{x}_1, \ldots, \mathbf{x}_p) \equiv \xi(\mathbf{x}_1, \ldots, \mathbf{x}_p)$ is usually referred to as the $p$-point correlation function. However, in general one may have $p$-point moments of order $r$ ($\mu_r(\mathbf{x}_1, \ldots, \mathbf{x}_p)$ with $p \leq r$) when some of the points of evaluation of the fields coincide with each other. Note that in (3.1), evaluation of all the fields at the same instant is understood. The moments can be obtained from the moment generating functional $Z(J)$ by functional differentiation:

$$\mu_p(\mathbf{x}_1, \ldots, \mathbf{x}_p) = \left[ \frac{(-1)^p \delta^p Z[J(\mathbf{x})]}{\delta J(\mathbf{x}_1) \ldots \delta J(\mathbf{x}_p)} \right]_{J=0}, \tag{3.2}$$

where:

$$Z[J(\mathbf{x})] \equiv \left\langle \exp\left[ -\int d^3x\, \delta(\mathbf{x}) J(\mathbf{x}) \right] \right\rangle = \int \mathcal{D}\delta(\mathbf{x}) P[\delta(\mathbf{x})] \exp\left[ -\int d^3x\, \delta(\mathbf{x}) J(\mathbf{x}) \right], \tag{3.3}$$

and $P[\delta(\mathbf{x})]$ is the one-point probability distribution functional of the density field fluctuations.

Under very general conditions, when all the moments exist, they characterize the underlying probability functional [60,61]. However, they are not the only set of functions with this property, or even the best. One inconvenience is that each moment contains correlations already present at lower orders. For that reason, one defines another set of functions, the cumulants $\kappa_p$, which specify the non-trivial correlations at each order. These are given in terms of the recursion formula (together with the condition $\mu_1 = \kappa_1 = \langle \delta \rangle = 0$):

$$\mu_p(\mathbf{x}_1, \ldots, \mathbf{x}_p) = \kappa_p(\mathbf{x}_1, \ldots, \mathbf{x}_p) + \sum_{\cup \mathcal{I}_\alpha = \mathcal{I}} \prod_\alpha \kappa_{|\mathcal{I}_\alpha|}(\mathcal{I}_\alpha), \tag{3.4}$$

where $\mathcal{I} \equiv \{\mathbf{x}_1, \ldots, \mathbf{x}_p\}$, $|\mathcal{I}_\alpha|$ is the number of elements of the subset $\mathcal{I}_\alpha$, and we sum over all the possible partitions of the set of arguments excluding the identity partition. Since all the trivial correlations have been taken away, the cumulants vanish when any subset of $\mathcal{I}$ is removed to infinite separation. The $p^{\text{th}}$ order, $p$-point cumulant $\kappa_p(\mathbf{x}_1, \ldots, \mathbf{x}_p)$ is usually referred to as the connected correlation function (in analogy with the connected Green's functions in quantum field theory [62]) and also denoted as $\xi_c(\mathbf{x}_1, \ldots, \mathbf{x}_p)$ or $\langle \delta(\mathbf{x}_1) \ldots \delta(\mathbf{x}_p) \rangle_c$. For isotropic and homogeneous random fields, the cumulants depend only on the relative distances $|\mathbf{x}_i - \mathbf{x}_j|$ for $i,j = 1, \ldots, p$ ($i \neq j$). In Fourier space, homogeneity implies that the configuration in k-space is closed, that is $\sum_{i=1}^p \mathbf{k}_i = 0$, whereas isotropy implies that the cumulants are invariant under reorientation of this closed configuration. For a Gaussian probability distribution functional, $\kappa_p = 0$ for $p > 2$.

The cumulant expansion theorem states that the cumulant generating function is simply $\ln Z[J(\mathbf{x})]$ (the proof is easily done by induction, see e.g., [63]), therefore:

$$\kappa_p(\mathbf{x}_1, \ldots, \mathbf{x}_p) = \left[ \frac{(-1)^p \delta^p \ln Z[J(\mathbf{x})]}{\delta J(\mathbf{x}_1) \ldots \delta J(\mathbf{x}_p)} \right]_{J=0}, \tag{3.5}$$

which gives a practical way of calculating cumulants when the probability functional is known.

In this paper, we adopt the conventional assumption that the primordial fluctuations are Gaussian distributed. While not universal, this is a generic prediction of the simplest inflationary models, in which $\delta_1$ is linearly related to an essentially free scalar field (the fluctuation of the inflaton field) in its vacuum state, described by a Gaussian wavefunctional. For Gaussian fluctuations, $\ln Z[J(\mathbf{x})]$ is quadratic in $J(\mathbf{x})$, and therefore $\kappa_p$ vanishes for $p > 2$ for the linear fluctuations. Since the equations of motion of a self-gravitating fluid are non-linear, higher order ($p > 2$) cumulants will be induced by dynamical evolution, driving the probability functional away from Gaussianity [6]. Once this occurs, the cumulant generating functional is no longer known *a priori*, and (3.5) cannot be employed to evaluate the cumulants (see, however, [19] for a calculation of the cumulant generating functional when the rms density fluctuation is vanishingly small).

In this work we focus on the 1-point cumulants of cosmological fields induced by non-linear gravitational dynamics. The linear fluctuations are characterized by the second order cumulant, the variance $\sigma_\ell^2 = \langle \delta_1^2 \rangle_c$, or equivalently, by the linear power spectrum $P_1(k, \tau)$, defined by:

$$\left\langle \delta_1(\mathbf{k}) \delta_1(\mathbf{k}') \right\rangle_c = \delta_D(\mathbf{k} + \mathbf{k}') P_1(k, \tau). \tag{3.6}$$

From (3.4), for Gaussian initial conditions we have:

$$\left\langle \delta_1(\mathbf{k}_1)\ldots\delta_1(\mathbf{k}_{2p})\right\rangle = \frac{1}{2^p p!}\sum_\pi \left\langle \delta_1(\mathbf{k}_{\pi(1)})\delta_1(\mathbf{k}_{\pi(2)})\right\rangle_c \ldots \left\langle \delta_1(\mathbf{k}_{\pi(2p-1)})\delta_1(\mathbf{k}_{\pi(2p)})\right\rangle_c \quad (3.7\text{a})$$

$$\left\langle \delta_1(\mathbf{k}_1)\ldots\delta_1(\mathbf{k}_{2p+1})\right\rangle = 0 \quad (3.7\text{b})$$

Due to homogeneity, one-point cumulants are independent of the spatial coordinate, and in Fourier space we have:

$$\langle \delta^p(\mathbf{x})\rangle_c \equiv \langle \delta^p\rangle_c = \int d^3k_1\ldots d^3k_p \left\langle \delta(\mathbf{k}_1)\ldots\delta(\mathbf{k}_p)\right\rangle_c . \quad (3.8)$$

Thus the $p^{\text{th}}$ order, one-point cumulants are integrals of the $p$-point spectra over the $p$ wave-vectors. According to the perturbation expansion (2.6), $\langle \delta^p\rangle_c$ can be written as:

$$\langle \delta^p\rangle_c = \sum_{r=p-1}^\infty \sum_{\mathcal{C}(2r,p)} \left\langle \prod_{i=1}^p \delta_{\mathcal{C}_i}\right\rangle_c, \quad (3.9)$$

where we sum over the arrangements $\mathcal{C}(2r,p)$ of $p$ positive integers that sum up to $2r = 2p-2, 2p, \ldots$ (because of the assumption of gaussian initial conditions, $2p-2$ is the minimum order in $\delta_1$ that we need to get a connected contribution [7]) and $\mathcal{C}_i$ denotes the $i^{\text{th}}$ component of a particular arrangement (i.e. an integer between 1 and $2r-p+1$). In Eq. (3.9), terms labeled by $r$ are contributions to $\langle \delta^p\rangle_c$ of order $\delta_1^{2r}$. For example, for $p=2,3,4$ and working to $\mathcal{O}(\delta_1^{10})$ we have (putting contributions of same order within square brackets; here and below superscript $(n)$ denotes the $n$-loop contribution):

$$\sigma^2 \equiv \langle \delta^2\rangle_c = \sigma_\ell^2\left(1 + s^{(1)}\sigma_\ell^2 + s^{(2)}\sigma_\ell^4 + s^{(3)}\sigma_\ell^6\right) + \mathcal{O}(\sigma_\ell^{10}), \quad (3.10\text{a})$$

$$s^{(1)} \equiv \left[2\langle \delta_1\delta_3\rangle_c + \langle \delta_2^2\rangle_c\right], \quad (3.10\text{b})$$

$$s^{(2)} \equiv \left[2\langle \delta_1\delta_5\rangle_c + 2\langle \delta_2\delta_4\rangle_c + \langle \delta_3^2\rangle_c\right], \quad (3.10\text{c})$$

$$s^{(3)} \equiv \left[2\langle \delta_1\delta_7\rangle_c + 2\langle \delta_2\delta_6\rangle_c + 2\langle \delta_3\delta_5\rangle_c + \langle \delta_4^2\rangle_c\right], \quad (3.10\text{d})$$

$$\langle \delta^3\rangle_c = 3\langle \delta_1^2\delta_2\rangle_c + \left[3\langle \delta_1^2\delta_4\rangle_c + 6\langle \delta_1\delta_2\delta_3\rangle_c + \langle \delta_2^3\rangle_c\right] + \left[3\langle \delta_1^2\delta_6\rangle_c + 6\langle \delta_1\delta_2\delta_5\rangle_c + 6\langle \delta_1\delta_3\delta_4\rangle_c + 3\langle \delta_2^2\delta_4\rangle_c \right.$$
$$\left. + 3\langle \delta_2\delta_3^2\rangle_c\right] + \mathcal{O}(\sigma_\ell^{10}), \quad (3.11)$$

$$\langle \delta^4\rangle_c = \left[4\langle \delta_1^3\delta_3\rangle_c + 6\langle \delta_1^2\delta_2^2\rangle_c\right] + \left[4\langle \delta_1^3\delta_5\rangle_c + 12\langle \delta_1^2\delta_2\delta_4\rangle_c + 6\langle \delta_1^2\delta_3^2\rangle_c + 12\langle \delta_1\delta_2^2\delta_3\rangle_c + \langle \delta_2^4\rangle_c\right] + \mathcal{O}(\sigma_\ell^{10}), \quad (3.12)$$

where the numerical factors come from the multinomial expansion of $\delta^p$ in $\delta_n$'s, and $\sigma_\ell$ is the linear rms density fluctuation given in terms of the linear power spectrum by:

$$\sigma_\ell^2 \equiv \langle \delta_1^2\rangle_c = \int d^3k\, P_1(k,\tau) \quad (3.13)$$

In view of (3.9), a convenient way to measure the departure from gaussianity due to dynamical evolution is to introduce the $S_p$ parameters ($p \geq 3$) defined by rescaling the 1-point cumulants [5]:

$$S_p \equiv \frac{\langle \delta^p\rangle_c}{\langle \delta^2\rangle_c^{p-1}}. \quad (3.14)$$

Therefore, the $S_p$ parameters are constants and independent of the normalization of the linear power spectrum to *lowest order* in perturbation theory. Note that since the $S_p$'s are constructed directly from one-point cumulants, each of these parameters contains statistical information independent of the others. $S_3$ is a measure of the skewness of the probability distribution (a positive value generally indicates that the upper tail of the distribution is the higher), whereas $S_4$ is a measure of the kurtosis (i.e., how "flat" or "peaked" the distribution is compared to a gaussian) [61]. From (3.10) we have

$$S_3 = S_3^{(0)} + S_3^{(1)} \sigma_\ell^2 + S_3^{(2)} \sigma_\ell^4 + \mathcal{O}(\sigma_\ell^6), \tag{3.15a}$$

$$S_4 = S_4^{(0)} + S_4^{(1)} \sigma_\ell^2 + \mathcal{O}(\sigma_\ell^4), \tag{3.15b}$$

where:

$$S_3^{(0)} \equiv \frac{3\langle \delta_1^2 \delta_2 \rangle_c}{\sigma_\ell^4}, \qquad S_3^{(1)} \equiv \frac{3\langle \delta_1^2 \delta_4 \rangle_c + 6\langle \delta_1 \delta_2 \delta_3 \rangle_c + \langle \delta_2^3 \rangle_c}{\sigma_\ell^6} - 2s^{(1)} S_3^{(0)}, \tag{3.16a}$$

$$S_3^{(2)} \equiv \frac{3\langle \delta_1^2 \delta_6 \rangle_c + 6\langle \delta_1 \delta_2 \delta_5 \rangle_c + 6\langle \delta_1 \delta_3 \delta_4 \rangle_c + 3\langle \delta_2^2 \delta_4 \rangle_c + 3\langle \delta_2 \delta_3^2 \rangle_c}{\sigma_\ell^8} - 2s^{(1)} S_3^{(1)} - \left( (s^{(1)})^2 + 2s^{(2)} \right) S_3^{(0)}, \tag{3.16b}$$

$$S_4^{(0)} \equiv \frac{4\langle \delta_1^3 \delta_3 \rangle_c + 6\langle \delta_1^2 \delta_2^2 \rangle_c}{\sigma_\ell^6}, \qquad S_4^{(1)} \equiv \frac{4\langle \delta_1^3 \delta_5 \rangle_c + 12\langle \delta_1^2 \delta_2 \delta_4 \rangle_c + 6\langle \delta_1^2 \delta_3^2 \rangle_c + 12\langle \delta_1 \delta_2^2 \delta_3 \rangle_c + \langle \delta_2^4 \rangle_c}{\sigma_\ell^8} - 3s^{(1)} S_4^{(0)}, \tag{3.16c}$$

and $s^{(1)}$ and $s^{(2)}$ come from the next-to-leading order corrections to $\langle \delta^2 \rangle_c$ in the denominator of Eq. (3.14) (see Eq. (3.10a)). To calculate the necessary ensemble averages, we use (2.7) to relate $\delta_n$ to the linear fluctuations and then ergodicity to relate averages over the present probability distribution to averages over the initial gaussian ensemble. In general we have:

$$\langle \delta^p \rangle_c = \sum_{r=p-1}^{\infty} \sum_{\mathcal{C}(2r,p)} \int d^3q_1 \ldots \int d^3q_{2r} \left[ F_{\mathcal{C}_1}^{(s)}(\mathbf{q}_1, \ldots \mathbf{q}_{\mathcal{C}_1}) \ldots F_{\mathcal{C}_p}^{(s)}(\mathbf{q}_{2r-\mathcal{C}_p}, \ldots \mathbf{q}_{2r}) \langle \delta_1(\mathbf{q}_1) \ldots \delta_1(\mathbf{q}_{2r}) \rangle \right]_c, \tag{3.17}$$

The next step is to reduce the moments of linear fluctuations in (3.17) according to their gaussian nature given by (3.7) and then take into account only the connected contributions to the integrals. The complexity of this procedure can be easily handled by using diagrammatic techniques.

## IV. DIAGRAMMATIC APPROACH TO PERTURBATION THEORY

### A. Diagrammatic Rules

A systematic framework for calculating correlations of cosmological fields in perturbation theory has been formulated using diagrammatic techniques similar to those in quantum field theory [62] and statistical mechanics [63]. In this approach [5,56], contributions to $p$-point cumulants of the density field come from connected diagrams with $p$ external (solid) lines and $r = p - 1, p, \ldots$ internal (dashed) lines. The perturbation expansion (3.17) leads to a collection of diagrams at each order, the leading order being tree-diagrams, the next to leading order 1-loop diagrams and so on. On the other hand, Eqns. (3.10), (3.15) indicate that the perturbation series for the one-point cumulants can be viewed as an expansion in powers of $\sigma_\ell^2$. Thus, in analogy with quantum field theory, $\sigma_\ell^2$ is the effective 'coupling constant' for the loop expansion, and we expect perturbative results to be sensible for small coupling, $\sigma_\ell^2 \ll 1$.

In each diagram, external lines represent the spectral components of the fields we are interested in (e.g., $\delta(\mathbf{k}, \tau)$). Each internal line is labeled by a wave-vector that is integrated over, and represents a linear power spectrum $P_1(q, \tau)$. Vertices of order $n$ (i.e., where $n$ internal lines join) represent a $n^{\text{th}}$ order perturbative solution $\delta_n$, and momentum conservation is imposed at each vertex. Figure 1 shows the factors associated with vertices and internal lines. To find the contribution of order $2r$ to the $p$-point spectrum of the density field proceed as follows:

- Draw all distinct connected diagrams containing $p$ vertices (with external lines labeled by $\mathbf{k}_1 \ldots \mathbf{k}_p$) joined by $r$ internal lines. Two diagrams are distinct if they cannot be deformed into each other by moving the vertices and lines without cutting any internal lines (sliding lines over other lines is allowed in the rearrangement process). For each of these diagrams:

1. Assign a factor of $\delta_D(\mathbf{k}_i - \mathbf{q}_1 - \ldots - \mathbf{q}_n) F_n^{(s)}(\mathbf{q}_1, \ldots, \mathbf{q}_n)$ to each vertex of order $n$ and external momentum $\mathbf{k}_i$ ($i = 1, \ldots, p$). For the arguments of $F_n^{(s)}$, we use the convention of assigning a positive sign to wave-vectors outgoing from the vertex.

2. Assign a factor of $P_1(q_j, \tau)$ to each internal line labeled by $\mathbf{q}_j$.

3. Integrate over all $\mathbf{q}_j$ ($j = 1, \ldots, r$).

4. Multiply by the symmetry factor of the graph, which is the number of permutations of linear fluctuations ($\delta_1$'s) that leaves the graph invariant.

5. Sum over distinct labelings of external lines, thus generating $p!/(n_1! \ldots n_{2r-p+1}!)$ diagrams (where $n_i$ denotes the number of vertices of order $i$).

- Add up the resulting expressions for all these diagrams.

To calculate 1-point cumulants, we have to integrate further over the $\mathbf{k}_i$ ($i = 1, \ldots p$) according to (3.8). These integrations are trivial because of the presence of delta functions given by rule 1, so we are left only with integrations over the $\mathbf{q}_j$'s. Also, since the $p!/(n_1! \ldots n_{2r-p+1}!)$ diagrams generated by rule 5 become equal contributions when the integration over external lines is performed, to find the contribution of order $2r$ to the $p^{th}$-order 1-point cumulant of the density field we replace rule 5 by the following:

- 5a. Integrate over $\mathbf{k}_i$ ($i = 1, \ldots p$) and multiply by the multinomial weight $p!/(n_1! \ldots n_{2r-p+1}!)$.

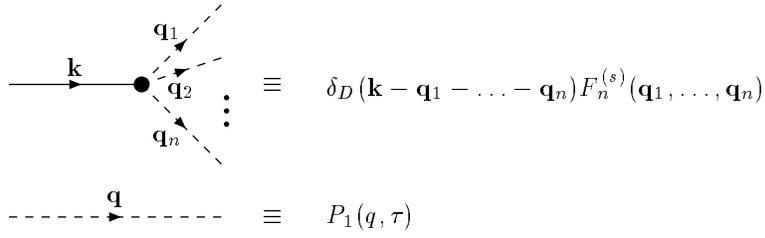

FIG. 1. Diagrammatic rules for vertices and internal lines for density field fluctuations. Equivalent rules hold for the velocity divergence replacing $F_n^{(s)}$ by $G_n^{(s)}$.

A few comments about these rules are in order. Rule 1 basically assigns a perturbation theory kernel of order $n$ to each vertex of that order. Rule 2 is a consequence of the assumed gaussian nature of the linear fluctuations, and describes the fact that the $\delta_1$'s coming out from each vertex "interact" with each other only through two-point functions. Rule 3 (together with rule 1) says that contributions to each external line with wave-vector $\mathbf{k}_i$ come from the mode coupling of the linear fields at wave vectors $\mathbf{q}_j$ whose sum is $\mathbf{k}_i$, as required by translation invariance. Rule 4 counts the number of ways we can join $\delta_1$'s into two-point correlations. Rule 5a comes from the multinomial expansion of $\delta^p$ into $\delta_n$'s according to the perturbation expansion (2.6). The same diagrammatic rules hold for the divergence of the velocity field, replacing $F_n^{(s)}$ by $G_n^{(s)}$. Using both sets of rules, one can construct diagrammatic expressions which represent contributions to arbitrary cross-correlations between density and divergence of the velocity fields. Finally, we note that these rules are somewhat unusual from the viewpoint of field theory, since new vertices appear at each order of perturbation theory.

It is convenient to introduce some bookkeeping notation in order to be able to refer to different diagrams. We shall denote by $\mathcal{D}_{i\ldots j}$ the amplitude given by the above rules for a diagram representing the contribution to $\langle \delta_i(\mathbf{q}_1) \ldots \delta_j(\mathbf{q}_p) \rangle_c$ (e.g., see Fig. 2), and define

$$\overline{\mathcal{D}_{i\ldots j}} \equiv \frac{\int d\sigma_\ell^2(\mathbf{q}_1) \ldots \int d\sigma_\ell^2(\mathbf{q}_p) \mathcal{D}_{i\ldots j}}{\int d\sigma_\ell^2(\mathbf{q}_1) \ldots \int d\sigma_\ell^2(\mathbf{q}_p)}, \qquad (4.1)$$

where the bar denotes the average of diagram amplitudes over the initial rms fluctuations, and:

$$d\sigma_\ell^2(\mathbf{q}) \equiv d^3 q P_1(q, \tau). \qquad (4.2)$$

We now turn to the explicit construction of the diagrammatic series for each cumulant to be considered.

## B. Loop Expansion for $\sigma^2$

The diagrams corresponding to the perturbative calculation of the variance (2nd order 1-point cumulant) are given in Figs. 2, 3, 4 and 5. According to the above rules, their contribution is:

$$\sigma^2 = \sigma_\ell^2 + \left[\int d\sigma_\ell^2(\mathbf{q}_1)\int d\sigma_\ell^2(\mathbf{q}_2)\left[\mathcal{D}_{22} + \mathcal{D}_{31}\right]\right] + \left[\int d\sigma_\ell^2(\mathbf{q}_1)\ldots\int d\sigma_\ell^2(\mathbf{q}_3)\left[\mathcal{D}_{33}^I + \mathcal{D}_{33}^{II} + \mathcal{D}_{42} + \mathcal{D}_{51}\right]\right]$$
$$+ \left[\int d\sigma_\ell^2(\mathbf{q}_1)\ldots\int d\sigma_\ell^2(\mathbf{q}_4)\left[\mathcal{D}_{44}^I + \mathcal{D}_{44}^{II} + \mathcal{D}_{53}^I + \mathcal{D}_{53}^{II} + \mathcal{D}_{62} + \mathcal{D}_{71}\right]\right] + \mathcal{O}(\sigma_\ell^{10}), \tag{4.3}$$

where:

$$\mathcal{D}_{22} \equiv 2F_2^{(s)}(\mathbf{q}_1, \mathbf{q}_2)F_2^{(s)}(-\mathbf{q}_1, -\mathbf{q}_2) = 2[F_2^{(s)}(\mathbf{q}_1, \mathbf{q}_2)]^2, \tag{4.4a}$$

$$\mathcal{D}_{31} \equiv 6F_3^{(s)}(\mathbf{q}_1, -\mathbf{q}_1, \mathbf{q}_2), \tag{4.4b}$$

are the one-loop contributions (see Fig. 2),

$$\mathcal{D}_{33}^I \equiv 6F_3^{(s)}(\mathbf{q}_1, \mathbf{q}_2, \mathbf{q}_3)F_3^{(s)}(-\mathbf{q}_1, -\mathbf{q}_2, -\mathbf{q}_3), \tag{4.5a}$$

$$\mathcal{D}_{33}^{II} \equiv 9F_3^{(s)}(\mathbf{q}_1, -\mathbf{q}_1, \mathbf{q}_2)F_3^{(s)}(-\mathbf{q}_2, \mathbf{q}_3, -\mathbf{q}_3), \tag{4.5b}$$

$$\mathcal{D}_{42} \equiv 24F_4^{(s)}(\mathbf{q}_1, -\mathbf{q}_1, \mathbf{q}_2, \mathbf{q}_3)F_2^{(s)}(-\mathbf{q}_2, -\mathbf{q}_3), \tag{4.6}$$

$$\mathcal{D}_{51} \equiv 30F_5^{(s)}(\mathbf{q}_1, -\mathbf{q}_1, \mathbf{q}_2, -\mathbf{q}_2, \mathbf{q}_3), \tag{4.7}$$

correspond to the two-loop diagrams (see Fig. 3), and

$$\mathcal{D}_{44}^I \equiv 24F_4^{(s)}(\mathbf{q}_1, \mathbf{q}_2, \mathbf{q}_3, \mathbf{q}_4)F_4^{(s)}(-\mathbf{q}_1, -\mathbf{q}_2, -\mathbf{q}_3, -\mathbf{q}_4), \tag{4.8a}$$

$$\mathcal{D}_{44}^{II} \equiv 72F_4^{(s)}(\mathbf{q}_1, -\mathbf{q}_1, \mathbf{q}_2, \mathbf{q}_3)F_4^{(s)}(-\mathbf{q}_2, -\mathbf{q}_3, \mathbf{q}_4, -\mathbf{q}_4), \tag{4.8b}$$

$$\mathcal{D}_{53}^I \equiv 120F_5^{(s)}(\mathbf{q}_1, -\mathbf{q}_1, \mathbf{q}_2, \mathbf{q}_3, \mathbf{q}_4)F_3^{(s)}(-\mathbf{q}_2, -\mathbf{q}_3, -\mathbf{q}_4), \tag{4.9a}$$

$$\mathcal{D}_{53}^{II} \equiv 90F_5^{(s)}(\mathbf{q}_1, -\mathbf{q}_1, \mathbf{q}_2, -\mathbf{q}_2, \mathbf{q}_3)F_3^{(s)}(-\mathbf{q}_3, \mathbf{q}_4, -\mathbf{q}_4), \tag{4.9b}$$

$$\mathcal{D}_{62} \equiv 180F_6^{(s)}(\mathbf{q}_1, -\mathbf{q}_1, \mathbf{q}_2, -\mathbf{q}_2, \mathbf{q}_3, \mathbf{q}_4)F_2^{(s)}(-\mathbf{q}_3, -\mathbf{q}_4), \tag{4.10}$$

$$\mathcal{D}_{71} \equiv 210F_7^{(s)}(\mathbf{q}_1, -\mathbf{q}_1, \mathbf{q}_2, -\mathbf{q}_2, \mathbf{q}_3, -\mathbf{q}_3, \mathbf{q}_4), \tag{4.11}$$

describe the 3-loop corrections (see Figs. 4 and 5).

The $\mathcal{D}_{22}$ diagram has a symmetry factor of 2, corresponding to the two ways in which the 4 $\delta_1$'s (2 in each vertex) can be paired. The $\mathcal{D}_{31}$ diagram has a symmetry factor of 3, since there are 3 ways in which we can choose one $\delta_1$ out of 3 $\delta_1$'s in the third order vertex to join with the only $\delta_1$ coming out from the first order vertex. There is also a multinomial factor of 2 in this case. The numerical factors in the 2-loop and 3-loop expressions are derived in a similar fashion.

Using Eq. (4.1), we can write (see Eq. (3.10)):

$$s^{(1)} = \overline{\mathcal{D}_{22}} + \overline{\mathcal{D}_{31}}, \tag{4.12}$$

$$s^{(2)} = \overline{\mathcal{D}_{33}^I} + \overline{\mathcal{D}_{33}^{II}} + \overline{\mathcal{D}_{42}} + \overline{\mathcal{D}_{51}}, \tag{4.13}$$

$$s^{(3)} = \overline{\mathcal{D}_{44}^I} + \overline{\mathcal{D}_{44}^{II}} + \overline{\mathcal{D}_{53}^I} + \overline{\mathcal{D}_{53}^{II}} + \overline{\mathcal{D}_{62}} + \overline{\mathcal{D}_{71}}, \tag{4.14}$$

which express loop correction coefficients in terms of averages of diagram amplitudes over the linear rms density fluctuations.

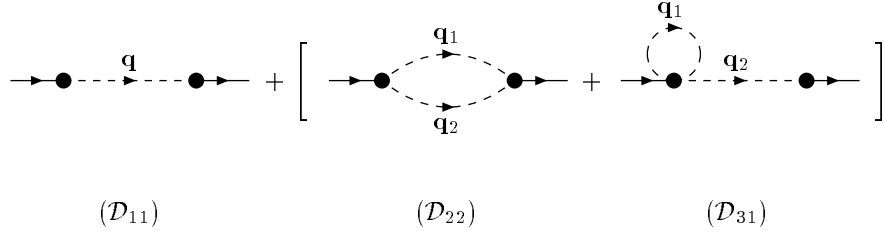

$(\mathcal{D}_{11})$  $(\mathcal{D}_{22})$  $(\mathcal{D}_{31})$

FIG. 2. Diagrams for $\sigma^2$ up to one loop. See Eqs. (4.4) for diagrams amplitudes.

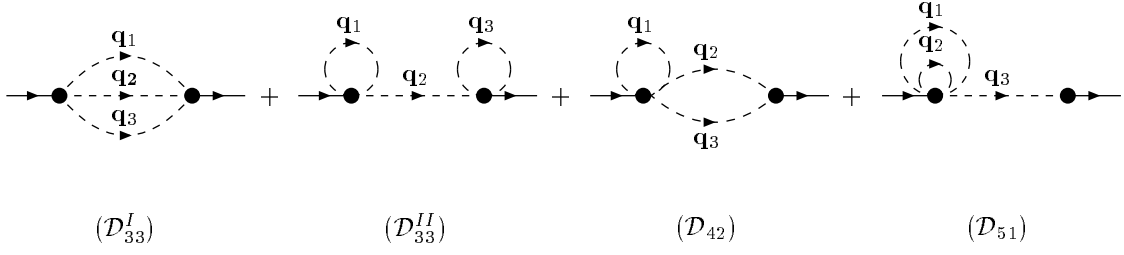

$(\mathcal{D}_{33}^{I})$  $(\mathcal{D}_{33}^{II})$  $(\mathcal{D}_{42})$  $(\mathcal{D}_{51})$

FIG. 3. Two-loop diagrams for $\sigma^2$. See Eqs. (4.5), (4.6) and (4.7) for diagrams amplitudes.

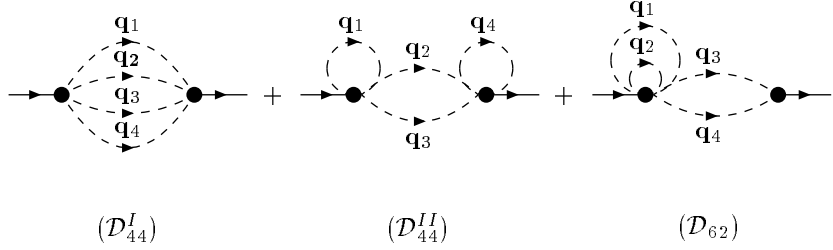

$(\mathcal{D}_{44}^{I})$  $(\mathcal{D}_{44}^{II})$  $(\mathcal{D}_{62})$

FIG. 4. Three-loop diagrams for $\sigma^2$ corresponding to $\mathcal{D}_{44}$ and $\mathcal{D}_{62}$. See Eqs. (4.8) and (4.10) for diagrams amplitudes.

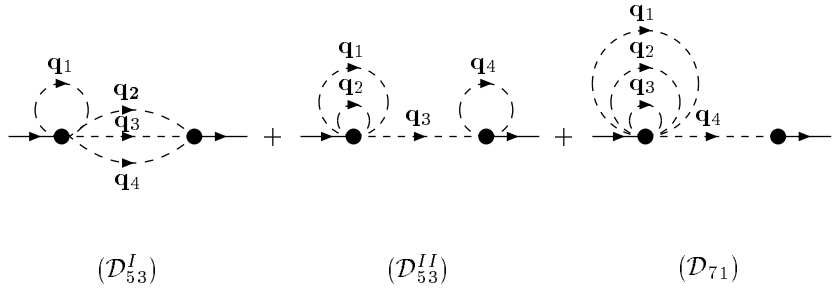

$(\mathcal{D}_{53}^{I})$  $(\mathcal{D}_{53}^{II})$  $(\mathcal{D}_{71})$

FIG. 5. Three-loop diagrams for $\sigma^2$ corresponding to $\mathcal{D}_{53}$ and $\mathcal{D}_{71}$. See Eqs. (4.9) and (4.11) for diagrams amplitudes.

### C. Loop Expansion for $S_3$

From the diagrammatic rules and Fig. 6 we have:

$$\mathcal{D}_{211} = 6F_2^{(s)}(\mathbf{q}_1, \mathbf{q}_2), \qquad (4.15)$$

for the tree-level contribution.

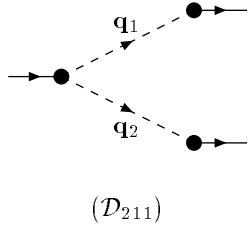

$(\mathcal{D}_{211})$

FIG. 6. Tree-level diagram for $\langle\delta^3\rangle_c$. The corresponding amplitude is given by Eq. (4.15).

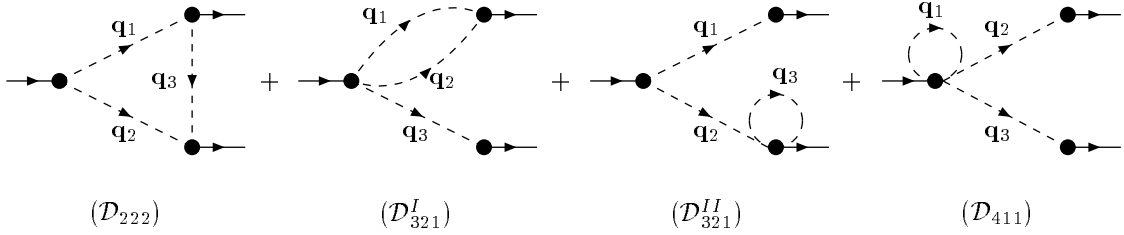

$(\mathcal{D}_{222})$ $\qquad$ $(\mathcal{D}_{321}^{I})$ $\qquad$ $(\mathcal{D}_{321}^{II})$ $\qquad$ $(\mathcal{D}_{411})$

FIG. 7. One-loop diagrams for $\langle\delta^3\rangle_c$. The corresponding amplitudes are given in Eq. (4.16).

For the 1-loop diagrams in Fig. 7 we get:

$$\mathcal{D}_{222} = 8 F_2^{(s)}(\mathbf{q}_1, \mathbf{q}_2) F_2^{(s)}(\mathbf{q}_2, \mathbf{q}_3) F_2^{(s)}(\mathbf{q}_3, -\mathbf{q}_1), \tag{4.16a}$$

$$\mathcal{D}_{321}^{I} = 36 F_3^{(s)}(\mathbf{q}_1, \mathbf{q}_2, \mathbf{q}_3) F_2^{(s)}(\mathbf{q}_1, \mathbf{q}_2), \tag{4.16b}$$

$$\mathcal{D}_{321}^{II} = 36 F_2^{(s)}(\mathbf{q}_1, \mathbf{q}_2) F_3^{(s)}(-\mathbf{q}_2, \mathbf{q}_3, -\mathbf{q}_3), \tag{4.16c}$$

$$\mathcal{D}_{411} = 36 F_4^{(s)}(\mathbf{q}_1, -\mathbf{q}_1, \mathbf{q}_2, \mathbf{q}_3). \tag{4.16d}$$

From Figs. 8, 9, and 10 we get the 2-loop contribution to the third order cumulant:

$$\mathcal{D}_{332}^{I} = 108 F_3^{(s)}(\mathbf{q}_1, \mathbf{q}_2, \mathbf{q}_3) F_3^{(s)}(-\mathbf{q}_1, -\mathbf{q}_2, \mathbf{q}_4) F_2^{(s)}(\mathbf{q}_3, \mathbf{q}_4), \tag{4.17a}$$

$$\mathcal{D}_{332}^{II} = 108 F_3^{(s)}(\mathbf{q}_1, \mathbf{q}_2, \mathbf{q}_3) F_3^{(s)}(-\mathbf{q}_3, \mathbf{q}_4, -\mathbf{q}_4) F_2^{(s)}(\mathbf{q}_1, \mathbf{q}_2), \tag{4.17b}$$

$$\mathcal{D}_{332}^{III} = 54 F_2^{(s)}(\mathbf{q}_1, \mathbf{q}_2) F_3^{(s)}(-\mathbf{q}_1, \mathbf{q}_3, -\mathbf{q}_3) F_3^{(s)}(-\mathbf{q}_2, \mathbf{q}_4, -\mathbf{q}_4), \tag{4.17c}$$

$$\mathcal{D}_{422}^{I} = 72 F_4^{(s)}(\mathbf{q}_1, \mathbf{q}_2, \mathbf{q}_3, \mathbf{q}_4) F_2^{(s)}(\mathbf{q}_1, \mathbf{q}_2) F_2^{(s)}(\mathbf{q}_3, \mathbf{q}_4), \tag{4.18a}$$

$$\mathcal{D}_{422}^{II} = 144 F_4^{(s)}(\mathbf{q}_1, -\mathbf{q}_1, \mathbf{q}_2, \mathbf{q}_3) F_2^{(s)}(\mathbf{q}_2, \mathbf{q}_4) F_2^{(s)}(-\mathbf{q}_3, \mathbf{q}_4), \tag{4.18b}$$

$$\mathcal{D}_{431}^{I} = 144 F_4^{(s)}(\mathbf{q}_1, \mathbf{q}_2, \mathbf{q}_3, \mathbf{q}_4) F_3^{(s)}(-\mathbf{q}_1, -\mathbf{q}_2, -\mathbf{q}_3), \tag{4.19a}$$

$$\mathcal{D}_{431}^{II} = 216 F_4^{(s)}(-\mathbf{q}_1, -\mathbf{q}_2, \mathbf{q}_4, -\mathbf{q}_4) F_3^{(s)}(\mathbf{q}_1, \mathbf{q}_2, \mathbf{q}_3), \tag{4.19b}$$

$$\mathcal{D}_{431}^{III} = 216 F_4^{(s)}(\mathbf{q}_1, -\mathbf{q}_1, \mathbf{q}_2, \mathbf{q}_3) F_3^{(s)}(-\mathbf{q}_3, \mathbf{q}_4, -\mathbf{q}_4), \tag{4.19c}$$

$$\mathcal{D}_{521}^{I} = 360 F_5^{(s)}(\mathbf{q}_1, -\mathbf{q}_1, \mathbf{q}_2, \mathbf{q}_3, \mathbf{q}_4) F_2^{(s)}(\mathbf{q}_2, \mathbf{q}_3), \tag{4.20a}$$

$$\mathcal{D}_{521}^{II} = 180 F_5^{(s)}(-\mathbf{q}_2, \mathbf{q}_3, -\mathbf{q}_3, \mathbf{q}_4, -\mathbf{q}_4) F_2^{(s)}(\mathbf{q}_1, \mathbf{q}_2), \tag{4.20b}$$

$$\mathcal{D}_{611} = 270 F_6^{(s)}(\mathbf{q}_1, -\mathbf{q}_1, \mathbf{q}_2, -\mathbf{q}_2, \mathbf{q}_3, \mathbf{q}_4). \tag{4.21}$$

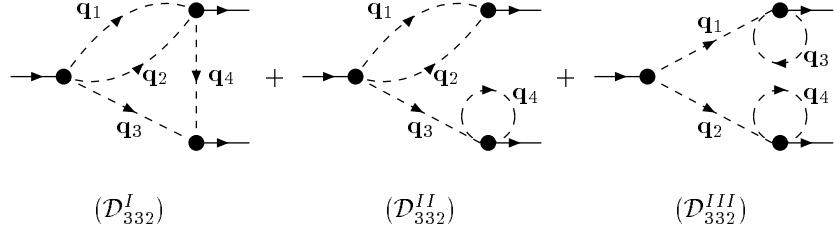

($\mathcal{D}^I_{332}$)　　　　　($\mathcal{D}^{II}_{332}$)　　　　　($\mathcal{D}^{III}_{332}$)

FIG. 8. Two-loop diagrams for $\langle \delta^3 \rangle_c$ corresponding to $\mathcal{D}_{332}$. See Eqs. (4.17) for diagrams amplitudes.

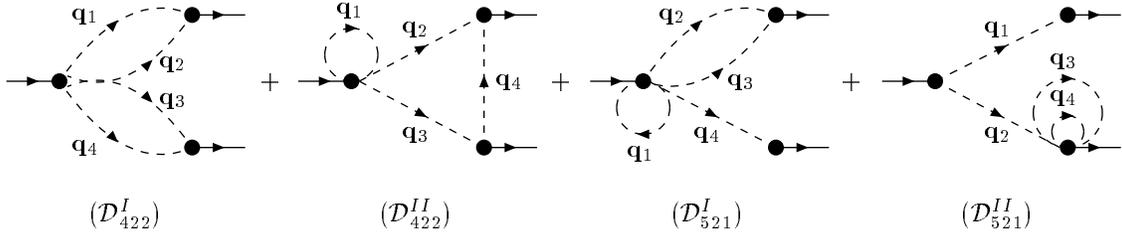

($\mathcal{D}^I_{422}$)　　　　($\mathcal{D}^{II}_{422}$)　　　　($\mathcal{D}^I_{521}$)　　　　($\mathcal{D}^{II}_{521}$)

FIG. 9. Two-loop diagrams for $\langle \delta^3 \rangle_c$ corresponding to $\mathcal{D}_{422}$ and $\mathcal{D}_{521}$. See Eqs. (4.18) and (4.20) for diagrams amplitudes.

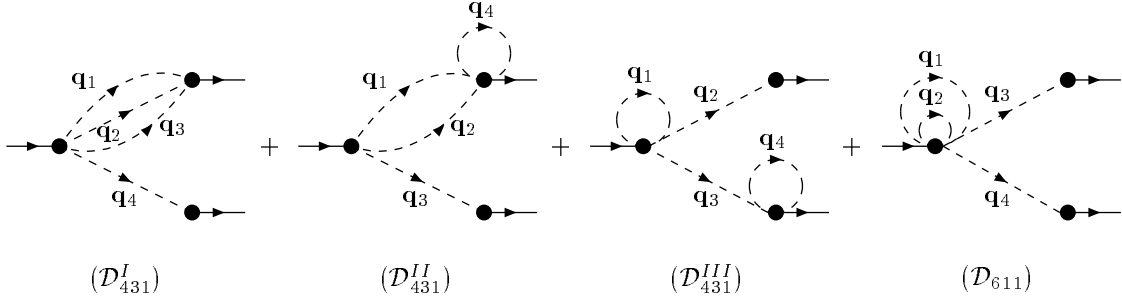

($\mathcal{D}^I_{431}$)　　　　($\mathcal{D}^{II}_{431}$)　　　　($\mathcal{D}^{III}_{431}$)　　　　($\mathcal{D}_{611}$)

FIG. 10. Two-loop diagrams for $\langle \delta^3 \rangle_c$ corresponding to $\mathcal{D}_{431}$ and $\mathcal{D}_{611}$. See Eqs. (4.19) and (4.21) for diagrams amplitudes.

Therefore:

$$\langle \delta^3 \rangle_c = \int d\sigma_\ell^2(\mathbf{q}_1) \int d\sigma_\ell^2(\mathbf{q}_2) \mathcal{D}_{211} + \left[ \int d\sigma_\ell^2(\mathbf{q}_1) \ldots \int d\sigma_\ell^2(\mathbf{q}_3) \left[ \mathcal{D}_{222} + \mathcal{D}^I_{321} + \mathcal{D}^{II}_{321} + \mathcal{D}_{411} \right] \right] + \left[ \int d\sigma_\ell^2(\mathbf{q}_1) \ldots \right.$$
$$\left. \times \int d\sigma_\ell^2(\mathbf{q}_4) \left[ \mathcal{D}^I_{332} + \mathcal{D}^{II}_{332} + \mathcal{D}^{III}_{332} + \mathcal{D}^I_{422} + \mathcal{D}^{II}_{422} + \mathcal{D}^I_{431} + \mathcal{D}^{II}_{431} + \mathcal{D}^{III}_{431} + \mathcal{D}^I_{521} + \mathcal{D}^{II}_{521} + \mathcal{D}_{611} \right] \right]$$
$$+ \mathcal{O}(\sigma_\ell^{10}), \tag{4.22}$$

which gives the contributions to the 3rd-order cumulant up to 2-loop corrections. Here, the tree-level diagram $\mathcal{D}_{211}$ (see Fig. 6) has a symmetry factor of 2! due to the permutations in the second order kernel. The one-loop diagrams (see Fig. 7) have symmetry factors of 8 for $\mathcal{D}_{222}$ (2! for each second order kernel), 6 for $\mathcal{D}^I_{321}$ (3 to choose one $\delta_1$ to connect the third order with the first order kernel, 2! from the remaining permutations with the second order one), 6 for $\mathcal{D}^{II}_{321}$ (3 from choosing one $\delta_1$ in the third order kernel, 2! from the second order one) and 12 for $\mathcal{D}_{411}$ (6 from choosing two $\delta_1$'s out of four in the fourth order kernel and 2 from the remaining permutation). Similar analysis gives the numerical coefficients in the two-loop diagrams.

We can now write Eq. (3.16a) as:

$$S_3^{(0)} = \overline{\mathcal{D}_{211}}, \tag{4.23a}$$
$$S_3^{(1)} = \overline{\mathcal{D}_{222}} + \overline{\mathcal{D}^I_{321}} + \overline{\mathcal{D}^{II}_{321}} + \overline{\mathcal{D}_{411}} - 2s^{(1)} S_3^{(0)} \tag{4.23b}$$

and Eq. (3.16b) as:

$$S_3^{(2)} = \overline{\mathcal{D}_{332}^{I}} + \overline{\mathcal{D}_{332}^{II}} + \overline{\mathcal{D}_{332}^{III}} + \overline{\mathcal{D}_{422}^{I}} + \overline{\mathcal{D}_{422}^{II}} + \overline{\mathcal{D}_{431}^{I}} + \overline{\mathcal{D}_{431}^{II}} + \overline{\mathcal{D}_{431}^{III}} + \overline{\mathcal{D}_{521}^{I}} + \overline{\mathcal{D}_{521}^{II}} + \overline{\mathcal{D}_{611}} - 2s^{(1)}S_3^{(1)}$$
$$- \left((s^{(1)})^2 + 2s^{(2)}\right)S_3^{(0)}. \tag{4.24}$$

### D. Loop Expansion for $S_4$

According to the diagrammatic rules in the case of the fourth order one-point cumulant we have:

$$\mathcal{D}_{2211} = 48 F_2^{(s)}(\mathbf{q}_1, \mathbf{q}_3) F_2^{(s)}(-\mathbf{q}_3, \mathbf{q}_2), \tag{4.25a}$$
$$\mathcal{D}_{3111} = 24 F_3^{(s)}(\mathbf{q}_1, \mathbf{q}_2, \mathbf{q}_3), \tag{4.25b}$$

for tree-level diagrams (see Fig. 11).

For one-loop diagrams (see Figs. 12, 13 and 14):

$$\mathcal{D}_{2222} = 48 F_2^{(s)}(\mathbf{q}_1, \mathbf{q}_2) F_2^{(s)}(-\mathbf{q}_2, \mathbf{q}_3) F_2^{(s)}(\mathbf{q}_3, \mathbf{q}_4) F_2^{(s)}(\mathbf{q}_4, -\mathbf{q}_1), \tag{4.26}$$

$$\mathcal{D}_{3221}^{I} = 288 F_2^{(s)}(\mathbf{q}_1, \mathbf{q}_2) F_3^{(s)}(-\mathbf{q}_2, \mathbf{q}_3, \mathbf{q}_4) F_2^{(s)}(\mathbf{q}_1, \mathbf{q}_3), \tag{4.27a}$$
$$\mathcal{D}_{3221}^{II} = 288 F_3^{(s)}(\mathbf{q}_1, \mathbf{q}_2, \mathbf{q}_3) F_2^{(s)}(\mathbf{q}_1, \mathbf{q}_2) F_2^{(s)}(-\mathbf{q}_3, \mathbf{q}_4), \tag{4.27b}$$
$$\mathcal{D}_{3221}^{III} = 288 F_2^{(s)}(\mathbf{q}_1, \mathbf{q}_2) F_3^{(s)}(-\mathbf{q}_1, \mathbf{q}_4, -\mathbf{q}_4) F_2^{(s)}(-\mathbf{q}_2, \mathbf{q}_3), \tag{4.27c}$$

$$\mathcal{D}_{3311}^{I} = 216 F_3^{(s)}(\mathbf{q}_1, \mathbf{q}_2, \mathbf{q}_3) F_3^{(s)}(\mathbf{q}_4, -\mathbf{q}_2, -\mathbf{q}_3), \tag{4.28a}$$
$$\mathcal{D}_{3311}^{II} = 216 F_3^{(s)}(\mathbf{q}_1, \mathbf{q}_2, \mathbf{q}_3) F_3^{(s)}(-\mathbf{q}_1, \mathbf{q}_4, -\mathbf{q}_4), \tag{4.28b}$$

$$\mathcal{D}_{4211}^{I} = 288 F_4^{(s)}(\mathbf{q}_1, \mathbf{q}_2, \mathbf{q}_3, \mathbf{q}_4) F_2^{(s)}(\mathbf{q}_1, \mathbf{q}_2), \tag{4.29a}$$
$$\mathcal{D}_{4211}^{II} = 576 F_4^{(s)}(\mathbf{q}_1, -\mathbf{q}_1, \mathbf{q}_2, \mathbf{q}_3) F_2^{(s)}(-\mathbf{q}_3, \mathbf{q}_4), \tag{4.29b}$$

$$\mathcal{D}_{5111} = 240 F_5^{(s)}(\mathbf{q}_1, -\mathbf{q}_1, \mathbf{q}_2, \mathbf{q}_3, \mathbf{q}_4), \tag{4.30}$$

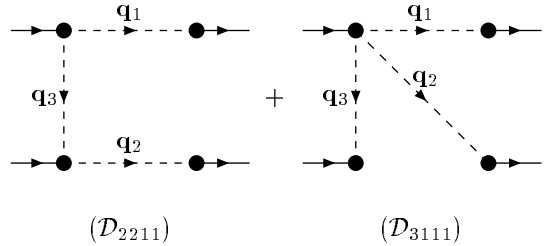

FIG. 11. Tree-level diagrams for $\langle \delta^4 \rangle_c$. The corresponding amplitudes are given in Eq. (4.25).

Therefore:

$$\langle \delta^4 \rangle_c = \int d\sigma_\ell^2(\mathbf{q}_1) \ldots \int d\sigma_\ell^2(\mathbf{q}_3) \Big[ \mathcal{D}_{2211} + \mathcal{D}_{3111} \Big] + \Big[ \int d\sigma_\ell^2(\mathbf{q}_1) \ldots \int d\sigma_\ell^2(\mathbf{q}_4) \Big[ \mathcal{D}_{2222} + \mathcal{D}_{3221}^I + \mathcal{D}_{3221}^{II}$$
$$+ \mathcal{D}_{3221}^{III} + \mathcal{D}_{3311}^I + \mathcal{D}_{3311}^{II} + \mathcal{D}_{4211}^I + \mathcal{D}_{4211}^{II} + \mathcal{D}_{5111} \Big] \Big] + \mathcal{O}(\sigma_\ell^{10}), \quad (4.31)$$

so we can write Eq. (3.16c) as:

$$S_4^{(0)} = \overline{\mathcal{D}_{2211}} + \overline{\mathcal{D}_{3111}}, \quad (4.32a)$$
$$S_4^{(1)} = \overline{\mathcal{D}_{2222}} + \overline{\mathcal{D}_{3221}^I} + \overline{\mathcal{D}_{3221}^{II}} + \overline{\mathcal{D}_{3221}^{III}} + \overline{\mathcal{D}_{3311}^I} + \overline{\mathcal{D}_{3311}^{II}} + \overline{\mathcal{D}_{4211}^I} + \overline{\mathcal{D}_{4211}^{II}} + \overline{\mathcal{D}_{5111}} - 3s^{(1)} S_4^{(0)}. \quad (4.32b)$$

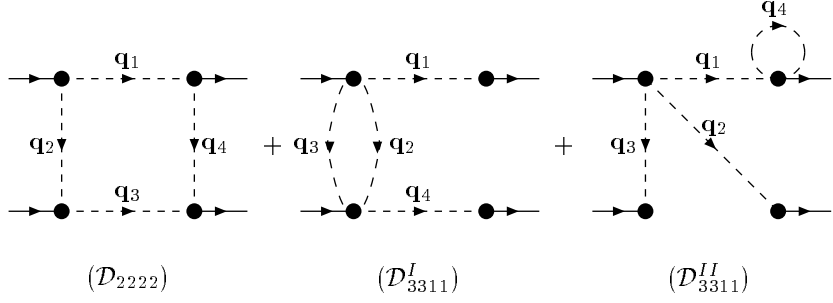

FIG. 12. One-loop diagrams for $\langle \delta^4 \rangle_c$ corresponding to $\mathcal{D}_{2222}$ and $\mathcal{D}_{3311}$ (see Eqs. (4.26) and (4.28)).

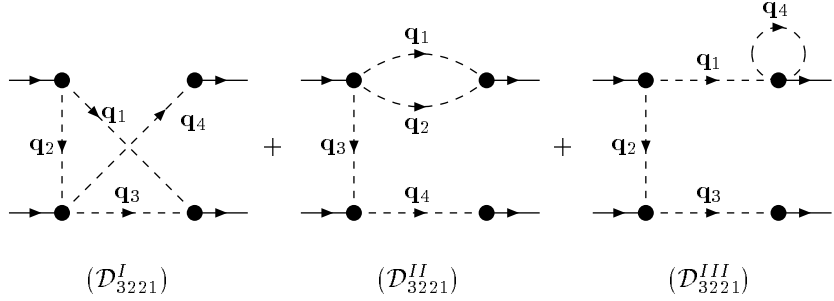

FIG. 13. One-loop diagrams for $\langle \delta^4 \rangle_c$ corresponding to $\mathcal{D}_{3221}$ (see Eq. (4.27)).

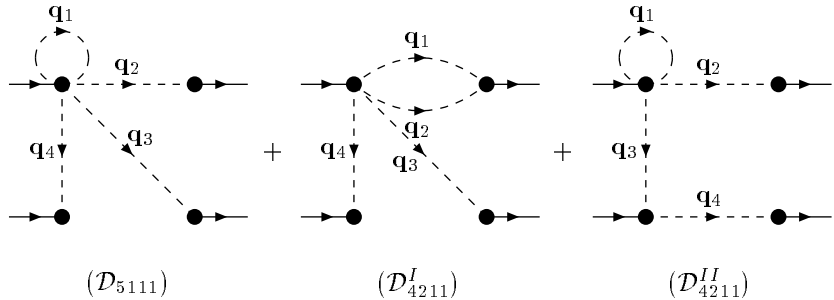

FIG. 14. One-loop diagrams for $\langle \delta^4 \rangle_c$ corresponding to $\mathcal{D}_{5111}$ and $\mathcal{D}_{4211}$ (see Eqs. (4.30) and (4.29)).

We now turn to the calculation of cumulants for the different perturbation theory kernels derived in Section II.

# V. APPLICATIONS: LOOP CORRECTIONS FOR DENSITY FIELD CUMULANTS

Using the results derived in Section IV, we can now calculate loop corrections of 1-point cumulants for the exact dynamics and for the different non-linear approximations. All we need to specify is the linear power spectrum $P_1(k,\tau)$ corresponding to the linear density fluctuations. We take this to be a truncated power-law:

$$P_1(k,\tau) \equiv \begin{cases} A\, a^2(\tau)\, k^n & \text{if } \epsilon \leq k \leq k_c, \\ 0 & \text{otherwise}, \end{cases} \quad (5.1)$$

where $A$ is a constant amplitude, $n$ is the spectral index, $\epsilon$ is an infinitesimal infrared cutoff and $k_c$ is an ultraviolet cutoff. The wave-number cutoffs are introduced in order to regularize the radial integrals. The ultraviolet cutoff $k_c$ is required because at high $k$ one reaches the strongly non-linear regime and perturbation theory breaks down. The spectral index is taken to be in the range $-2 \leq n \leq 2$. The lower limit $n = -2$ is chosen because for $n = -3$ there is an infrared (logarithmic) divergence in the expansion parameter $\sigma_\ell$. The upper limit $n = 2$ is taken to cover a range in $n$ of physical interest. We have also considered other schemes for regularizing the integrals, such as dimensional regularization.

For the exact dynamics and the Zel'dovich approximation, it turns out that the results for the scaled one-point cumulants $S_p$ can be written in terms of $\sigma_\ell^2$, independent of the cutoffs aside from the implicit cutoff-dependence of $\sigma_\ell$ itself. In other words, the $S_p$ parameters for these cases are infrared-finite, and their ultraviolet divergences can be absorbed into the expansion parameter $\sigma_\ell^2 \sim k_c^{n+3}$.

## A. Loop Corrections for Non-Linear Approximations

None of the kernels in the different non-linear approximations derived in Section II C involve scalar products of wave-vectors in their denominators. Thus, the relevant angular integrals are:

$$\int d\Omega_1 \int d\Omega_2 \int d\Omega_3\, (\mathbf{q}_1 \cdot \mathbf{q}_2)^a (\mathbf{q}_2 \cdot \mathbf{q}_3)^b (\mathbf{q}_3 \cdot \mathbf{q}_1)^c, \quad (5.2)$$

for $\mathcal{O}(\sigma_\ell^6)$ corrections and:

$$\int d\Omega_1 \int d\Omega_2 \int d\Omega_3 \int d\Omega_4\, (\mathbf{q}_1 \cdot \mathbf{q}_2)^a (\mathbf{q}_2 \cdot \mathbf{q}_3)^b (\mathbf{q}_3 \cdot \mathbf{q}_1)^c (\mathbf{q}_1 \cdot \mathbf{q}_4)^d (\mathbf{q}_2 \cdot \mathbf{q}_4)^e (\mathbf{q}_3 \cdot \mathbf{q}_4)^f, \quad (5.3)$$

for $\mathcal{O}(\sigma_\ell^8)$ corrections. Even though these integrals are straightforward to evaluate analytically, the number of terms to be integrated makes calculation by hand not feasible for $p \geq 3$ in the 1-loop case. For example, in the ZA, calculation of 1-loop corrections to $S_3$ and $S_4$ involve 150 and 2431 terms to be integrated, respectively. The situation worsens considerably for higher $p$ and/or higher number of loops. Therefore, it is essential to perform the calculation with the aid of a Computer Algebra System. We have used *Mathematica* [64] to generate the different perturbation theory kernels according to their recursion relations, and then developed a symbolic integration routine that uses the results derived in Appendix A (see Eq. (A2) and Eq. (A3)). The resulting radial integrals are straightforward, since they only involve powers of radial components of wave-vectors and therefore decouple into a product of one-dimensional integrations. The final result is then expanded in a Laurent series of $\epsilon/k_c$ and the limit $\epsilon \to 0$ is taken afterwards, except for the divergent terms.

### 1. Loop Corrections for $\sigma^2$

We now proceed to calculate one-loop corrections to the variance of the density field. Let:

$$I_{nm} \equiv \int \frac{d\Omega_1}{4\pi} \int \frac{d\Omega_2}{4\pi}\, \mathcal{D}_{nm}. \quad (5.4)$$

Using the expressions for the perturbation theory kernels and doing the angular integrals using Eq. (A2) and Eq. (A3) we get (see Eqns. (4.4)):

$$I_{22}^{Z} = \frac{19}{15} + \frac{1}{6}\frac{q_1^2}{q_2^2} + \frac{1}{6}\frac{q_2^2}{q_1^2}, \tag{5.5}$$

$$I_{22}^{LP} = \frac{1219}{1500} + \frac{49}{600}\frac{q_1^2}{q_2^2} + \frac{49}{600}\frac{q_2^2}{q_1^2}, \tag{5.6}$$

$$I_{22}^{FF} = \frac{7}{12} + \frac{1}{24}\frac{q_1^2}{q_2^2} + \frac{1}{24}\frac{q_2^2}{q_1^2}, \tag{5.7}$$

and:

$$I_{31}^{Z} = -\frac{1}{3}\frac{q_2^2}{q_1^2}, \tag{5.8}$$

$$I_{31}^{LP} = -\frac{1}{15} - \frac{19}{105}\frac{q_2^2}{q_1^2}, \tag{5.9}$$

$$I_{31}^{FF} = -\frac{1}{9} - \frac{1}{9}\frac{q_2^2}{q_1^2}. \tag{5.10}$$

Now, integrating over radial components of wavevectors and dividing by $\sigma_\ell^4$ we get:

$$\overline{\mathcal{D}_{22}^{Z}} = \frac{19}{15} + \frac{1}{3}\overline{k^2}\,\overline{k^{-2}}, \tag{5.11}$$

$$\overline{\mathcal{D}_{22}^{LP}} = \frac{1219}{1500} + \frac{49}{300}\overline{k^2}\,\overline{k^{-2}}, \tag{5.12}$$

$$\overline{\mathcal{D}_{22}^{FF}} = \frac{7}{12} + \frac{1}{12}\overline{k^2}\,\overline{k^{-2}}, \tag{5.13}$$

and:

$$\overline{\mathcal{D}_{31}^{Z}} = -\frac{1}{3}\overline{k^2}\,\overline{k^{-2}}, \tag{5.14}$$

$$\overline{\mathcal{D}_{31}^{LP}} = -\frac{1}{15} - \frac{19}{105}\overline{k^2}\,\overline{k^{-2}}, \tag{5.15}$$

$$\overline{\mathcal{D}_{31}^{FF}} = -\frac{1}{9} - \frac{1}{9}\overline{k^2}\,\overline{k^{-2}}, \tag{5.16}$$

where we have defined:

$$\overline{k^m} \equiv \frac{1}{\sigma_\ell^2} \int d\sigma_\ell^2(\mathbf{q})\; q^m. \tag{5.17}$$

Summing the two diagrams we obtain the final result:

$$s_Z^{(1)} = \frac{19}{15}, \tag{5.18}$$

$$s_{LP}^{(1)} = \frac{373}{500} - \frac{37}{2100}\overline{k^2}\,\overline{k^{-2}}, \tag{5.19}$$

$$s_{FF}^{(1)} = \frac{17}{36} - \frac{1}{36}\overline{k^2}\,\overline{k^{-2}}. \tag{5.20}$$

Since:

$$\overline{k^2}\,\overline{k^{-2}} \equiv \begin{cases} (\Lambda + 1)/3 + \mathcal{O}(\Lambda^{-1}) & \text{if } n = -2, \\ \ln \Lambda + \mathcal{O}(\Lambda^{-2} \ln \Lambda) & \text{if } n = -1, \\ (n+3)^2/(n+1)(n+5) + \mathcal{O}(\Lambda^{-n-1}) & \text{if } n \geq 0, \end{cases} \tag{5.21}$$

where $\Lambda \equiv k_c/\epsilon$, from Eq. (5.11), Eq. (5.14), and Eq. (5.21), we see that for ZA the infrared divergences present in $\overline{\mathcal{D}_{31}}$ and $\overline{\mathcal{D}_{22}}$ for $n = -2, -1$ as $\epsilon \to 0$ exactly cancel each other, leaving a finite correction which is independent of spectral index. However, for the LPA and FFA, infrared divergences do not cancel and the 1-loop correction $s^{(1)}$ diverges for $n = -2, -1$. We postpone a discussion of the physical reason behind cancellation of infrared divergences to Sec. VII. Note on the other hand that for $n = 0, 1, 2$ the corrections are finite and almost independent of spectral index. In these cases, the loop corrections to $\sigma^2$ are most important for the ZA, followed by the LPA, with FFA having the smallest corrections. This trend in fact agrees with the behavior of tree-level calculations [52,53,51], although the inclusion of smoothing of the fields changes this situation somewhat [51]. Tables I, II, and III summarize the results for 1-loop corrections to $\sigma^2$ in the different non-linear approximations as a function of spectral index.

TABLE I. One-loop corrections to $\sigma^2$ in the Zel'dovich Approximation ($\Lambda \equiv k_c/\epsilon$)

| $n$ | $\overline{\mathcal{D}_{31}}$ | $\overline{\mathcal{D}_{22}}$ | $s^{(1)} = \overline{\mathcal{D}_{31}} + \overline{\mathcal{D}_{22}}$ |
|---|---|---|---|
| $-2$ | $-(1/9)\Lambda - 1/9$ | $(1/9)\Lambda + 62/45$ | $19/15 \approx 1.266$ |
| $-1$ | $-(1/3)\ln\Lambda$ | $(1/3)\ln\Lambda + 19/15$ | $19/15 \approx 1.266$ |
| $0$ | $-3/5$ | $28/15$ | $19/15 \approx 1.266$ |
| $1$ | $-4/9$ | $77/45$ | $19/15 \approx 1.266$ |
| $2$ | $-25/63$ | $524/315$ | $19/15 \approx 1.266$ |

TABLE II. One-loop corrections to $\sigma^2$ in the Linear Potential Approximation ($\Lambda \equiv k_c/\epsilon$)

| $n$ | $\overline{\mathcal{D}_{31}}$ | $\overline{\mathcal{D}_{22}}$ | $s^{(1)} = \overline{\mathcal{D}_{31}} + \overline{\mathcal{D}_{22}}$ |
|---|---|---|---|
| $-2$ | $-(19/315)\Lambda - 8/63$ | $(49/900)\Lambda + 1951/2250$ | $-(37/6300)\Lambda + 11657/15750$ |
| $-1$ | $-(19/105)\ln\Lambda - 1/15$ | $(49/300)\ln\Lambda + 1219/1500$ | $-(37/2100)\ln\Lambda + 373/500$ |
| $0$ | $-206/525$ | $83/75$ | $5/7 \approx 0.714$ |
| $1$ | $-97/315$ | $4637/4500$ | $22759/31500 \approx 0.722$ |
| $2$ | $-622/2205$ | $1133/1125$ | $39967/55125 \approx 0.725$ |

TABLE III. One-loop corrections to $\sigma^2$ in the Frozen Flow Approximation ($\Lambda \equiv k_c/\epsilon$)

| $n$ | $\overline{\mathcal{D}_{31}}$ | $\overline{\mathcal{D}_{22}}$ | $s^{(1)} = \overline{\mathcal{D}_{31}} + \overline{\mathcal{D}_{22}}$ |
|---|---|---|---|
| $-2$ | $-(1/27)\Lambda - 4/27$ | $(1/36)\Lambda + 11/18$ | $-(1/108)\Lambda + 25/54$ |
| $-1$ | $-(1/9)\ln\Lambda - 1/9$ | $(1/12)\ln\Lambda + 7/12$ | $-(1/36)\ln\Lambda + 17/36$ |
| $0$ | $-14/45$ | $11/15$ | $19/45 \approx 0.422$ |
| $1$ | $-7/27$ | $25/36$ | $47/108 \approx 0.435$ |
| $2$ | $-46/189$ | $43/63$ | $83/189 \approx 0.439$ |

For ZA, we also compute two and three loop corrections to the variance. The calculations follow along the same lines as shown above for the one-loop case. Divergences from different diagrams present for $n = -2, -1$ also cancel in this case, leaving a finite correction independent of spectral index. These results are shown in tables IV and V as a function of spectral index.

TABLE IV. Two-loop corrections to $\sigma^2$ in the Zel'dovich Approximation ($\Lambda \equiv k_c/\epsilon$)

| $n$ | $\overline{\mathcal{D}_{24}}$ | $\overline{\mathcal{D}_{33}^I}$ | $\overline{\mathcal{D}_{33}^{II}} = \overline{\mathcal{D}_{51}}$ | $s^{(2)}$ |
|---|---|---|---|---|
| $-2$ | $-(1/45)\Lambda^2 - (26/45)\Lambda - 26/45$ | $(1/90)\Lambda^2 + (17/30)\Lambda + 233/90$ | $(1/180)\Lambda^2 + (1/180)\Lambda + 1/180$ | $91/45 \approx 2.022$ |
| $-1$ | $-(4/27)\ln^2 \Lambda - (5/3)\ln \Lambda$ | $(2/27)\ln^2 \Lambda + (5/3)\ln \Lambda + 91/45$ | $(1/27)\ln^2 \Lambda$ | $91/45 \approx 2.022$ |
| $0$ | $-24/7$ | $3299/630$ | $3/28$ | $91/45 \approx 2.022$ |
| $1$ | $-22/9$ | $196/45$ | $1/18$ | $91/45 \approx 2.022$ |
| $2$ | $-11000/5103$ | $208819/51030$ | $125/2916$ | $91/45 \approx 2.022$ |

TABLE V. Three-loop corrections to $\sigma^2$ in the Zel'dovich Approximation ($\Lambda \equiv k_c/\epsilon$)

| $n$ | $\overline{\mathcal{D}_{62}} = \overline{\mathcal{D}_{44}^{II}}$ | $\overline{\mathcal{D}_{71}}$ |
|---|---|---|
| $-2$ | $(1/756)\Lambda^3 + (27/350)\Lambda^2 + (289/2700)\Lambda + 647/4725$ | $-(\Lambda^3 + \Lambda^2 + \Lambda + 1)/4536$ |
| $-1$ | $(1/54)\ln^3 \Lambda + (6523/11340)\ln^2 \Lambda$ | $-(1/324)\ln^3 \Lambda$ |
| $0$ | $4832/2625$ | $-1/72$ |
| $1$ | $914/945$ | $-2/405$ |
| $2$ | $6231200/8251551$ | $-625/192456$ |
| $n$ | $\overline{\mathcal{D}_{44}^I}$ | $\overline{\mathcal{D}_{53}^{II}}$ |
| $-2$ | $(1/1134)\Lambda^3 + (2162/14175)\Lambda^2 + (2791/1134)\Lambda + 65404/10125$ | $-(\Lambda^3 + \Lambda^2 + \Lambda + 1)/1512$ |
| $-1$ | $(1/81)\ln^3 \Lambda + (6523/5670)\ln^2 \Lambda + (506/75)\ln \Lambda + 1477/375$ | $-(1/108)\ln^3 \Lambda$ |
| $0$ | $154768/7875$ | $-1/24$ |
| $1$ | $1051013/70875$ | $-2/135$ |
| $2$ | $4625937328/343814625$ | $-625/64152$ |
| $n$ | $\overline{\mathcal{D}_{53}^I}$ | $s^{(3)}$ |
| $-2$ | $-(1/378)\Lambda^3 - (413/1350)\Lambda^2 - (25273/9450)\Lambda - 8801/3150$ | $1477/375 \approx 3.939$ |
| $-1$ | $-(1/27)\ln^3 \Lambda - (6523/2835)\ln^2 \Lambda - (506/75)\ln \Lambda$ | $1477/375 \approx 3.939$ |
| $0$ | $-101537/5250$ | $1477/375 \approx 3.939$ |
| $1$ | $-20168/1575$ | $1477/375 \approx 3.939$ |
| $2$ | $-181755199/16503102$ | $1477/375 \approx 3.939$ |

We can summarize the results for the variance by writing:

$$\sigma^2 = \sigma_\ell^2 + \frac{19}{15}\sigma_\ell^4 + \frac{91}{45}\sigma_\ell^6 + \frac{1477}{375}\sigma_\ell^8 + \mathcal{O}(\sigma_\ell^{10}) \approx \sigma_\ell^2 + 1.266\,\sigma_\ell^4 + 2.022\,\sigma_\ell^6 + 3.939\,\sigma_\ell^8 + \mathcal{O}(\sigma_\ell^{10}), \quad (5.22)$$

for the ZA,

$$\sigma^2 \approx \sigma_\ell^2 + 0.720\,\sigma_\ell^4 + \mathcal{O}(\sigma_\ell^6), \quad (5.23)$$

for LPA (within 1% approximately independent of spectral index when $n > -1$), and

$$\sigma^2 \approx \sigma_\ell^2 + 0.431\,\sigma_\ell^4 + \mathcal{O}(\sigma_\ell^6), \quad (5.24)$$

for FFA (within 2% approximately independent of spectral index when $n > -1$). These results are displayed in Fig. 15.

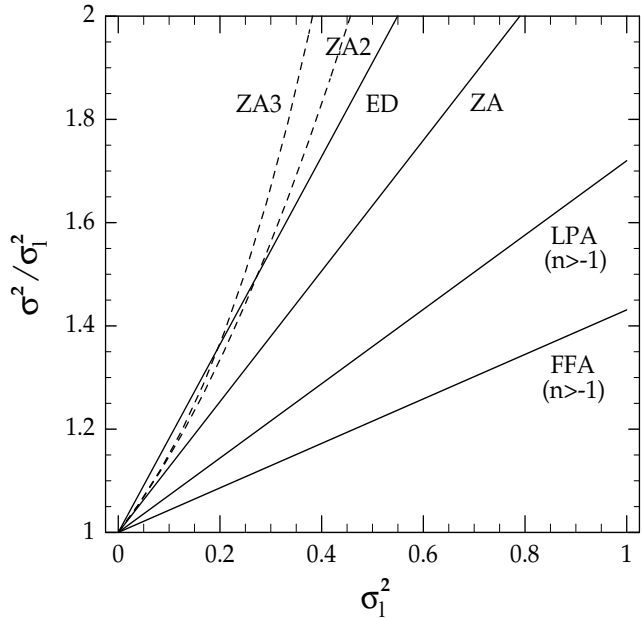

FIG. 15. The ratio of the non-linear to the linear variance $\sigma^2/\sigma_\ell^2$ as a function of the linear variance $\sigma_\ell^2$ for the exact dynamics (ED) to one loop, the Zel'dovich approximation at one (ZA), two (ZA2), and three loops (ZA3), and the LPA and FFA approximations to one loop.

### 2. Loop Corrections for $S_3$

Table VI presents the results of tree-level $S_3$ for different dynamics. One-loop corrections to $S_3$ for the non-linear approximations are calculated similarly to the variance case, the only complication being the increase in the number of terms to be integrated. The results of these calculations are presented in tables VIII, VII and IX for ZA, LPA, and FFA respectively. The general trend seen for corrections to $\sigma^2$ is also valid here, namely, ZA has the largest corrections followed by LPA and then FFA. In this case, however, infrared divergences in the FFA $<\delta^3>_c$ cancel with the divergences in the denominator ($\sigma^4$), making the coefficient $S_3^{(1)}$ finite (see table IX).

For the ZA we also compute two-loop corrections, shown in table X. We can summarize the results for ZA as follows:

$$S_3 = 4 + \frac{352}{75}\sigma_\ell^2 + \frac{5224}{375}\sigma_\ell^4 + \mathcal{O}(\sigma_\ell^6) \approx 4 + 4.693\ \sigma_\ell^2 + 13.931\ \sigma_\ell^4 + \mathcal{O}(\sigma_\ell^6), \tag{5.25}$$

or by inverting Eq. (5.22):

$$S_3(\sigma^2) = 4 + \frac{352}{75}\sigma^2 + \frac{8984}{1125}\sigma^4 + \mathcal{O}(\sigma^6) \approx 4 + 4.693\ \sigma^2 + 7.986\ \sigma^4 + \mathcal{O}(\sigma^6), \tag{5.26}$$

as a function of the non-linear variance. For the ZA, the $\mathcal{O}(\sigma^2)$ correction $S_3^{(1)}$ was previously estimated numerically in [20] to be $S_3^{(1)} \approx 4.6$, in very good agreement with Eq. (5.25). For LPA we have:

$$S_3 \approx 3.400 + 2.124\ \sigma_\ell^2 + \mathcal{O}(\sigma_\ell^4), \tag{5.27}$$

($n > -1$), and for FFA we obtain:

$$S_3 = 3 + \sigma_\ell^2 + \mathcal{O}(\sigma_\ell^4). \tag{5.28}$$

Fig. 16 gives a visual summary of these results.

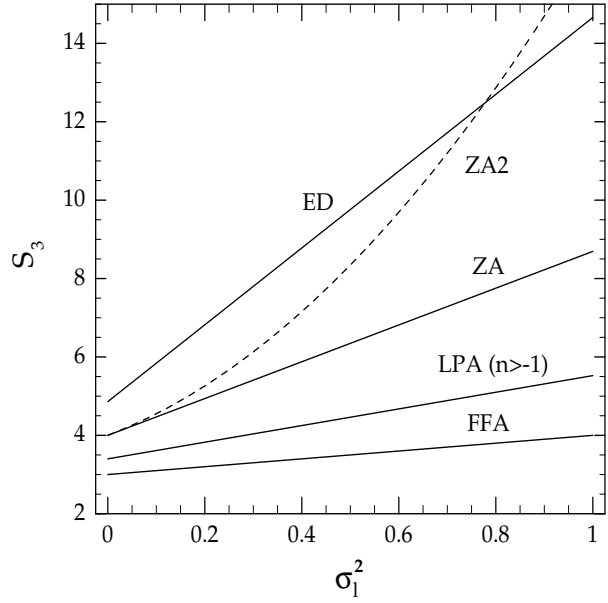

FIG. 16. The skewness $S_3$ as a function of the linear variance $\sigma_\ell^2$ for the exact dynamics (ED) to one loop, the Zel'dovich approximation at one (ZA) and two loops (ZA2), and the LPA and FFA approximations to one loop.

TABLE VI. Tree-level $S_3$ and $S_4$ for different dynamics in Perturbation Theory

| Dynamics | $S_3^{(0)} = \overline{\mathcal{D}_{211}}$ | $\overline{\mathcal{D}_{2211}}$ | $\overline{\mathcal{D}_{3111}}$ | $S_4^{(0)} = \overline{\mathcal{D}_{2211}} + \overline{\mathcal{D}_{3111}}$ |
|---|---|---|---|---|
| ED | $34/7 \approx 4.857$ | $4624/147$ | $2728/189$ | $60712/1323 \approx 45.890$ |
| ZA | 4 | $64/3$ | $80/9$ | $272/9 \approx 30.222$ |
| LPA | $17/5 = 3.4$ | $1156/75$ | $1828/315$ | $33416/1575 \approx 21.216$ |
| FFA | 3 | 12 | 4 | 16 |

TABLE VII. One-loop corrections to $S_3$ in the Linear Potential Approximation ($\Lambda \equiv k_c/\epsilon$)

| $n$ | $\overline{\mathcal{D}_{222}}$ | $\overline{\mathcal{D}_{321}^I}$ | $\overline{\mathcal{D}_{321}^{II}}$ |
|---|---|---|---|
| $-2$ | $-(343/4500)\Lambda + 73351/56250$ | $(118/225)\Lambda + 53834/7875$ | $-(323/1575)\Lambda - 136/315$ |
| $-1$ | $-(343/1500)\ln\Lambda + 17523/12500$ | $(118/75)\ln\Lambda + 16568/2625$ | $-(323/525)\ln\Lambda - 17/75$ |
| 0 | $3027/3125$ | $24002/2625$ | $-3502/2625$ |
| 1 | $120997/112500$ | $66224/7875$ | $-1649/1575$ |
| 2 | $31163/28125$ | $64454/7875$ | $-10574/11025$ |
| $n$ | $\overline{\mathcal{D}_{411}}$ | $-2s^{(1)}S_3^{(0)}$ | $S_3^{(1)}$ |
| $-2$ | $-(5353/18900)\Lambda - 5207/9450$ | $(629/15750)\Lambda - 198169/39375$ | $-(1/6750)\Lambda + 1254763/590625$ |
| $-1$ | $-(5353/6300)\ln\Lambda - 241/900$ | $(629/5250)\ln\Lambda - 6341/1250$ | $-(1/2250)\ln\Lambda + 836567/393750$ |
| 0 | $-14153/7875$ | $-34/7$ | $418126/196875 \approx 2.123$ |
| 1 | $-26473/18900$ | $-386903/78750$ | $2509001/1181250 \approx 2.124$ |
| 2 | $-42313/33075$ | $-1358878/275625$ | $1254538/590625 \approx 2.124$ |

TABLE VIII. One-loop corrections to $S_3$ in the Zel'dovich Approximation ($\Lambda \equiv k_c/\epsilon$)

| $n$ | $\overline{\mathcal{D}_{222}}$ | $\overline{\mathcal{D}^I_{321}}$ | $\overline{\mathcal{D}^{II}_{321}}$ | $\overline{\mathcal{D}_{411}}$ | $S_3^{(1)\,\text{a}}$ |
|---|---|---|---|---|---|
| $-2$ | $-(2/9)\Lambda + 436/225$ | $(4/3)\Lambda + 14$ | $-(4/9)\Lambda - 4/9$ | $-(2/3)\Lambda - 2/3$ | $352/75 \approx 4.693$ |
| $-1$ | $-(2/3)\ln\Lambda + 54/25$ | $4\ln\Lambda + 38/3$ | $-(4/3)\ln\Lambda$ | $-2\ln\Lambda$ | $352/75 \approx 4.693$ |
| $0$ | $24/25$ | $298/15$ | $-12/5$ | $-18/5$ | $352/75 \approx 4.693$ |
| $1$ | $286/225$ | $18$ | $-16/9$ | $-8/3$ | $352/75 \approx 4.693$ |
| $2$ | $2152/1575$ | $122/7$ | $-100/63$ | $-50/21$ | $352/75 \approx 4.693$ |

[a] See Eq. (4.23b). In this case $2s^{(1)} S_3^{(0)} = 152/15$ independent of spectral index.

TABLE IX. One-loop corrections to $S_3$ in the Frozen Flow Approximation ($\Lambda \equiv k_c/\epsilon$)

| $n$ | $\overline{\mathcal{D}_{222}}$ | $\overline{\mathcal{D}^I_{321}}$ | $\overline{\mathcal{D}^{II}_{321}}$ |
|---|---|---|---|
| $-2$ | $-(1/36)\Lambda + 17/18$ | $(2/9)\Lambda + 34/9$ | $-(1/9)\Lambda - 4/9$ |
| $-1$ | $-(1/12)\ln\Lambda + 35/36$ | $(2/3)\ln\Lambda + 32/9$ | $-(1/3)\ln\Lambda - 1/3$ |
| $0$ | $37/45$ | $214/45$ | $-14/15$ |
| $1$ | $31/36$ | $40/9$ | $-7/9$ |
| $2$ | $55/63$ | $274/63$ | $-46/63$ |
| $n$ | $\overline{\mathcal{D}_{411}}$ | $-2s^{(1)} S_3^{(0)}$ | $S_3^{(1)}$ |
| $-2$ | $-(5/36)\Lambda - 1/2$ | $(1/18)\Lambda - 25/9$ | $1$ |
| $-1$ | $-(5/12)\ln\Lambda - 13/36$ | $(1/6)\ln\Lambda - 17/6$ | $1$ |
| $0$ | $-10/9$ | $-38/15$ | $1$ |
| $1$ | $-11/12$ | $-47/18$ | $1$ |
| $2$ | $-6/7$ | $-166/63$ | $1$ |

TABLE X. Two-loop corrections to $S_3$ in the Zel'dovich Approximation ($\Lambda \equiv k_c/\epsilon$)

| $n$ | $\overline{\mathcal{D}^I_{332}}$ | $\overline{\mathcal{D}^{II}_{332}}$ | $\overline{\mathcal{D}^{III}_{332}}$ |
|---|---|---|---|
| $-2$ | $-(8/81)\Lambda^2 - (2242/2025)\Lambda + 36898/3375$ | $-(34/405)\Lambda^2 - (344/405)\Lambda - 41/45$ | $(1/81)\Lambda^2 + (2/81)\Lambda + 1/27$ |
| $-1$ | $-(106/135)\ln^2\Lambda - (644/225)\ln\Lambda + 13606/1125$ | $-(19/27)\ln^2\Lambda - (19/9)\ln\Lambda$ | $(1/9)\ln^2\Lambda$ |
| $0$ | $35446/7875$ | $-211/35$ | $9/25$ |
| $1$ | $70514/10125$ | $-326/81$ | $16/81$ |
| $2$ | $34162714/4465125$ | $-124025/35721$ | $625/3969$ |
| $n$ | $\overline{\mathcal{D}^I_{422}}$ | $\overline{\mathcal{D}^{II}_{422}}$ | $\overline{\mathcal{D}_{611}}$ |
| $-2$ | $(56/405)\Lambda^2 + (7016/2025)\Lambda + 52696/3375$ | $(4/45)\Lambda^2 - (116/225)\Lambda - 52/75$ | $(32/405)\Lambda^2 + (46/405)\Lambda + 4/27$ |
| $-1$ | $(154/135)\ln^2\Lambda + (2182/225)\ln\Lambda + 13562/1125$ | $(94/135)\ln^2\Lambda - (196/175)\ln\Lambda$ | $(82/135)\ln^2\Lambda$ |
| $0$ | $260696/7875$ | $-2232/875$ | $1632/875$ |
| $1$ | $272678/10125$ | $-524/225$ | $404/405$ |
| $2$ | $112252328/4465125$ | $-78632/35721$ | $4000/5103$ |
| $n$ | $\overline{\mathcal{D}^I_{431}}$ | $\overline{\mathcal{D}^{II}_{431}}$ | $\overline{\mathcal{D}^{III}_{431}}$ |
| $-2$ | $(8/45)\Lambda^2 + (1754/225)\Lambda + 2408/75$ | $-(10/81)\Lambda^2 - (1219/405)\Lambda - 136/45$ | $(14/135)\Lambda^2 + (19/135)\Lambda + 8/45$ |
| $-1$ | $(58/45)\ln^2\Lambda + (568/25)\ln\Lambda + 364/15$ | $-(23/27)\ln^2\Lambda - (388/45)\ln\Lambda$ | $(7/9)\ln^2\Lambda$ |
| $0$ | $181204/2625$ | $-3154/175$ | $414/175$ |
| $1$ | $12736/225$ | $-5186/405$ | $34/27$ |
| $2$ | $3149996/59535$ | $-402910/35721$ | $11750/11907$ |
| $n$ | $\overline{\mathcal{D}^I_{521}}$ | $\overline{\mathcal{D}^{II}_{521}}$ | $S_3^{(2)\,\text{a}}$ |
| $-2$ | $-(128/405)\Lambda^2 - (11861/2025)\Lambda - 1349/225$ | $(1/45)\Lambda^2 + (1/45)\Lambda + 1/45$ | $5224/375 \approx 13.931$ |
| $-1$ | $-(328/135)\ln^2\Lambda - (3647/225)\ln\Lambda$ | $(4/27)\ln^2\Lambda$ | $5224/375 \approx 13.931$ |
| $0$ | $-32057/875$ | $3/7$ | $5224/375 \approx 13.931$ |
| $1$ | $-51844/2025$ | $2/9$ | $5224/375 \approx 13.931$ |
| $2$ | $-114469/5103$ | $125/729$ | $5224/375 \approx 13.931$ |

[a] See Eq. (4.24). In this case $2s^{(1)} S_3^{(1)} + \left((s^{(1)})^2 + 2s^{(2)}\right) S_3^{(0)} = 12932/375$ independent of spectral index.

### 3. Loop Corrections for $S_4$

Table VI summarizes the tree-level $S_4$ values for the different dynamics considered here. Results of one-loop corrections to $S_4$ for the non-linear approximations are shown in tables XI, XII, and XIII for LPA, ZA, and FFA respectively. Again, the same comments apply in this case about the size of loop corrections in the different approximations and the cancellation of infrared divergences. We can summarize the results for ZA as follows:

$$S_4 = \frac{272}{9} + \frac{110822}{1125}\sigma_\ell^2 + \mathcal{O}(\sigma_\ell^4) \approx 30.222 + 98.508\,\sigma_\ell^2 + \mathcal{O}(\sigma_\ell^4). \quad (5.29)$$

For LPA and FFA we have, respectively ($n > -1$):

$$S_4 \approx 21.216 + 37.120\,\sigma_\ell^2 + \mathcal{O}(\sigma_\ell^4), \quad (5.30)$$
$$S_4 \approx 16 + 15.00\,\sigma_\ell^2 + \mathcal{O}(\sigma_\ell^4). \quad (5.31)$$

For the ZA, $S_4^{(1)}$ was previously estimated numerically in [20] to be $S_4^{(1)} \approx 100$, in very good agreement with Eq. (5.29). The $S_4$ results are displayed graphically in Fig. 17.

TABLE XI. One-loop corrections to $S_4$ in the Linear Potential Approximation ($\Lambda \equiv k_c/\epsilon$)

| $n$ | $\overline{\mathcal{D}}_{5111}$ | $\mathcal{D}^I_{4211}$ |
|---|---|---|
| $-2$ | $-(761962/779625)\Lambda - 197752/111375$ | $(4163/2250)\Lambda + 2384653/118125$ |
| $-1$ | $-(761962/259875)\ln\Lambda - 207434/259875$ | $(4163/750)\ln\Lambda + 4332191/236250$ |
| $0$ | $-7894828/1299375$ | $3346306/118125$ |
| $1$ | $-13346/2835$ | $6080651/236250$ |
| $2$ | $-23405164/5457375$ | $2946658/118125$ |

| $n$ | $\mathcal{D}^I_{3221}$ | $\mathcal{D}^{II}_{4211}$ |
|---|---|---|
| $-2$ | $-(133/10125)\Lambda^2 - (13937/10125)\Lambda + 11666324/590625$ | $-(182002/70875)\Lambda - 354076/70785$ |
| $-1$ | $-(133/1125)\ln^2\Lambda - (1519/375)\ln\Lambda + 4162358/196875$ | $-(182002/23625)\ln\Lambda - 8194/3375$ |
| $0$ | $2651492/196875$ | $-1924808/118125$ |
| $1$ | $27519122/1771875$ | $-900082/70875$ |
| $2$ | $28619972/1771875$ | $-5754568/496125$ |

| $n$ | $\mathcal{D}^{II}_{3221}$ | $\mathcal{D}^{III}_{3221}$ |
|---|---|---|
| $-2$ | $(8024/3375)\Lambda + 3660712/118125$ | $-(21964/23625)\Lambda - 9248/4725$ |
| $-1$ | $(8024/1125)\Lambda + 1126624/39375$ | $-(21964/7875)\ln\Lambda - 1156/1125$ |
| $0$ | $1632136/39375$ | $-238136/39375$ |
| $1$ | $4503232/118125$ | $-112132/23625$ |
| $2$ | $4382872/118125$ | $-719032/165375$ |

| $n$ | $\mathcal{D}^I_{3311}$ | $\mathcal{D}^{II}_{3311}$ |
|---|---|---|
| $-2$ | $(722/99225)\Lambda^2 + (6568/3969)\Lambda + 425476/23625$ | $-(17366/33075)\Lambda - 7312/6615$ |
| $-1$ | $(722/11025)\ln^2\Lambda + (6028/1225)\ln\Lambda + 901154/55125$ | $-(17366/11025)\ln\Lambda - 914/1575$ |
| $0$ | $7005592/275625$ | $-188284/55125$ |
| $1$ | $11423266/496125$ | $-88658/33075$ |
| $2$ | $542076664/24310125$ | $-568508/231525$ |

| $n$ | $\overline{\mathcal{D}}_{2222}$ | $-3s^{(1)}S_4^{(0)}$ |
|---|---|---|
| $-2$ | $(2401/405000)\Lambda^2 + (2401/20250)\Lambda + 43262149/8437500$ | $(309098/275625)\Lambda - 194765156/4134375$ |
| $-1$ | $(2401/45000)\ln^2\Lambda + (2401/7500)\ln\Lambda + 28141141/5625000$ | $(309098/275625)\ln\Lambda - 3116042/65625$ |
| $0$ | $4044362/703125$ | $-33416/735$ |
| $1$ | $279681269/50625000$ | $-190128686/4134375$ |
| $2$ | $34549018/6328125$ | $-1335537272/28940625$ |

| $n$ | $S_4^{(1)}$ |
|---|---|
| $-2$ | $(1369/19845000)\Lambda^2 - (43621/27286875)\Lambda + 56278351337/1515937500$ |
| $-1$ | $(1369/2205000)\ln^2\Lambda - (63181/12127500)\ln\Lambda + 112561340299/3031875000$ |
| $0$ | $14067375968/378984375 \approx 37.119$ |
| $1$ | $92081148881/2480625000 \approx 37.120$ |
| $2$ | $6204054381248/167132109375 \approx 37.121$ |

TABLE XII. One-loop corrections to $S_4$ in the Zel'dovich Approximation ($\Lambda \equiv k_c/\epsilon$)

| $n$ | $\overline{\mathcal{D}_{5111}}$ | $\overline{\mathcal{D}^{I}_{4211}}$ | $\overline{\mathcal{D}^{II}_{4211}}$ | $\overline{\mathcal{D}^{I}_{3221}}$ |
|---|---|---|---|---|
| $-2$ | $-(28/9)\Lambda - 28/9$ | $(56/9)\Lambda + 512/9$ | $-(64/9)\Lambda - 64/9$ | $-(4/81)\Lambda^2 - (404/81)\Lambda - 5152/135$ |
| $-1$ | $-(28/3)\ln\Lambda$ | $(56/3)\ln\Lambda + 152/3$ | $-(64/3)\ln\Lambda$ | $-(4/9)\ln^2\Lambda - (44/3)\ln\Lambda + 216/5$ |
| $0$ | $-84/5$ | $1264/15$ | $-192/5$ | $384/25$ |
| $1$ | $-112/9$ | $680/9$ | $-256/9$ | $9256/405$ |
| $2$ | $-100/9$ | $656/9$ | $-1600/63$ | $498304/19845$ |
| $n$ | $\overline{\mathcal{D}^{II}_{3221}}$ | $\overline{\mathcal{D}^{III}_{3221}}$ | $\overline{\mathcal{D}^{I}_{3311}}$ | $\overline{\mathcal{D}^{II}_{3311}}$ |
| $-2$ | $(64/9)\Lambda + 224/3$ | $-(64/27)\Lambda - 64/27$ | $(2/81)\Lambda^2 + (424/81)\Lambda + 1287/27$ | $-(40/27)\Lambda - 40/27$ |
| $-1$ | $(64/3)\ln\Lambda + 608/9$ | $-(64/9)\Lambda$ | $(2/9)\ln^2\Lambda + (140/9)\ln\Lambda + 380/9$ | $-(40/9)\ln\Lambda$ |
| $0$ | $4768/45$ | $-64/5$ | $15962/225$ | $-8$ |
| $1$ | $96$ | $-256/27$ | $5132/81$ | $-160/27$ |
| $2$ | $1952/21$ | $-1600/189$ | $242330/3969$ | $-1000/189$ |
| $n$ |  | $\overline{\mathcal{D}_{2222}}$ | $S_4^{(1)}$ [a] |  |
| $-2$ |  | $(2/81)\Lambda^2 + (40/81)\Lambda + 34516/3375$ | $110822/1125 \approx 98.508$ |  |
| $-1$ |  | $(2/9)\ln^2\Lambda + (4/3)\ln\Lambda + 10922/1125$ | $110822/1125 \approx 98.508$ |  |
| $0$ |  | $14432/1125$ | $110822/1125 \approx 98.508$ |  |
| $1$ |  | $120298/10125$ | $110822/1125 \approx 98.508$ |  |
| $2$ |  | $5760352/496125$ | $110822/1125 \approx 98.508$ |  |

[a] See Eq. (4.32b). In this case $3s^{(1)}S_4^{(0)} = 5168/45$ independent of spectral index.

TABLE XIII. One-loop corrections to $S_4$ in the Frozen-Flow Approximation ($\Lambda \equiv k_c/\epsilon$)

| $n$ | $\overline{\mathcal{D}_{5111}}$ | $\overline{\mathcal{D}^{I}_{4211}}$ | $\overline{\mathcal{D}^{II}_{4211}}$ | $\overline{\mathcal{D}^{I}_{3221}}$ |
|---|---|---|---|---|
| $-2$ | $-(142/405)\Lambda - 92/81$ | $(97/162)\Lambda + 635/81$ | $-(10/9)\Lambda - 4$ | $-(1/243)\Lambda^2 - (101/243)\Lambda + 908/81$ |
| $-1$ | $-(142/135)\ln\Lambda - 106/135$ | $(97/54)\ln\Lambda + 391/54$ | $-(10/3)\ln\Lambda - 26/9$ | $-(1/27)\ln^2\Lambda - (11/9)\ln\Lambda + 314/27$ |
| $0$ | $-1808/675$ | $1414/135$ | $-80/9$ | $6284/675$ |
| $1$ | $-886/405$ | $1561/162$ | $-22/3$ | $2414/243$ |
| $2$ | $-5776/2835$ | $5318/567$ | $-48/7$ | $120524/11907$ |
| $n$ | $\overline{\mathcal{D}^{II}_{3221}}$ | $\overline{\mathcal{D}^{III}_{3221}}$ | $\overline{\mathcal{D}^{II}_{3311}}$ | $\overline{\mathcal{D}^{I}_{3311}}$ |
| $-2$ | $(8/9)\Lambda + 136/9$ | $-(4/9)\Lambda - 16/9$ | $-(2/9)\Lambda - 8/9$ | $(2/729)\Lambda^2 + (424/729)\Lambda + 1900/243$ |
| $-1$ | $(8/3)\ln\Lambda + 128/9$ | $-(4/3)\Lambda - 4/3$ | $-(2/3)\ln\Lambda - 2/3$ | $(2/81)\ln^2\Lambda + (140/81)\ln\Lambda + 586/81$ |
| $0$ | $856/45$ | $-56/15$ | $-28/15$ | $21112/2025$ |
| $1$ | $160/9$ | $-28/9$ | $-14/9$ | $6986/729$ |
| $2$ | $1096/63$ | $-184/63$ | $-92/63$ | $333176/35721$ |
| $n$ | $\overline{\mathcal{D}_{2222}}$ | $-3s^{(1)}S_4^{(0)}$ |  | $S_4^{(1)}$ |
| $-2$ | $(1/648)\Lambda^2 + (5/162)\Lambda + 329/108$ | $(4/9)\Lambda - 200/9$ |  | $(1/5832)\Lambda^2 + (2/3645)\Lambda + 14581/972$ |
| $-1$ | $(1/72)\ln^2\Lambda + (1/12)\ln\Lambda + 217/72$ | $(4/3)\ln\Lambda - 68/3$ |  | $(1/648)\ln^2\Lambda + (1/1620)\ln\Lambda + 48601/3240$ |
| $0$ | $722/225$ | $-304/15$ |  | $30388/2025 \approx 15.006$ |
| $1$ | $2041/648$ | $-188/9$ |  | $437513/29160 \approx 15.004$ |
| $2$ | $12434/3969$ | $-1328/63$ |  | $2679652/178605 \approx 15.003$ |

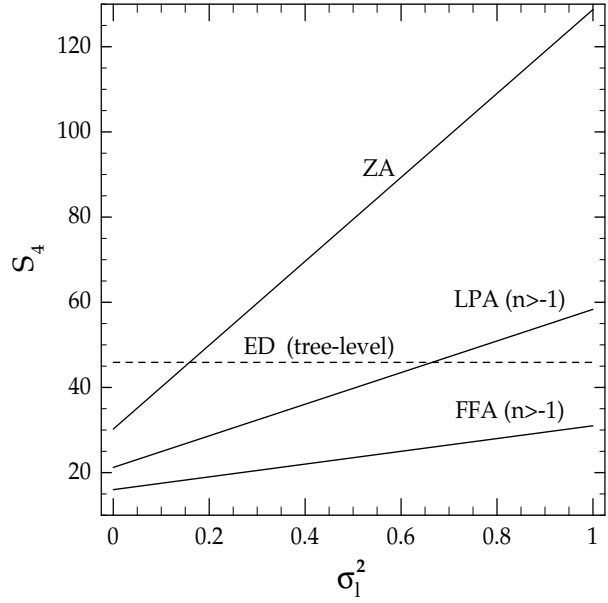

FIG. 17. The kurtosis $S_4$ as a function of the linear variance $\sigma_\ell^2$ for the exact dynamics (ED) at tree level, the Zel'dovich approximation at one loop (ZA), and the LPA and FFA approximations to one loop.

### B. Loop Corrections for the Exact Dynamics

The calculation of loop corrections in this case is technically much more complicated than in the non-linear approximations, due to the appearance of logarithmic terms coming from solving the exact Poisson equation. Since the Green's function of the Poisson equation is non-local, this introduces a weak dependence of the loop corrections on the spectral index $n$. This is the fundamental difference between exact gravitational instability and non-linear approximations [65,66]. The principles of calculation, however, are exactly the same, and we refer the reader to Appendices A and B for the details. Infrared divergences present in individual diagrams for $n = -2, -1$ are canceled when the sum over diagrams contributing to a given correction is done. The mechanism behind this cancellation is discussed in Sec. VII.

#### 1. Loop Corrections for $\sigma^2$

The results of angular integration of Eqs. (4.4a) and (4.4b) in this case are:

$$I_{22} = \frac{1219}{735} + \frac{1}{6}\frac{q_1^2}{q_2^2} + \frac{1}{6}\frac{q_2^2}{q_1^2}, \tag{5.32}$$

$$I_{31} = \frac{25}{126} + \frac{1}{42}\frac{q_2^4}{q_1^4} - \frac{1}{12}\frac{q_1^2}{q_2^2} - \frac{79}{252}\frac{q_2^2}{q_1^2} + \frac{(q_1^2 - q_2^2)^3(7q_1^2 + 2q_2^2)}{168 q_1^5 q_2^3} \ln\frac{|q_1 + q_2|}{|q_1 - q_2|}. \tag{5.33}$$

Now, integrating over radial components of wavevectors and dividing by $\sigma_\ell^4$ we get:

$$\overline{\mathcal{D}_{22}} = \frac{1219}{735} + \frac{1}{3}\overline{k^2}\,\overline{k^{-2}}, \tag{5.34}$$

$$\overline{\mathcal{D}_{31}} = \frac{25}{126} - \frac{25}{63}\overline{k^2}\,\overline{k^{-2}} + \frac{1}{42}\overline{k^4}\,\overline{k^{-4}} - \frac{1}{42}\overline{\ln(k, k^{-1})} + \frac{1}{28}\overline{\ln(k^3, k^{-3})} - \frac{1}{84}\overline{\ln(k^5, k^{-5})}, \tag{5.35}$$

where:

$$\overline{\ln(k^a, k^b)} \equiv \frac{1}{\sigma_\ell^4} \int d\sigma_\ell^2(\mathbf{q}_1) \int d\sigma_\ell^2(\mathbf{q}_2)\, q_1^a\, q_2^b\, \ln\frac{|q_1 + q_2|}{|q_1 - q_2|}. \tag{5.36}$$

In this case the cancellation of infrared divergences can be readily checked by expanding Eq. (5.33) for small $q_1$:

$$I_{31} = \frac{116}{315} - \frac{1}{3}\frac{q_2^2}{q_1^2} - \frac{188}{735}\frac{q_1^2}{q_2^2} + \mathcal{O}(q_1/q_2)^4, \tag{5.37}$$

with $I_{31}$ being convergent as $q_2 \to 0$ (in fact, $I_{31} \propto q_2^2$ as $q_2 \to 0$, see Eq. (4.4b) and property (1) in Sec. II B). In Table XIV we show the results for individual diagrams and their sum. We see that the logarithmic terms in $I_{31}$ are responsible for the very weak $n$-dependence in the final results. This effect does not happen at the tree-level, because tree diagrams correspond to taking the angular average of the perturbation theory kernels, which results in constants independent of the $q_i$'s [19] and therefore independent of the amount of power at a given scale. Recall that the loop corrections in the ZA were independent of $n$. This was due to the absence of logarithmic terms, which come from the angular dependence in the denominator of perturbation theory kernels and appear only when solving the exact Poisson equation.

From Eqns. (5.17) and (5.36) it follows that $\overline{k^m\,k^{-m}}$ and $\overline{\ln(k^m, k^{-m})}$ are invariant under $n \to -(n+6)$. This implies that for the exact dynamics the results for the 1-loop variance are symmetric about $n = -3$. The large $n$ asymptotes can be determined from the fact that:

$$\overline{k^m\,k^{-m}} \approx \frac{(n+3)^2}{(n+3)^2 - m^2}, \tag{5.38}$$

for $n+3 \neq \pm m$. Therefore, $\overline{k^m\,k^{-m}}$ becomes unity as $\pm n$ increases. Similarly, using the expansion:

$$\ln\frac{|q_1+q_2|}{|q_1-q_2|} = \sum_{k=0}^{\infty} \frac{2}{2k+1}\left(\frac{q_<}{q_>}\right)^{2k+1}, \tag{5.39}$$

where $q_<$ ($q_>$) is the smaller (larger) of $\{q_1, q_2\}$, we obtain:

$$\overline{\ln(k^m, k^{-m})} = \sum_{k=0}^{\infty} \frac{2}{2k+1}\,\overline{k^{m+2k+1}\,k^{-m-2k-1}}, \tag{5.40}$$

and therefore the logarithmic terms in Eq. (5.33) cancel each other in the large $\pm n$ limit. This leaves us with the one loop result:

$$\sigma^2 = \sigma_\ell^2 + \frac{4007}{2205}\,\sigma_\ell^4 + \mathcal{O}(\sigma_\ell^6) \approx \sigma_\ell^2 + 1.817\,\sigma_\ell^4 + \mathcal{O}(\sigma_\ell^6), \tag{5.41}$$

in the large $\pm n$ limit.

We can therefore summarize the results for the one-loop corrections to the variance:

$$\sigma^2 = \sigma_\ell^2 + 1.82\,\sigma_\ell^4 + \mathcal{O}(\sigma_\ell^6), \tag{5.42}$$

which is independent of $n$ to better than 1%. This analytic result was recently confirmed by numerical integrations given in [32]. Comparing the ED result to the equivalent corrections obtained in the non-linear approximations, we see that the latter underestimate the exact results by 30 % (ZA), 60 % (LPA), and 76 % (FFA) (see Fig. 15).

TABLE XIV. One-loop corrections to $\sigma^2$ in the Exact Dynamics ($\Lambda \equiv k_c/\epsilon$)

| $n$ | $\overline{\mathcal{D}_{31}}$ | $\overline{\mathcal{D}_{22}}$ | $s^{(1)} = \overline{\mathcal{D}_{31}} + \overline{\mathcal{D}_{22}}$ |
|---|---|---|---|
| $-2$ | $-(1/9)\Lambda + (17/189 - \pi^2/336)$ | $(1/9)\Lambda + 3902/2205$ | $12301/6615 - \pi^2/336 \approx 1.830$ |
| $-1$ | $-(1/3)\ln\Lambda + (1634/6615 - (\ln 2)256/2205)$ | $(1/3)\ln\Lambda + 1219/735$ | $2521/1323 - (\ln 2)256/2205 \approx 1.825$ |
| $0$ | $-179/315 + 3\pi^2/224$ | $332/147$ | $3727/2205 + 3\pi^2/224 \approx 1.822$ |
| $1$ | $-7726/19845 + (\ln 2)1024/6615$ | $4637/2205$ | $34007/19845 + (\ln 2)1024/6615 \approx 1.821$ |
| $2$ | $-17/105 - 5\pi^2/672$ | $4532/2205$ | $835/441 - 5\pi^2/672 \approx 1.820$ |

## 2. Loop Corrections for $S_3$

The one-loop corrections to $S_3$ are calculated using the expressions obtained in Sec. IV C and the results of integrals given in the Appendices. Although the following results are much more complex than in the case of the variance, their structure is similar, in the sense that logarithmic terms introduce a small variation of $S_3^{(1)}$ with $n$ and infrared divergences in individual diagrams (for $n = -2, -1$) also cancel when summing all the contributions. For $n = -2$ we have:

$$\overline{\mathcal{D}_{222}} = -\frac{2}{9}\Lambda + \frac{293404}{77175}, \tag{5.43}$$

$$\overline{\mathcal{D}_{321}^I} = \frac{32}{21}\Lambda + \frac{3\pi^2}{112}\ln(\Lambda) + \frac{25223}{1080} - \frac{\pi^2}{1470} - \frac{3}{16}\zeta(3), \tag{5.44}$$

$$\overline{\mathcal{D}_{321}^{II}} = -\frac{34}{63}\Lambda + \frac{578}{1323} - \frac{17\pi^2}{1176}, \tag{5.45}$$

$$\overline{\mathcal{D}_{411}} = -\frac{16}{21}\Lambda - \frac{3\pi^2}{112}\ln(\Lambda) + \frac{4951280852713}{6571901952000} + \frac{1035171259\pi^2}{25035816960} + \frac{9\pi^4}{68992} + \frac{1576973}{20697600}\ln(2) - \frac{32\pi^2}{8085}\ln(2) - \frac{544}{40425}\ln^2(2)$$
$$- \frac{1}{1617}\ln^3(2) + \frac{50392659319}{10709766144000}\ln(3) - \frac{7577\pi^2}{37255680}\ln(3) + \frac{275115257}{10431590400}\ln(2)\ln(3) - \frac{275115257}{20863180800}\ln^2(3)$$
$$+ \frac{7577}{37255680}\ln^3(3) - \frac{275115257}{10431590400}\text{Li}_2(1/3) - \frac{13}{4928}\text{Li}_3(-1/2) + \frac{7577}{6209280}\text{Li}_3(-1/3) - \frac{7577}{3104640}\text{Li}_3(1/3)$$
$$+ \frac{219}{34496}\text{Li}_3(1/2) + \frac{2231753}{12418560}\zeta(3), \tag{5.46}$$

$$-2s^{(1)}S_3^{(0)} = -\frac{836468}{46305} + \frac{17\pi^2}{588}, \tag{5.47}$$

where $\text{Li}_n(x)$ denotes the polylogarithm of order $n$ (see Eq. (B8) ) and $\zeta(x)$ is the Riemann zeta function. Therefore,

$$S_3^{(1)}(-2) = \frac{6757496010073}{6571901952000} + \frac{690290107\pi^2}{25035816960} + \frac{9\pi^4}{68992} + \frac{1576973}{20697600}\ln(2) - \frac{32\pi^2}{8085}\ln(2) - \frac{544}{40425}\ln^2(2) - \frac{1}{1617}\ln^3(2)$$
$$+ \frac{50392659319}{10709766144000}\ln(3) - \frac{7577\pi^2}{37255680}\ln(3) + \frac{275115257}{10431590400}\ln(2)\ln(3) - \frac{275115257}{20863180800}\ln^2(3)$$
$$+ \frac{7577}{37255680}\ln^3(3) - \frac{275115257}{10431590400}\text{Li}_2(1/3) - \frac{13}{4928}\text{Li}_3(-1/2) + \frac{7577}{6209280}\text{Li}_3(-1/3) - \frac{7577}{3104640}\text{Li}_3(1/3)$$
$$+ \frac{219}{34496}\text{Li}_3(1/2) + \frac{96727}{12418560}\zeta(3)$$
$$\approx 10.034 \tag{5.48}$$

For $n = -1$ we have:

$$\overline{\mathcal{D}_{222}} = -\frac{2}{3}\ln(\Lambda) + \frac{34506}{8575}, \tag{5.49}$$

$$\overline{\mathcal{D}_{321}^I} = \frac{32}{7}\ln(\Lambda) + \frac{5010232}{231525} + \frac{68512}{77175}\ln(2), \tag{5.50}$$

$$\overline{\mathcal{D}_{321}^{II}} = -\frac{34}{21}\ln(\Lambda) + \frac{55556}{46305} - \frac{8704}{15435}\ln(2), \tag{5.51}$$

$$\overline{\mathcal{D}_{411}} = -\frac{16}{7}\ln(\Lambda) + \frac{286837213}{142619400} - \frac{73963\pi^2}{95079600} - \frac{1120128}{660275}\ln(2) - \frac{27\pi^2}{7546}\ln(2) + \frac{19968}{660275}\ln^2(2) - \frac{3}{539}\ln^3(2)$$
$$- \frac{324178}{1037575}\ln(3) + \frac{8}{8085}\ln(2)\ln(3) - \frac{4}{8085}\ln^2(3) - \frac{8}{8085}\text{Li}_2(1/3) + \frac{18}{539}\text{Li}_3(1/2) - \frac{249}{8624}\zeta(3), \tag{5.52}$$

$$-2s^{(1)}S_3^{(0)} = -\frac{171428}{9261} + \frac{17408}{15435}\ln(2), \tag{5.53}$$

and therefore:

$$\begin{aligned}S_3^{(1)}(-1) =\ & \frac{1478165197}{142619400} - \frac{73963\pi^2}{95079600} - \frac{161632}{660275}\ln(2) - \frac{27\pi^2}{7546}\ln(2) + \frac{19968}{660275}\ln^2(2) - \frac{3}{539}\ln^3(2) - \frac{324178}{1037575}\ln(3) \\ & + \frac{8}{8085}\ln(2)\ln(3) - \frac{4}{8085}\ln^2(3) - \frac{8}{8085}\mathrm{Li}_2(1/3) + \frac{18}{539}\mathrm{Li}_3(1/2) - \frac{249}{8624}\zeta(3) \\ \approx\ & 9.815 \end{aligned} \tag{5.54}$$

For $n = 0$ we have:

$$\overline{\mathcal{D}_{222}} = \frac{24216}{8575}, \tag{5.55}$$

$$\overline{\mathcal{D}_{321}^{I}} = \frac{68608}{2205} - \frac{543\pi^2}{7840}, \tag{5.56}$$

$$\overline{\mathcal{D}_{321}^{II}} = -\frac{6086}{2205} + \frac{51\pi^2}{784}, \tag{5.57}$$

$$\begin{aligned}\overline{\mathcal{D}_{411}} =\ & -\frac{350978055455171}{66266678016000} + \frac{105842425\pi^2}{818107136} + \frac{513\pi^4}{1103872} - \frac{54676}{471625}\ln(2) + \frac{799\pi^2}{150920}\ln(2) - \frac{192}{8575}\ln^2(2) - \frac{131}{37730}\ln^3(2) \\ & + \frac{8360129690981}{35996713984000}\ln(3) + \frac{893\pi^2}{7589120}\ln(3) - \frac{369823547}{2191358400}\ln(2)\ln(3) + \frac{369823547}{4382716800}\ln^2(3) - \frac{893}{7589120}\ln^3(3) \\ & + \frac{369823547}{2191358400}\mathrm{Li}_2(1/3) - \frac{2679}{3794560}\mathrm{Li}_3(-1/3) + \frac{2679}{1897280}\mathrm{Li}_3(1/3) + \frac{393}{18865}\mathrm{Li}_3(1/2) - \frac{44817}{7589120}\zeta(3), \end{aligned} \tag{5.58}$$

$$-2s^{(1)}S_3^{(0)} = -\frac{171428}{9261} + \frac{17408}{15435}\ln(2), \tag{5.59}$$

and therefore:

$$\begin{aligned}S_3^{(1)}(0) =\ & \frac{627058864885309}{66266678016000} - \frac{20192731\pi^2}{4090535680} + \frac{513\pi^4}{1103872} - \frac{54676}{471625}\ln(2) + \frac{799\pi^2}{150920}\ln(2) - \frac{192}{8575}\ln^2(2) - \frac{131}{37730}\ln^3(2) \\ & + \frac{8360129690981}{35996713984000}\ln(3) + \frac{893\pi^2}{7589120}\ln(3) - \frac{369823547}{2191358400}\ln(2)\ln(3) + \frac{369823547}{4382716800}\ln^2(3) - \frac{893}{7589120}\ln^3(3) \\ & + \frac{369823547}{2191358400}\mathrm{Li}_2(1/3) - \frac{2679}{3794560}\mathrm{Li}_3(-1/3) + \frac{2679}{1897280}\mathrm{Li}_3(1/3) + \frac{393}{18865}\mathrm{Li}_3(1/2) - \frac{44817}{7589120}\zeta(3) \\ \approx\ & 9.699 \end{aligned} \tag{5.60}$$

For $n = 1$ we have:

$$\overline{\mathcal{D}_{222}} = \frac{241954}{77175}, \tag{5.61}$$

$$\overline{\mathcal{D}_{321}^{I}} = \frac{19963372}{694575} - \frac{153088}{231525}\ln(2), \tag{5.62}$$

$$\overline{\mathcal{D}_{321}^{II}} = -\frac{262684}{138915} + \frac{34816}{46305}\ln(2), \tag{5.63}$$

$$\begin{aligned}\overline{\mathcal{D}_{411}} =\ & -\frac{90306554812}{26473726125} - \frac{475876271\pi^2}{22187503800} - \frac{13563074048}{344158439625}\ln(2) + \frac{1286\pi^2}{1867635}\ln(2) + \frac{1859584}{28014525}\ln^2(2) - \frac{4}{231}\zeta(3) \\ & - \frac{52}{266805}\ln^3(2) + \frac{43036964}{52026975}\ln(3) - \frac{64}{31185}\ln(2)\ln(3) + \frac{32}{31185}\ln^2(3) + \frac{64}{31185}\mathrm{Li}_2(1/3) + \frac{104}{88935}\mathrm{Li}_3(1/2), \end{aligned} \tag{5.64}$$

$$-2s^{(1)}S_3^{(0)} = -\frac{2312476}{138915} - \frac{69632}{46305}\ln(2), \tag{5.65}$$

and therefore:

$$\begin{aligned}S_3^{(1)}(1) &= \frac{262834942358}{26473726125} - \frac{475876271\pi^2}{22187503800} - \frac{499893398528}{344158439625}\ln(2) + \frac{1286\pi^2}{1867635}\ln(2) + \frac{1859584}{28014525}\ln^2(2) - \frac{4}{231}\zeta(3) \\ &\quad - \frac{52}{266805}\ln^3(2) + \frac{43036964}{52026975}\ln(3) - \frac{64}{31185}\ln(2)\ln(3) + \frac{32}{31185}\ln^2(3) + \frac{64}{31185}\mathrm{Li}_2(1/3) + \frac{104}{88935}\mathrm{Li}_3(1/2) \\ &\approx 9.635\end{aligned} \tag{5.66}$$

For $n = 2$ we have:

$$\overline{\mathcal{D}_{222}} = \frac{249304}{77175}, \tag{5.67}$$

$$\overline{\mathcal{D}_{321}^{I}} = \frac{301492}{11025} + \frac{1061\pi^2}{37632}, \tag{5.68}$$

$$\overline{\mathcal{D}_{321}^{II}} = -\frac{578}{735} - \frac{85\pi^2}{2352}, \tag{5.69}$$

$$\begin{aligned}\overline{\mathcal{D}_{411}} &= -\frac{667518526399}{707011905600} - \frac{909575099\pi^2}{8701684992} - \frac{276209456}{509864355}\ln(2) - \frac{19808}{3776773}\ln^2(2) - \frac{161239541}{881760880}\ln(3) + \frac{17807}{66066}\ln(2)\ln(3) \\ &\quad - \frac{17807}{132132}\ln^2(3) - \frac{17807}{66066}\mathrm{Li}_2(1/3) + \frac{8055}{68992}\zeta(3),\end{aligned} \tag{5.70}$$

$$-2s^{(1)}S_3^{(0)} = -\frac{56780}{3087} + \frac{85\pi^2}{1176}, \tag{5.71}$$

and therefore:

$$\begin{aligned}S_3^{(1)}(2) &= \frac{7390246004417}{707011905600} - \frac{21860303\pi^2}{543855312} - \frac{276209456}{509864355}\ln(2) - \frac{19808}{3776773}\ln^2(2) - \frac{161239541}{881760880}\ln(3) + \frac{17807}{66066}\ln(2)\ln(3) \\ &\quad - \frac{17807}{132132}\ln^2(3) - \frac{17807}{66066}\mathrm{Li}_2(1/3) + \frac{8055}{68992}\zeta(3) \\ &\approx 9.561\end{aligned} \tag{5.72}$$

We can summarize the results for the one loop corrections to the skewness factor in the exact dynamics writing:

$$S_3 \approx 4.86 + 9.80\ \sigma_\ell^2 + \mathcal{O}(\sigma_\ell^4), \tag{5.73}$$

which is independent of $n$ to better than 3% when $-2 \leq n \leq 2$. Comparing to the equivalent corrections given in the non-linear approximations, we see that these underestimate the exact results by 52 % (ZA), 78 % (LPA), and 90 % (FFA) (see Fig. 16).

## VI. APPLICATIONS: LOOP CORRECTIONS FOR VELOCITY FIELD DIVERGENCE CUMULANTS

In order to simplify the expressions for the moments of the divergence of the velocity field it is convenient to introduce the normalized velocity divergence or expansion scalar [6] given by:

$$\Theta(\mathbf{k}, \tau) \equiv \frac{\theta(\mathbf{k}, \tau)}{\mathcal{H}(\tau)}, \tag{6.1}$$

so that to tree-level the variance of $\Theta$ is simply $\sigma_\ell^2$ (see Eq. (2.6)). The same expressions derived in Section IV for the density field are valid for the normalized velocity divergence upon replacing the kernels $F_n$'s by $G_n$'s. Loop expansion coefficients for one-point cumulants in this case follow analogous notation, namely, $t^{(1)}$ gives the one-loop correction to $\sigma_\Theta^2$, $T_3^{(1)}$ the one loop correction to $T_3$ and so on. Also, we shall refer to the amplitude of diagrams given by the diagrammatic rules for $\Theta$ by $\mathcal{T}_{i\ldots j}$, in similar notation to $\mathcal{D}_{i\ldots j}$ in the case of the density field. We now proceed to present the results of loop corrections for the normalized velocity divergence.

## A. Loop Corrections for Non-Linear Approximations

### 1. Loop Corrections for $\sigma_\Theta^2$

One-loop corrections to the variance of the normalized velocity divergence are presented in Table XV (ZA) and Table XVI (LPA). Since the FFA assumes that the velocity field remains linear, all the loop corrections to one-point cumulants of the normalized velocity divergence vanish in this approximation. For the ZA, tables XVII and XVIII present results for two and three loop corrections to $\sigma_\Theta$ respectively. We can summarize these results for ZA by writing:

$$\sigma_\Theta^2 = \sigma_\ell^2 + \frac{11}{15}\sigma_\ell^4 + \frac{13}{15}\sigma_\ell^6 + \frac{4733}{3375}\sigma_\ell^8 + \mathcal{O}(\sigma_\ell^{10}) \approx \sigma_\ell^2 + 0.733\,\sigma_\ell^4 + 0.866\,\sigma_\ell^6 + 1.402\,\sigma_\ell^8 + \mathcal{O}(\sigma_\ell^{10}), \tag{6.2}$$

or, by inverting Eq. (5.22):

$$\sigma_\Theta^2(\sigma^2) = \sigma^2 - \frac{8}{15}\sigma^4 + \frac{44}{225}\sigma^6 - \frac{4}{15}\sigma^8 + \mathcal{O}(\sigma^{10}) \approx \sigma^2 - 0.533\,\sigma^4 + 0.195\,\sigma^6 - 0.266\,\sigma^8 + \mathcal{O}(\sigma^{10}). \tag{6.3}$$

as a function of the non-linear variance. Similarly, for LPA we have:

$$\sigma_\Theta^2 \approx \sigma_\ell^2 + 0.08\,\sigma_\ell^4 + \mathcal{O}(\sigma_\ell^6), \tag{6.4}$$

which is approximately valid when $n > -1$. For $n = -2, -1$, the one-loop correction is infrared divergent, as it was in the case of the variance of the density field. These results are shown in Fig. 18.

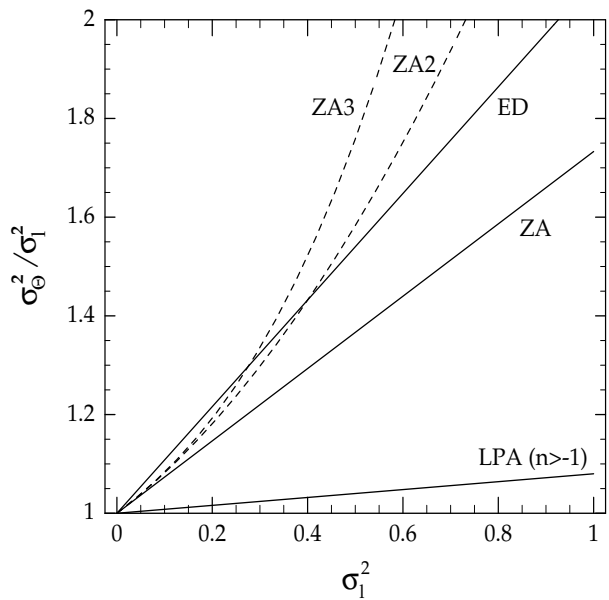

FIG. 18. Same as Fig. 15 but for the variance of the velocity divergence.

TABLE XV. One-loop corrections to $\sigma_\Theta^2$ in the Zel'dovich Approximation ($\Lambda \equiv k_c/\epsilon$)

| $n$ | $\overline{\mathcal{T}_{31}}$ | $\overline{\mathcal{T}_{22}}$ | $t^{(1)} = \overline{\mathcal{T}_{31}} + \overline{\mathcal{T}_{22}}$ |
|---|---|---|---|
| $-2$ | $-(1/9)\Lambda - 1/9$ | $(1/9)\Lambda + 38/45$ | $11/15 \approx 0.733$ |
| $-1$ | $-(1/3)\ln\Lambda$ | $(1/3)\ln\Lambda + 11/15$ | $11/15 \approx 0.733$ |
| $0$ | $-3/5$ | $4/3$ | $11/15 \approx 0.733$ |
| $1$ | $-4/9$ | $53/45$ | $11/15 \approx 0.733$ |
| $2$ | $-25/63$ | $356/315$ | $11/15 \approx 0.733$ |

TABLE XVI. One-loop corrections to $\sigma_\Theta^2$ in the Linear Potential Approximation ($\Lambda \equiv k_c/\epsilon$)

| $n$ | $\overline{\mathcal{T}_{31}}$ | $\overline{\mathcal{T}_{22}}$ | $t^{(1)} = \overline{\mathcal{T}_{31}} + \overline{\mathcal{T}_{22}}$ |
|---|---|---|---|
| $-2$ | $-(8/315)\Lambda - 8/315$ | $(4/225)\Lambda + 152/1125$ | $-(4/525)\Lambda + 96/875$ |
| $-1$ | $-(8/105)\ln\Lambda$ | $(4/75)\ln\Lambda + 44/375$ | $(4/175)\ln\Lambda + 44/375$ |
| $0$ | $-24/175$ | $16/75$ | $8/105 \approx 0.076$ |
| $1$ | $-32/315$ | $212/1125$ | $76/875 \approx 0.087$ |
| $2$ | $-40/441$ | $1424/7875$ | $552/6125 \approx 0.090$ |

TABLE XVII. Two-loop corrections to $\sigma_\Theta^2$ in the Zel'dovich Approximation ($\Lambda \equiv k_c/\epsilon$)

| $n$ | $\overline{\mathcal{T}_{24}}$ | $\overline{\mathcal{T}_{33}^I}$ | $\overline{\mathcal{T}_{33}^{II}} = \overline{\mathcal{T}_{51}}$ | $t^{(2)}$ |
|---|---|---|---|---|
| $-2$ | $-(1/45)\Lambda^2 - (2/5)\Lambda - 2/5$ | $(1/90)\Lambda^2 + (7/18)\Lambda + 113/90$ | $(1/180)\Lambda^2 + (1/180)\Lambda + 1/180$ | $13/15 \approx 0.866$ |
| $-1$ | $-(4/27)\ln^2\Lambda - (17/15)\ln\Lambda$ | $(2/27)\ln^2\Lambda + (17/15)\ln\Lambda + 13/15$ | $(1/27)\ln^2\Lambda$ | $13/15 \approx 0.866$ |
| $0$ | $-432/175$ | $3277/1050$ | $3/28$ | $13/15 \approx 0.866$ |
| $1$ | $-26/15$ | $112/45$ | $1/18$ | $13/15 \approx 0.866$ |
| $2$ | $-7760/5103$ | $117451/51030$ | $125/2916$ | $13/15 \approx 0.866$ |

TABLE XVIII. Three-loop corrections to $\sigma_\Theta^2$ in the Zel'dovich Approximation ($\Lambda \equiv k_c/\epsilon$)

| $n$ | $\overline{\mathcal{T}_{62}} = \overline{\mathcal{T}_{44}^{II}}$ | $\overline{\mathcal{T}_{71}}$ |
|---|---|---|
| $-2$ | $(1/756)\Lambda^3 + (179/3150)\Lambda^2 + (493/6300)\Lambda + 157/1575$ | $-(\Lambda^3 + \Lambda^2 + \Lambda + 1)/4536$ |
| $-1$ | $(1/54)\ln^3\Lambda + (4763/11340)\ln^2\Lambda$ | $-(1/324)\ln^3\Lambda$ |
| $0$ | $512/375$ | $-1/72$ |
| $1$ | $674/945$ | $-2/405$ |
| $2$ | $4585600/8251551$ | $-625/192456$ |

| $n$ | $\overline{\mathcal{T}_{44}^I}$ | $\overline{\mathcal{T}_{53}^{II}}$ |
|---|---|---|
| $-2$ | $(1/1134)\Lambda^3 + (1586/14175)\Lambda^2 + (38239/28350)\Lambda + 198028/70875$ | $-(\Lambda^3 + \Lambda^2 + \Lambda + 1)/1512$ |
| $-1$ | $(1/81)\ln^3\Lambda + (4763/5670)\ln^2\Lambda + (806/225)\ln\Lambda + 4733/3375$ | $-(1/108)\ln^3\Lambda$ |
| $0$ | $35336/3375$ | $-1/24$ |
| $1$ | $536213/70875$ | $-2/135$ |
| $2$ | $6964713704/1031443875$ | $-625/64152$ |

| $n$ | $\overline{\mathcal{T}_{53}^I}$ | $t^{(3)}$ |
|---|---|---|
| $-2$ | $-(1/378)\Lambda^3 - (2123/9450)\Lambda^2 - (677/450)\Lambda - 5009/3150$ | $4733/3375 \approx 1.402$ |
| $-1$ | $-(1/27)\ln^3\Lambda - (4763/2835)\ln^2\Lambda - (806/225)\ln\Lambda$ | $4733/3375 \approx 1.402$ |
| $0$ | $-8807/750$ | $4733/3375 \approx 1.402$ |
| $1$ | $-35768/4725$ | $4733/3375 \approx 1.402$ |
| $2$ | $-106419983/16503102$ | $4733/3375 \approx 1.402$ |

## 2. Loop Corrections for $T_3$

Table XIX shows the results of tree-level calculations of $T_3$ for different dynamics in NLCPT. One-loop corrections to $T_3$ are shown in Table XX for the ZA and Table XXI for the LPA, and Table XXII presents the results of two-loop corrections to $T_3$ for ZA. We can summarize the results for ZA as follows:

$$T_3 = -2 - \frac{368}{225}\sigma_\ell^2 - \frac{12916}{3375}\sigma_\ell^4 + \mathcal{O}(\sigma_\ell^6) \approx -2 - 1.635\ \sigma_\ell^2 - 3.827\ \sigma_\ell^4 + \mathcal{O}(\sigma_\ell^6), \tag{6.5}$$

or, by inverting Eq. (5.22):

$$T_3(\sigma^2) = -2 - \frac{368}{225}\sigma^2 - \frac{5924}{3375}\sigma^4 + \mathcal{O}(\sigma^6) \approx -2 - 1.635\ \sigma^2 - 1.755\ \sigma^4 + \mathcal{O}(\sigma^6), \tag{6.6}$$

as a function of the non-linear variance. For LPA we have:

$$T_3 \approx -0.9 - 0.19\ \sigma_\ell^2 + \mathcal{O}(\sigma_\ell^4), \tag{6.7}$$

approximately independent of $n$ when $n > -1$. Fig. 19 shows these results.

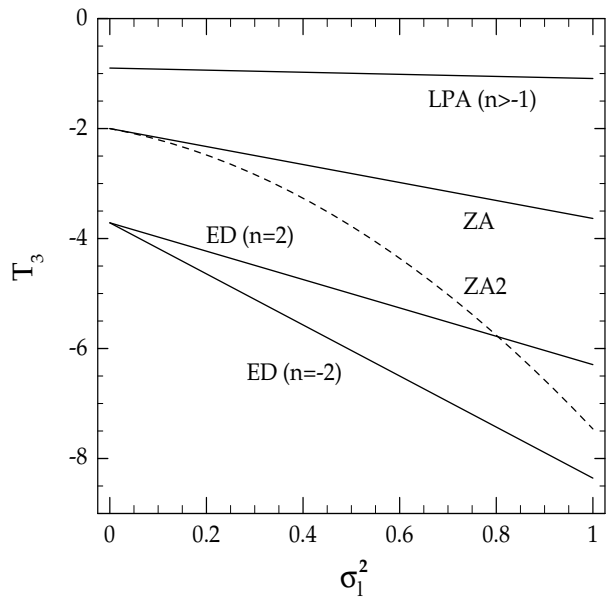

FIG. 19. Same as Fig. 16 but for the skewness of the velocity divergence.

TABLE XIX. Tree-level $T_3$ and $T_4$ for different dynamics in Perturbation Theory

| Dynamics | $T_3^{(0)} = \overline{\mathcal{T}_{211}}$ | $\overline{\mathcal{T}_{2211}}$ | $\overline{\mathcal{T}_{3111}}$ | $T_4^{(0)} = \overline{\mathcal{T}_{2211}} + \overline{\mathcal{T}_{3111}}$ |
|---|---|---|---|---|
| ED | $-26/7 \approx -3.714$ | $2704/147$ | $568/63$ | $12088/441 \approx 27.410$ |
| ZA | $-2$ | $16/3$ | $8/3$ | $8$ |
| LPA | $-4/5 = -0.9$ | $64/75$ | $64/105$ | $256/175 \approx 1.463$ |
| FFA | $0$ | $0$ | $0$ | $0$ |

TABLE XX. One-loop corrections to $T_3$ in the Zel'dovich Approximation ($\Lambda \equiv k_c/\epsilon$)

| $n$ | $\overline{\mathcal{T}_{222}}$ | $\overline{\mathcal{T}_{321}^{I}}$ | $\overline{\mathcal{T}_{321}^{II}}$ | $\overline{\mathcal{T}_{411}}$ | $T_3^{(1)\,\text{a}}$ |
|---|---|---|---|---|---|
| $-2$ | $(2/9)\Lambda + 4/75$ | $-(8/9)\Lambda - 238/45$ | $(2/9)\Lambda + 2/9$ | $(4/9)\Lambda + 4/9$ | $-368/225 \approx -1.635$ |
| $-1$ | $(2/3)\ln\Lambda - 38/225$ | $-(8/3)\ln\Lambda - 22/5$ | $(2/3)\ln\Lambda$ | $(4/3)\ln\Lambda$ | $-368/225 \approx -1.635$ |
| $0$ | $232/225$ | $-46/5$ | $6/15$ | $12/15$ | $-368/225 \approx -1.635$ |
| $1$ | $18/25$ | $-358/45$ | $8/9$ | $16/9$ | $-368/225 \approx -1.635$ |
| $2$ | $328/525$ | $-2386/315$ | $50/63$ | $100/63$ | $-368/225 \approx -1.635$ |

[a] See Eq. (4.23b). In this case $2t^{(1)}T_3^{(0)} = -44/15$ independent of spectral index.

TABLE XXI. One-loop corrections to $T_3$ in the Linear Potential Approximation ($\Lambda \equiv k_c/\epsilon$)

| $n$ | $\overline{\mathcal{T}_{222}}$ | $\overline{\mathcal{T}_{321}^{I}}$ | $\overline{\mathcal{T}_{321}^{II}}$ |
|---|---|---|---|
| $-2$ | $(16/1125)\Lambda + 32/9375$ | $-(128/1575)\Lambda - 544/1125$ | $(32/1575)\Lambda + 32/1575$ |
| $-1$ | $(16/375)\ln\Lambda - 304/875$ | $-(128/525)\ln\Lambda - 352/2625$ | $(32/525)\ln\Lambda$ |
| $0$ | $1856/28125$ | $-736/875$ | $96/875$ |
| $1$ | $144/3125$ | $-5728/7875$ | $128/1575$ |
| $2$ | $2624/65625$ | $-38176/55125$ | $32/441$ |

| $n$ | $\overline{\mathcal{T}_{411}}$ | $-2t^{(1)}T_3^{(0)}$ | $T_3^{(1)}$ |
|---|---|---|---|
| $-2$ | $(304/4725)\Lambda + 352/4725$ | $-(32/2625)\Lambda + 768/4375$ | $(128/23625)\Lambda - 123904/590625$ |
| $-1$ | $(304/1575)\ln\Lambda + 16/1575$ | $-(32/875)\ln\Lambda + 352/1875$ | $(128/7875)\ln\Lambda - 42368/196875$ |
| $0$ | $2816/7875$ | $64/525$ | $-36608/196875 \approx -0.186$ |
| $1$ | $1264/4725$ | $608/4375$ | $-114304/590625 \approx -0.193$ |
| $2$ | $7936/33075$ | $4416/30625$ | $-809728/4134375 \approx -0.196$ |

TABLE XXII. Two-loop corrections to $T_3$ in the Zel'dovich Approximation ($\Lambda \equiv k_c/\epsilon$)

| $n$ | $\overline{\mathcal{T}_{332}^{I}}$ | $\overline{\mathcal{T}_{332}^{II}}$ | $\overline{\mathcal{T}_{332}^{III}}$ |
|---|---|---|---|
| $-2$ | $(37/405)\Lambda^2 + (2626/2025)\Lambda + 1511/3375$ | $(8/135)\Lambda^2 + (46/135)\Lambda + 17/45$ | $-(1/162)\Lambda^2 - (1/81)\Lambda - 1/54$ |
| $-1$ | $(97/135)\ln^2\Lambda + (782/225)\ln\Lambda - 112/125$ | $(13/27)\ln^2\Lambda + (11/15)\ln\Lambda$ | $-(1/18)\ln^2\Lambda$ |
| $0$ | $6637/875$ | $99/35$ | $-9/50$ |
| $1$ | $49948/10125$ | $242/135$ | $-8/81$ |
| $2$ | $18677123/4465125$ | $54185/35721$ | $-625/7938$ |

| $n$ | $\overline{\mathcal{T}_{422}^{I}}$ | $\overline{\mathcal{T}_{422}^{II}}$ | $\overline{\mathcal{T}_{611}}$ |
|---|---|---|---|
| $-2$ | $-(44/405)\Lambda^2 - (3644/2025)\Lambda - 16208/3375$ | $-(34/405)\Lambda^2 - (292/2025)\Lambda - 124/675$ | $-(8/135)\Lambda^2 - (23/270)\Lambda - 1/9$ |
| $-1$ | $-(118/135)\ln^2\Lambda - (122/25)\ln\Lambda - 1102/375$ | $-(88/135)\ln^2\Lambda - (14/225)\ln\Lambda$ | $-(41/90)\ln^2\Lambda$ |
| $0$ | $-37936/2625$ | $-1856/875$ | $-1224/875$ |
| $1$ | $-110534/10125$ | $-2348/2025$ | $-101/135$ |
| $2$ | $-44249264/4465125$ | $-32896/35721$ | $-1000/1701$ |

| $n$ | $\overline{\mathcal{T}_{431}^{I}}$ | $\overline{\mathcal{T}_{431}^{II}}$ | $\overline{\mathcal{T}_{431}^{III}}$ |
|---|---|---|---|
| $-2$ | $-(52/405)\Lambda^2 - (7564/2025)\Lambda - 7228/675$ | $(32/405)\Lambda^2 + (529/405)\Lambda + 178/135$ | $-(28/405)\Lambda^2 - (38/405)\Lambda - 16/135$ |
| $-1$ | $-(128/135)\ln^2\Lambda - (2408/225)\ln\Lambda - 104/15$ | $(5/9)\ln^2\Lambda + (164/45)\ln\Lambda$ | $-(14/27)\ln^2\Lambda$ |
| $0$ | $-76292/2625$ | $1436/175$ | $-276/175$ |
| $1$ | $-46016/2025$ | $2318/405$ | $-68/81$ |
| $2$ | $-3726388/178605$ | $59660/11907$ | $-23500/35721$ |

| $n$ | $\overline{\mathcal{T}_{521}^{I}}$ | $\overline{\mathcal{T}_{521}^{II}}$ | $T_3^{(2)\,\text{a}}$ |
|---|---|---|---|
| $-2$ | $(32/135)\Lambda^2 + (661/225)\Lambda + 2053/675$ | $-(1/90)\Lambda^2 - (1/90)\Lambda - 1/90$ | $-12916/3375 \approx -3.827$ |
| $-1$ | $(82/45)\ln^2\Lambda + (1753/225)\ln\Lambda$ | $-(2/27)\ln^2\Lambda$ | $-12916/3375 \approx -3.827$ |
| $0$ | $17167/875$ | $-3/14$ | $-12916/3375 \approx -3.827$ |
| $1$ | $9032/675$ | $-1/9$ | $-12916/3375 \approx -3.827$ |
| $2$ | $19777/1701$ | $-125/1458$ | $-12916/3375 \approx -3.827$ |

[a] See Eq. (4.24). In this case $2t^{(1)}T_3^{(1)} + \left((t^{(1)})^2 + 2t^{(2)}\right)T_3^{(0)} = -23426/3375$ independent of spectral index.

### 3. Loop Corrections for $T_4$

One-loop corrections to $T_4$ are given for the ZA in Table XXIII and for LPA in Table XXIV. Using the tree-level values given in Table XIX, we can summarize the results for ZA as follows:

$$T_4 = 8 + \frac{2198}{125}\sigma_\ell^2 + \mathcal{O}(\sigma_\ell^4) \approx 8 + 17.584\,\sigma_\ell^2 + \mathcal{O}(\sigma_\ell^4). \tag{6.8}$$

For LPA we have ($n > -1$):

$$T_4 \approx 1.463 + \sigma_\ell^2 + \mathcal{O}(\sigma_\ell^4). \tag{6.9}$$

The kurtosis of the velocity divergence in these approximations is shown in Fig. 20.

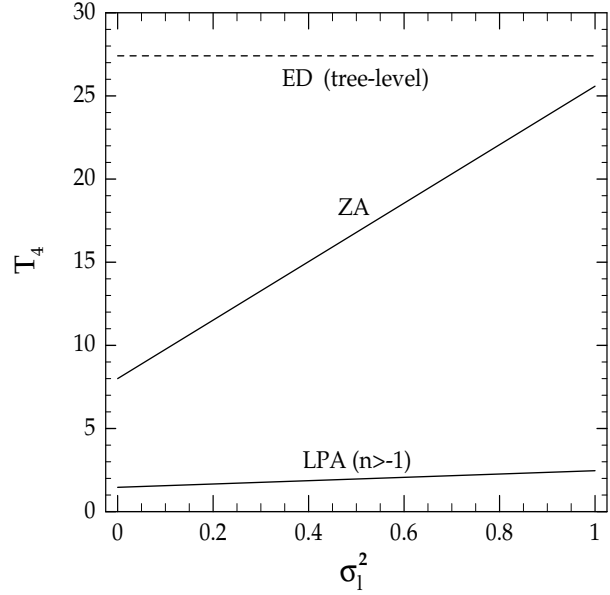

FIG. 20. Same as Fig. 16 but for the kurtosis of the velocity divergence.

TABLE XXIII. One-loop corrections to $T_4$ in the Zel'dovich Approximation ($\Lambda \equiv k_c/\epsilon$)

| $n$ | $\overline{\mathcal{T}_{5111}}$ | $\overline{\mathcal{T}_{4211}^{I}}$ | $\overline{\mathcal{T}_{4211}^{II}}$ | $\overline{\mathcal{T}_{3221}^{I}}$ |
|---|---|---|---|---|
| $-2$ | $-(40/27)\Lambda - 40/27$ | $-(64/27)\Lambda - 64/27$ | $(80/27)\Lambda + 1984/135$ | $-(4/81)\Lambda^2 - (260/81)\Lambda - 832/675$ |
| $-1$ | $-(40/9)\ln\Lambda$ | $-(64/9)\ln\Lambda$ | $(80/9)\ln\Lambda + 176/15$ | $-(4/9)\ln^2\Lambda - (28/3)\ln\Lambda + 152/75$ |
| $0$ | $-8$ | $-64/5$ | $416/15$ | $-1216/75$ |
| $1$ | $-160/27$ | $-256/27$ | $3184/135$ | $-22696/2025$ |
| $2$ | $-1000/189$ | $-1600/189$ | $21088/945$ | $-963904/99225$ |
| $n$ | $\overline{\mathcal{T}_{3221}^{II}}$ | $\overline{\mathcal{T}_{3221}^{III}}$ | $\overline{\mathcal{T}_{3311}^{I}}$ | $\overline{\mathcal{T}_{3311}^{II}}$ |
| $-2$ | $(64/27)\Lambda + 1904/135$ | $-(16/27)\Lambda - 16/27$ | $(2/81)\Lambda^2 + (184/81)\Lambda + 1498/135$ | $-(4/9)\Lambda - 4/9$ |
| $-1$ | $(64/9)\ln\Lambda + 176/15$ | $-(16/9)\Lambda$ | $(2/9)\ln^2\Lambda + (20/3)\ln\Lambda + 44/5$ | $-(4/3)\ln\Lambda$ |
| $0$ | $368/15$ | $-16/5$ | $538/25$ | $-12/5$ |
| $1$ | $2864/135$ | $-64/27$ | $7324/405$ | $-16/9$ |
| $2$ | $19088/945$ | $-400/189$ | $338386/19845$ | $-100/63$ |
| $n$ | | | $\overline{\mathcal{T}_{2222}}$ | $T_4^{(1)\,\text{a}}$ |
| $-2$ | | | $(2/81)\Lambda^2 + (40/81)\Lambda + 4756/3375$ | $2198/125 \approx 17.584$ |
| $-1$ | | | $(2/9)\ln^2\Lambda + (4/3)\ln\Lambda + 334/375$ | $2198/125 \approx 17.584$ |
| $0$ | | | $1504/375$ | $2198/125 \approx 17.584$ |
| $1$ | | | $31018/10125$ | $2198/125 \approx 17.584$ |
| $2$ | | | $1385632/496125$ | $2198/125 \approx 17.584$ |

[a]See Eq. (4.32b). In this case $3t^{(1)}T_4^{(0)} = 88/5$ independent of spectral index.

TABLE XXIV. One-loop corrections to $T_4$ in the Linear Potential Approximation ($\Lambda \equiv k_c/\epsilon$)

| $n$ | $\overline{\mathcal{T}_{5111}}$ | $\overline{\mathcal{T}_{4211}^{I}}$ |
|---|---|---|
| $-2$ | $-(21376/155925)\Lambda - 3712/22275$ | $-(9728/70875)\Lambda - 11264/70875$ |
| $-1$ | $-(21376/51975)\ln\Lambda - 512/17325$ | $-(9728/23625)\ln\Lambda - 512/23625$ |
| $0$ | $-66688/86625$ | $-90112/118125$ |
| $1$ | $-8192/14175$ | $-40448/70875$ |
| $2$ | $-566656/1091475$ | $-253952/496125$ |

| $n$ | $\overline{\mathcal{T}_{4211}^{II}}$ | $\overline{\mathcal{T}_{3221}^{I}}$ |
|---|---|---|
| $-2$ | $(3968/23625)\Lambda + 96512/118125$ | $-(1664/14175)\Lambda - 26624/590625$ |
| $-1$ | $(3968/7875)\ln\Lambda + 76672/118125$ | $-(128/7875)\ln^2\Lambda - (128/375)\ln\Lambda + 4864/65625$ |
| $0$ | $183808/118125$ | $-38912/65625$ |
| $1$ | $156032/118125$ | $-726272/1771875$ |
| $2$ | $1032704/826875$ | $-30844928/86821875$ |

| $n$ | $\overline{\mathcal{T}_{3221}^{II}}$ | $\overline{\mathcal{T}_{3221}^{III}}$ |
|---|---|---|
| $-2$ | $(2048/23625)\Lambda + 8704/16875$ | $-(512/23625)\Lambda - 512/23625$ |
| $-1$ | $(2048/7875)\Lambda + 5632/13125$ | $-(512/7875)\ln\Lambda$ |
| $0$ | $11776/13125$ | $-512/4375$ |
| $1$ | $91648/118125$ | $-2048/23625$ |
| $2$ | $610816/826875$ | $-512/6615$ |

| $n$ | $\overline{\mathcal{T}_{3311}^{I}}$ | $\overline{\mathcal{T}_{3311}^{II}}$ |
|---|---|---|
| $-2$ | $(128/99225)\Lambda^2 + (11776/99225)\Lambda + 13696/23625$ | $-(256/11025)\Lambda - 256/11025$ |
| $-1$ | $(128/11025)\ln^2\Lambda + (256/735)\ln\Lambda + 2816/6125$ | $-(256/3675)\ln\Lambda$ |
| $0$ | $34432/30625$ | $-768/6125$ |
| $1$ | $468736/496125$ | $-1024/11025$ |
| $2$ | $21656704/24310125$ | $-256/3087$ |

| $n$ | $\overline{\mathcal{T}_{2222}}$ | $-3t^{(1)}T_4^{(0)}$ |
|---|---|---|
| $-2$ | $(32/50625)\Lambda^2 + (128/10125)\Lambda + 76096/2109375$ | $(1024/30625)\Lambda - 73728/153125$ |
| $-1$ | $(32/5625)\ln^2\Lambda + (64/1875)\ln\Lambda + 5344/234375$ | $(3072/30625)\ln\Lambda - 11264/21875$ |
| $0$ | $24064/234375$ | $-2048/6125$ |
| $1$ | $496288/6328125$ | $-58368/153125$ |
| $2$ | $22170112/310078125$ | $-423936/1071875$ |

| $n$ | | $T_4^{(1)}$ |
|---|---|---|
| $-2$ | | $(32/275625)\Lambda^2 - (17408/1010625)\Lambda + 132846016/126328125$ |
| $-1$ | | $(32/30625)\ln^2\Lambda - (52928/1010625)\ln\Lambda + 405022048/378984375$ |
| $0$ | | $370578688/378984375 \approx 0.978$ |
| $1$ | | $34478368/34453125 \approx 1.001$ |
| $2$ | | $187157803 52/185702 34375 \approx 1.008$ |

TABLE XXV. One-loop corrections to $\sigma_\odot^2$ in the Exact Dynamics ($\Lambda \equiv k_c/\epsilon$)

| $n$ | $\overline{\mathcal{T}_{31}}$ | $\overline{\mathcal{T}_{22}}$ | $t^{(1)} = \overline{\mathcal{T}_{31}} + \overline{\mathcal{T}_{22}}$ |
|---|---|---|---|
| $-2$ | $-(1/9)\Lambda - (5/63 + \pi^2/112)$ | $(1/9)\Lambda + 2798/2205$ | $2623/2205 - \pi^2/112 \approx 1.101$ |
| $-1$ | $-(1/3)\ln\Lambda + (374/2205 - (\ln 2)256/735)$ | $(1/3)\ln\Lambda + 851/735$ | $2927/2205 - (\ln 2)256/735 \approx 1.086$ |
| $0$ | $-113/105 + 9\pi^2/224$ | $1292/735$ | $167/245 + 9\pi^2/224 \approx 1.078$ |
| $1$ | $-5626/6615 + (\ln 2)1024/2205$ | $3533/2205$ | $4973/6615 + (\ln 2)1024/2205 \approx 1.074$ |
| $2$ | $-83/315 - 5\pi^2/224$ | $3428/2205$ | $949/735 - 5\pi^2/224 \approx 1.071$ |

## B. Loop Corrections for the Exact Dynamics

We now present the results of one-loop corrections to one-point cumulants of the normalized velocity divergence for the exact dynamics. The same remarks made above for the density field apply here, namely, the logarithmic terms coming from the non-locality of the Poisson Green's function induce a small variation of the corrections with spectral index, and infrared divergences present for $n = -2, -1$ are exactly canceled when the sum over diagrams is done.

### 1. Loop Corrections for $\sigma_\Theta^2$

Table XXV (see previous page) gives the results of one-loop corrections to the variance of $\Theta$ for the exact dynamics. Comparing with the density field correction, we see that $\sigma_\Theta$ is less affected by non-linearities than $\sigma$, at least in the weakly non-linear regime. We can summarize the results of Table XXV by:

$$\sigma_\Theta^2 \approx \sigma_\ell^2 + 1.08\ \sigma_\ell^4 + \mathcal{O}(\sigma_\ell^6), \tag{6.10}$$

or, by inverting Eq. (5.42):

$$\sigma_\Theta^2(\sigma^2) \approx \sigma^2 - 0.74\ \sigma^4 + \mathcal{O}(\sigma^6). \tag{6.11}$$

Comparison with the one-loop corrections in the non-linear approximations indicates that ZA underestimates the correct results by 32 % and LPA by 92 % (see Fig. 18).

### 2. Loop Corrections for $T_3$

We now present the results of one-loop corrections to $T_3$ in the exact dynamics. For $n = -2$ we have:

$$\overline{\mathcal{T}_{222}} = \frac{2}{9}\Lambda - \frac{37844}{25725}, \tag{6.12}$$

$$\overline{\mathcal{T}_{321}^I} = -\frac{80}{63}\Lambda - \frac{3\pi^2}{112}\ln(\Lambda) - \frac{79711}{5880} + \frac{\pi^2}{245} + \frac{3}{16}\zeta(3), \tag{6.13}$$

$$\overline{\mathcal{T}_{321}^{II}} = \frac{26}{63}\Lambda + \frac{130}{441} + \frac{13\pi^2}{392}, \tag{6.14}$$

$$\begin{aligned}
\overline{\mathcal{T}_{411}} =\ & \frac{40}{63}\Lambda + \frac{3\pi^2}{112}\ln(\Lambda) - \frac{708960735827}{1642975488000} - \frac{810641107\pi^2}{6258954240} + \frac{9\pi^4}{17248} + \frac{1576973}{5174400}\ln(2) - \frac{128\pi^2}{8085}\ln(2) - \frac{2176}{40425}\ln^2(2) \\
& - \frac{4}{1617}\ln^3(2) - \frac{175638489101}{2677441536000}\ln(3) + \frac{73781\pi^2}{18627840}\ln(3) + \frac{178450997}{2607897600}\ln(2)\ln(3) - \frac{178450997}{5215795200}\ln^2(3) \\
& - \frac{73781}{18627840}\ln^3(3) - \frac{178450997}{2607897600}\mathrm{Li}_2(1/3) - \frac{13}{1232}\mathrm{Li}_3(-1/2) - \frac{73781}{3104640}\mathrm{Li}_3(-1/3) + \frac{73781}{1552320}\mathrm{Li}_3(1/3) \\
& + \frac{219}{8624}\mathrm{Li}_3(1/2) + \frac{1415461}{6209280}\zeta(3),
\end{aligned} \tag{6.15}$$

$$-2t^{(1)}T_3^{(0)} = \frac{136396}{15435} - \frac{13\pi^2}{196}, \tag{6.16}$$

and therefore:

$$\begin{aligned}
T_3^{(1)}(-2) =\ & -\frac{8977705003693}{1642975488000} - \frac{628620499\pi^2}{6258954240} + \frac{9\pi^4}{17248} + \frac{1576973}{5174400}\ln(2) - \frac{128\pi^2}{8085}\ln(2) - \frac{2176}{40425}\ln^2(2) \\
& - \frac{4}{1617}\ln^3(2) - \frac{175638489101}{2677441536000}\ln(3) + \frac{73781\pi^2}{18627840}\ln(3) + \frac{178450997}{2607897600}\ln(2)\ln(3) - \frac{178450997}{5215795200}\ln^2(3) \\
& - \frac{73781}{18627840}\ln^3(3) - \frac{178450997}{2607897600}\mathrm{Li}_2(1/3) - \frac{13}{1232}\mathrm{Li}_3(-1/2) - \frac{73781}{3104640}\mathrm{Li}_3(-1/3) + \frac{73781}{1552320}\mathrm{Li}_3(1/3) \\
& + \frac{219}{8624}\mathrm{Li}_3(1/2) - \frac{251221}{6209280}\zeta(3) \\
\approx\ & -4.643
\end{aligned} \tag{6.17}$$

For $n = -1$ we have:

$$\overline{\mathcal{T}_{222}} = \frac{2}{3}\ln(\Lambda) - \frac{130682}{77175}, \tag{6.18}$$

$$\overline{\mathcal{T}_{321}^{I}} = -\frac{80}{21}\ln(\Lambda) - \frac{937544}{77175} - \frac{19424}{25725}\ln(2), \tag{6.19}$$

$$\overline{\mathcal{T}_{321}^{II}} = \frac{26}{21}\ln(\Lambda) - \frac{9724}{15435} + \frac{6656}{5145}\ln(2), \tag{6.20}$$

$$\overline{\mathcal{T}_{411}} = \frac{40}{21}\ln(\Lambda) - \frac{32159713}{35654850} + \frac{73963\pi^2}{23769900} + \frac{5248736}{1980825}\ln(2) + \frac{54\pi^2}{3773}\ln(2) - \frac{79872}{660275}\ln^2(2) + \frac{12}{539}\ln^3(2)$$
$$+ \frac{1296712}{1037575}\ln(3) - \frac{32}{8085}\ln(2)\ln(3) + \frac{16}{8085}\ln^2(3) + \frac{32}{8085}\mathrm{Li}_2(1/3) - \frac{72}{539}\mathrm{Li}_3(1/2) + \frac{249}{2156}\zeta(3), \tag{6.21}$$

$$-2t^{(1)}T_3^{(0)} = \frac{152204}{15435} - \frac{13312}{5145}\ln(2), \tag{6.22}$$

and therefore:

$$T_3^{(1)}(-1) = -\frac{7862053}{1426194} + \frac{73963\pi^2}{23769900} + \frac{1190528}{1980825}\ln(2) + \frac{54\pi^2}{3773}\ln(2) - \frac{79872}{660275}\ln^2(2) + \frac{12}{539}\ln^3(2)$$
$$+ \frac{1296712}{1037575}\ln(3) - \frac{32}{8085}\ln(2)\ln(3) + \frac{16}{8085}\ln^2(3) + \frac{32}{8085}\mathrm{Li}_2(1/3) - \frac{72}{539}\mathrm{Li}_3(1/2) + \frac{249}{2156}\zeta(3)$$
$$\approx -3.577 \tag{6.23}$$

For $n = 0$ we have:

$$\overline{\mathcal{T}_{222}} = -\frac{38072}{77175}, \tag{6.24}$$

$$\overline{\mathcal{T}_{321}^{I}} = -\frac{14704}{735} + \frac{423\pi^2}{7840}, \tag{6.25}$$

$$\overline{\mathcal{T}_{321}^{II}} = \frac{2938}{735} - \frac{117\pi^2}{784}, \tag{6.26}$$

$$\overline{\mathcal{T}_{411}} = -\frac{146302676485571}{16566669504000} + \frac{142985207\pi^2}{1022633920} + \frac{513\pi^4}{275968} - \frac{218704}{471625}\ln(2) + \frac{799\pi^2}{37730}\ln(2) - \frac{768}{8575}\ln^2(2)$$
$$- \frac{262}{18865}\ln^3(2) + \frac{8360129690981}{8999178496000}\ln(3) + \frac{893\pi^2}{1897280}\ln(3) - \frac{369823547}{547839600}\ln(2)\ln(3) + \frac{369823547}{1095679200}\ln^2(3)$$
$$- \frac{893}{1897280}\ln^3(3) + \frac{369823547}{547839600}\mathrm{Li}_2(1/3) - \frac{2679}{948640}\mathrm{Li}_3(-1/3) + \frac{2679}{474320}\mathrm{Li}_3(1/3) + \frac{1572}{18865}\mathrm{Li}_3(1/2)$$
$$- \frac{44817}{1897280}\zeta(3), \tag{6.27}$$

$$-2t^{(1)}T_3^{(0)} = \frac{8684}{1715} + \frac{117\pi^2}{392}, \tag{6.28}$$

and therefore:

$$T_3^{(1)}(0) = -\frac{43185676451389}{16566669504000} + \frac{64802527\pi^2}{1022633920} + \frac{513\pi^4}{275968} - \frac{218704}{471625}\ln(2) + \frac{799\pi^2}{37730}\ln(2) - \frac{768}{8575}\ln^2(2)$$
$$-\frac{262}{18865}\ln^3(2) + \frac{8360129690981}{8999178496000}\ln(3) + \frac{893\pi^2}{1897280}\ln(3) - \frac{369823547}{547839600}\ln(2)\ln(3) + \frac{369823547}{1095679200}\ln^2(3)$$
$$-\frac{893}{1897280}\ln^3(3) + \frac{369823547}{547839600}\text{Li}_2(1/3) - \frac{2679}{948640}\text{Li}_3(-1/3) + \frac{2679}{474320}\text{Li}_3(1/3) + \frac{1572}{18865}\text{Li}_3(1/2)$$
$$-\frac{44817}{1897280}\zeta(3)$$
$$\approx -3.123, \tag{6.29}$$

For $n = 1$ we have:
$$\overline{\mathcal{T}_{222}} = -\frac{6898}{8575}, \tag{6.30}$$

$$\overline{\mathcal{T}_{321}^I} = -\frac{4170884}{231525} + \frac{37376}{77175}\ln(2), \tag{6.31}$$

$$\overline{\mathcal{T}_{321}^{II}} = \frac{146276}{46305} - \frac{26624}{15435}\ln(2), \tag{6.32}$$

$$\overline{\mathcal{T}_{411}} = \frac{150452937298}{26473726125} + \frac{475876271\pi^2}{5546875950} + \frac{1412780667392}{344158439625}\ln(2) - \frac{5144\pi^2}{1867635}\ln(2) - \frac{7438336}{28014525}\ln^2(2) + \frac{16}{231}\zeta(3)$$
$$+\frac{208}{266805}\ln^3(2) - \frac{172147856}{52026975}\ln(3) + \frac{256}{31185}\ln(2)\ln(3) - \frac{128}{31185}\ln^2(3) - \frac{256}{31185}\text{Li}_2(1/3) - \frac{416}{88935}\text{Li}_3(1/2),$$
$$\tag{6.33}$$

$$-2t^{(1)}T_3^{(0)} = \frac{258596}{46305} + \frac{53248}{15435}\ln(2), \tag{6.34}$$

and therefore:
$$T_3^{(1)}(1) = -\frac{116287648352}{26473726125} + \frac{475876271\pi^2}{5546875950} + \frac{2173099907072}{344158439625}\ln(2) - \frac{5144\pi^2}{1867635}\ln(2) - \frac{7438336}{28014525}\ln^2(2) + \frac{16}{231}\zeta(3)$$
$$+\frac{208}{266805}\ln^3(2) - \frac{172147856}{52026975}\ln(3) + \frac{256}{31185}\ln(2)\ln(3) - \frac{128}{31185}\ln^2(3) - \frac{256}{31185}\text{Li}_2(1/3) - \frac{416}{88935}\text{Li}_3(1/2)$$
$$\approx -2.871 \tag{6.35}$$

For $n = 2$ we have:
$$\overline{\mathcal{T}_{222}} = -\frac{23144}{25725}, \tag{6.36}$$

$$\overline{\mathcal{T}_{321}^I} = -\frac{186692}{11025} - \frac{247\pi^2}{12544}, \tag{6.37}$$

$$\overline{\mathcal{T}_{321}^{II}} = \frac{2158}{2205} + \frac{65\pi^2}{784}, \tag{6.38}$$

$$\overline{\mathcal{T}_{411}} = \frac{239944659679}{176752976400} + \frac{2080959611\pi^2}{8701684992} + \frac{1104837824}{509864355}\ln(2) + \frac{79232}{3776773}\ln^2(2) + \frac{161239541}{220440220}\ln(3) - \frac{35614}{33033}\ln(2)\ln(3)$$
$$+\frac{17807}{33033}\ln^2(3) + \frac{35614}{33033}\text{Li}_2(1/3) - \frac{8055}{17248}\zeta(3), \tag{6.39}$$

$$-2t^{(1)}T_3^{(0)} = \frac{49348}{5145} - \frac{65\pi^2}{392}, \tag{6.40}$$

and therefore:

$$T_3^{(1)}(2) = -\frac{1043821311809}{176752976400} + \frac{74261045\pi^2}{543855312} + \frac{1104837824}{509864355}\ln(2) + \frac{79232}{3776773}\ln^2(2) + \frac{161239541}{220440220}\ln(3) - \frac{35614}{33033}\ln(2)$$
$$\times \ln(3) + \frac{17807}{33033}\ln^2(3) + \frac{35614}{33033}\text{Li}_2(1/3) - \frac{8055}{17248}\zeta(3)$$
$$\approx -2.579 \quad (6.41)$$

We see that the dependence of the velocity divergence skewness upon $n$ is stronger than for the density field skewness. Taking $n = 1$ as a reference, we find that ZA underestimates the one-loop correction by 43 % and LPA by 93 % (see Fig. 19).

## VII. CANCELLATION OF INFRARED DIVERGENCES AND GALILEAN INVARIANCE

As we saw in the previous sections, individual loop diagrams contain infrared divergences for spectral indices $n = -2, -1$. These divergences cancel when the total contribution to a given cumulant is computed (the diagrams are summed over) for the exact dynamics and the Zel'dovich approximation. However, for LPA and FFA, infrared divergences remain in the final answer. The issue of cancellation of infrared divergences has been considered recently by [67], who showed that *leading* infrared divergences always cancel in the contribution to the power spectrum to *arbitrary* number of loops in the exact dynamics. Although this does not prove the cancellation of subdominant divergences, it suggests that there is an underlying mechanism for these cancellations. In this section, we show that this behavior can be related to the properties of the corresponding equations of motion under a Galilean transformation. This is analogous to what happens in quantum field theory, where symmetries such as gauge invariance play a critical role in the structure and cancellation of divergences. In fact, symmetries lead to relations between correlation functions (the Ward Identities) which are crucial to the proof of renormalizability [68].

The Newtonian equations of motion of gravitational instability constitute a classical non-relativistic field theory and therefore are invariant under Galilean transformations (GT). In order to understand the behavior of the exact dynamics and the non-linear approximations under GT, we recall some basic properties of GT in comoving coordinates. Under a GT, the proper coordinates $(\mathbf{r}, t)$ transform as:

$$\mathbf{r}' = \mathbf{r} - \mathbf{u}t \quad (7.1)$$
$$\mathbf{V}'(\mathbf{r}') = \mathbf{V}(\mathbf{r}) - \mathbf{u}, \quad (7.2)$$

where $\mathbf{V} \equiv d\mathbf{r}/dt$, and $\mathbf{u}$ is a uniform velocity. In terms of comoving conformal coordinates $(\mathbf{x}, \tau)$, we have $\mathbf{r} = a\mathbf{x}$ and $\mathbf{V} = \mathcal{H}\mathbf{x} + \mathbf{v}$, where $\mathbf{v} = d\mathbf{x}/d\tau$, and the GT in comoving coordinates becomes:

$$\mathbf{x}' = \mathbf{x} - \mathbf{u}T \quad (7.3)$$
$$\mathbf{v}'(\mathbf{x}') = \mathbf{v}(\mathbf{x}) - \mathbf{u}(1 - \mathcal{H}T), \quad (7.4)$$

where

$$T(\tau) \equiv \frac{1}{a(\tau)} \int_0^\tau a(\tau') \, d\tau'. \quad (7.5)$$

Consider the transformation of individual terms in the equations of motion of the exact dynamics (see Eqs. (2.1)). Since for any scalar field $\Psi$ we have $\Psi'(\mathbf{x}') = \Psi(\mathbf{x}) = \Psi(\mathbf{x}' + \mathbf{u}T)$ (except for the velocity field which also undergoes a transformation of its homogeneous mode), the partial time derivative transforms as:

$$\frac{\partial}{\partial \tau} \longrightarrow \frac{\partial}{\partial \tau} + (1 - \mathcal{H}T)\mathbf{u} \cdot \nabla, \quad (7.6)$$

where we have used the fact that $\partial T/\partial \tau = (1 - \mathcal{H}T)$. This means that the operator

$$\frac{d}{d\tau} \equiv \frac{\partial}{\partial \tau} + \mathbf{v} \cdot \nabla \quad (7.7)$$

is Galilean invariant, which in turn implies the invariance of mass conservation, Eq. (2.1a). The Galilean invariance of the momentum conservation equation can also be demonstrated; under a GT we have:

$$\frac{d\mathbf{v}}{d\tau} \longrightarrow \frac{d\mathbf{v}}{d\tau} + T\frac{\partial \mathcal{H}}{\partial \tau}\mathbf{u} + \mathcal{H}(1 - \mathcal{H}T)\mathbf{u}, \tag{7.8a}$$

$$\mathcal{H}\mathbf{v} \longrightarrow \mathcal{H}\mathbf{v} - \mathcal{H}(1 - \mathcal{H}T)\mathbf{u}, \tag{7.8b}$$

$$\nabla\Phi \longrightarrow \nabla\Phi - T\frac{\partial \mathcal{H}}{\partial \tau}\mathbf{u}, \tag{7.8c}$$

where the last expression arises because $\Phi$ is the gravitational potential of the density *fluctuations*, related to the gravitational potential of the total density $\varphi$ by $\Phi = \varphi + (1/2)x^2 \partial \mathcal{H}/\partial \tau$ [6]. Substituting Eq. (7.8) into the Euler equation, Eq. (2.1b), demonstrates the Galilean invariance of momentum conservation in a general homogeneous and isotropic background. For the Einstein-de Sitter model, we have $T = \tau/3$ and letting $\mathbf{u} \to 3\mathbf{u}$ the GT becomes

$$\frac{d\mathbf{v}}{d\tau} \longrightarrow \frac{d\mathbf{v}}{d\tau}, \tag{7.9}$$

$$\mathcal{H}\mathbf{v} \longrightarrow \mathcal{H}\mathbf{v} - \mathcal{H}\mathbf{u}, \tag{7.10}$$

$$\nabla\Phi \longrightarrow \nabla\Phi + \mathcal{H}\mathbf{u}. \tag{7.11}$$

Thus, in the Einstein-de Sitter case, the GT-dependent terms in the gravitational force and the Hubble expansion drag cancel each other.

It is also useful to view the invariance properties of the equations of motion in momentum space. In Fourier space, the density and velocity fields transform under a GT (for the Einstein-de Sitter case) as:

$$\delta(\mathbf{k}) \longrightarrow \exp(i\tau\mathbf{u}\cdot\mathbf{k})\delta(\mathbf{k}) \tag{7.12}$$

$$\mathbf{v}(\mathbf{k}) \longrightarrow \exp(i\tau\mathbf{u}\cdot\mathbf{k})\mathbf{v}(\mathbf{k}) - \mathbf{u}\delta_D(\mathbf{k}). \tag{7.13}$$

Infrared divergences come from the $\mathbf{v}\cdot\nabla$ terms in the equations of motion [67]. These "convective" terms represent the time variation of fields due to the transport of fluid elements in Eulerian space. This transport is dominated by the homogeneous mode of the velocity field, as can be seen from the infrared behavior of the mode coupling functions $\alpha(\mathbf{k},\mathbf{k}_1)$ and $\beta(\mathbf{k},\mathbf{k}_1,\mathbf{k}_2)$ (see Eq. (2.5)). Thus these divergences are just a kinematical effect due to the fact that the linear fluctuations are characterized by the density field power spectrum, and therefore the rms velocity field gets divergent contributions from the homogeneous mode for spectral indices $n = -2, -1$ [67]. To see this, since we are interested in the infrared behavior, we can use linear theory to relate the velocity field power spectrum to the density power spectrum:

$$<v_i(\mathbf{k})v_j(\mathbf{k}')> \equiv \delta_D(\mathbf{k}+\mathbf{k}')\ B_{ij}(\mathbf{k}) \approx -\mathcal{H}^2 \frac{k_i k_j}{k^4} P_1(\mathbf{k})\delta_D(\mathbf{k}+\mathbf{k}'), \tag{7.14}$$

which shows that the rms velocity, $v_{rms}$, becomes infrared divergent for spectral indices $n = -2, -1$:

$$v_{rms}^2 \equiv \int d^3k\, B_{ii}(\mathbf{k}) \approx -\mathcal{H}^2 \int d^3k \frac{P_1(\mathbf{k})}{k^2} \propto \int dk\, k^n. \tag{7.15}$$

Since the infrared divergences represent a kinematic effect, they should not appear when calculating Galilean-invariant quantities, such as the density field (equal-time) $p$-point cumulants. From (7.12), under a GT these transform as

$$<\delta(\mathbf{k}_1)\ldots\delta(\mathbf{k}_p)> \longrightarrow \exp[i\tau\mathbf{u}\cdot(\mathbf{k}_1+\ldots+\mathbf{k}_p)]<\delta(\mathbf{k}_1)\ldots\delta(\mathbf{k}_p)>, \tag{7.16}$$

and are Galilean-invariant due to translation invariance, which requires $\sum_{i=1}^{p}\mathbf{k}_i = 0$. Similar analysis applies to the velocity field divergence. Non-equal-time $p$-point cumulants are *not* invariant under GT, since the "distance" between points changes under a GT when measured non-simultaneously. In this case, it is straightforward to check that infrared divergences do not cancel in either ED or ZA. A similar effect happens when computing the phase shift of the density field, which is found to be infrared divergent [67], as expected from the fact that it is not a Galilean invariant quantity (see Eq. (7.12)).

We now examine the properties of the different non-linear approximation dynamics under a GT, in particular the momentum conservation equation, which is the one being modified from the ED in each case. For FFA, we can write the evolution for $\theta(\mathbf{x},\tau)$ as:

$$\frac{\partial \theta(\mathbf{x},\tau)}{\partial \tau} - \frac{\mathcal{H}(\tau)}{2}\theta(\mathbf{x},\tau) = 0, \tag{7.17}$$

which is obviously non-invariant under GT. We can view the non-cancellation of infrared divergences in FFA as a result of the non-cancellation of the convective terms in the infrared region, since these are absent in the momentum equation. Therefore, the breaking of Galilean invariance in this approximation leads to infrared-divergent results.

The case of LPA is a little more subtle. The velocity divergence equation is:

$$\frac{\partial \theta(\mathbf{x}, \tau)}{\partial \tau} + \mathcal{H}(\tau) \ \theta(\mathbf{x}, \tau) + \frac{3}{2}\mathcal{H}^2(\tau)\delta_1(\mathbf{x}, \tau) = -\nabla \cdot \{[\mathbf{v}(\mathbf{x}, \tau) \cdot \nabla]\mathbf{v}(\mathbf{x}, \tau)\}, \tag{7.18}$$

the only difference with the exact dynamics being the Poisson term on the left hand side. This term is only present in linear theory, being replaced by zero in higher orders. Therefore, LPA effectively treats the dynamics in an "external field" $\delta_1(\mathbf{x}, \tau)$ that is fixed by the initial conditions. Even though Eq. (7.18) is invariant under GT, Galilean invariance is "spontaneously broken" because the external force leads to a non-zero bulk velocity. By dimensional analysis, the induced rms flow is $v_{rms} \approx |\nabla \phi_1|/\mathcal{H} \approx \phi_1/(\mathcal{H}R)$, where $R$ is the characteristic comoving distance between minima of $\phi_1$ set by the initial conditions. In fact, in LPA, particles oscillate about the minima of the linear gravitational potential with a characteristic velocity which can be derived for simple potentials [49].

In the ZA, the equation of motion for $\theta(\mathbf{x}, \tau)$ is:

$$\frac{\partial \theta(\mathbf{x}, \tau)}{\partial \tau} - \frac{\mathcal{H}(\tau)}{2} \ \theta(\mathbf{x}, \tau) = -\nabla \cdot \{[\mathbf{v}(\mathbf{x}, \tau) \cdot \nabla]\mathbf{v}(\mathbf{x}, \tau)\}, \tag{7.19}$$

which is invariant under GT, consistent with the fact that loop corrections in ZA are well behaved in the infrared. Note that at the level of the equation of motion for $\mathbf{v}(\mathbf{x}, \tau)$ (i.e., Eq. (2.12)), ZA is *not* invariant under GT; however, invariance is recovered at the velocity divergence level, which is what matters for this discussion.

One can now ask whether the exact equations of motion could be modified in a different way from ZA, so as to yield a different non-linear approximation with infrared-convergent loop corrections. Starting from the exact dynamics, one can write a general class of non-linear dynamical theories of gravitational evolution in the form:

$$\lambda_1 \frac{\partial \delta(\mathbf{x}, \tau)}{\partial \tau} + \lambda_2 [\mathbf{v}(\mathbf{x}, \tau) \cdot \nabla]\delta(\mathbf{x}, \tau) + \lambda_3 [1 + \delta(\mathbf{x}, \tau)]\theta(\mathbf{x}, \tau) = 0, \tag{7.20}$$

$$\gamma_0 \frac{\partial \theta(\mathbf{x}, \tau)}{\partial \tau} + \gamma_1 \nabla \cdot \{[\mathbf{v}(\mathbf{x}, \tau) \cdot \nabla]\mathbf{v}(\mathbf{x}, \tau)\} + \gamma_2 \mathcal{H}(\tau) \ \theta(\mathbf{x}, \tau) + \frac{3}{2}\mathcal{H}^2(\tau)\gamma_3 \delta(\mathbf{x}, \tau) = 0, \tag{7.21}$$

where $\gamma_\nu$ ($\nu = 0, 1, 2, 3$) and $\lambda_i$ ($i = 1, 2, 3$) are arbitrary coefficients that characterize the dynamical model. ED corresponds to $\gamma_\nu = (1, 1, 1, 1)$, ZA to $\gamma_\nu = (1, 1, -1/2, 0)$ and FFA to $\gamma_\nu = (1, 0, -1/2, 0)$. All of them have $\lambda_i = (1, 1, 1)$ (LPA cannot be cast into this form). By dimensional analysis, one could add new terms to the equations of motion, such as $\mathcal{H}\theta\delta$ to Eq. (7.21), but their physical origin is not clear, so we do not include them. Without loss of generality, we can set $\gamma_0 = \lambda_3 \equiv 1$. Galilean invariance then is equivalent to the conditions

$$\gamma_1 = 1, \qquad \lambda_1 = \lambda_2. \tag{7.22}$$

From the generalized equations of motion (7.20), (7.21), we can derive the corresponding perturbation theory kernels by following the procedure of Section II B. In order for this family of theories to correctly describe linear gravitational instability in the limit $\delta, \theta \ll 1$, we must have:

$$\lambda_1 = 1, \qquad 2\gamma_2 - 3\gamma_3 = -1. \tag{7.23}$$

Another condition comes from the fact that $\delta(\mathbf{x}, \tau)$ and $\theta(\mathbf{x}, \tau)$ vanish at the homogeneous mode, that is $\delta(\mathbf{k} = 0) = \theta(\mathbf{k} = 0) = 0$. This plus translation invariance (which guarantees that momentum is conserved at each diagram vertex) implies that, e.g., $F_2(\mathbf{q}, -\mathbf{q}) = 0$ and $G_2(\mathbf{q}, -\mathbf{q}) = 0$. Using the perturbation theory kernels for this theory, these two conditions lead to, respectively:

$$\frac{(9\gamma_3 - 4\lambda_1\gamma_2)(\lambda_2 - 1)}{(8\gamma_2\lambda_1 - 15\gamma_3)} = 0, \tag{7.24}$$

$$\frac{\gamma_3(\lambda_2 - 1)}{(8\gamma_2\lambda_1 - 15\gamma_3)} = 0, \tag{7.25}$$

which imply that $\lambda_2 = 1$. This in turn guarantees the Galilean invariance of mass conservation.

We now calculate the divergent contribution to the one-loop variance of the density field (setting $\lambda_1 = 1$ but keeping $\lambda_2$ arbitrary for the moment):

$$\frac{4(9\gamma_3 - 4\gamma_2)(\gamma_1 + 2\gamma_2\lambda_2 - 3\gamma_3\lambda_2)^2}{27(8\gamma_2 - 15\gamma_3)^2(4\gamma_2 - 7\gamma_3)}\Lambda, \tag{7.26}$$

whereas for the divergent part of $t^{(1)}$ we get:

$$\frac{8\gamma_3(\gamma_1 + 2\gamma_2\lambda_2 - 3\gamma_3\lambda_2)^2}{9(8\gamma_2 - 15\gamma_3)^2(4\gamma_2 - 7\gamma_3)}\Lambda, \tag{7.27}$$

(a similar calculation for the cross-correlation $<\delta\Theta>$ to one loop does not add a new condition). Note that the poles of Eqs. (7.26) and (7.27), $\gamma_3 = -4/3, -2$ correspond to repulsive gravity and therefore are not relevant to the present discussion. From Eqs. (7.22),(7.23), (7.26), and (7.27), we conclude that under the assumptions made above, *Galilean invariance is a necessary and sufficient condition for the cancellation of infrared divergences* in non-linear theories that have the correct linear limit. These equations also show the general structure of the terms that cancel each other. Note that all the terms in Eqs. (7.26) and (7.27) are proportional to either $\lambda_2$ or $\gamma_1$, which are the convective terms. While this discussion does not constitute a rigorous proof to all orders in perturbation theory, it does demonstrate that Galilean invariance plays a key role in the cancellation of infrared divergences.

We can now answer the question of whether it is possible to find a different non-linear approximation from ZA which is well behaved in the infrared. The restrictions imposed by Galilean invariance, translation invariance, and linear gravitational instability leave us with a family of non-linear theories which satisfy the constraints $\lambda_i = 1$, $\gamma_0 = \gamma_1 = 1$, and $2\gamma_2 - 3\gamma_3 = -1$. Taking $\gamma_3 \neq 0$ means solving the Poisson equation (with a renormalized Newton's constant $G' \equiv \gamma_3 G$). As we discussed above, this implies introducing a fundamental non-locality which greatly complicates the calculation of loop diagrams because of logarithmic corrections. Therefore, if $\gamma_3 \neq 0$, there is no reason to consider cases other than $\gamma_3 = 1$, which corresponds to ED. On the other hand, if we take $\gamma_3 = 0$, we arrive at the ZA. In this sense, ZA is the only (simple) non-linear approximation which is well-behaved in the infrared.

## VIII. SUMMARY AND CONCLUSIONS

We explicitly developed the general diagrammatic loop expansion for one-point cumulants of cosmological fields in NLCPT. We applied this formalism to calculate one-loop corrections to the variance and skewness of the density and velocity divergence fields in the exact dynamics of gravitational instability for scale-free initial power spectra. One-loop corrections for the unsmoothed density field dominate over tree-level contributions when $\sigma_\ell^2 \approx 1/2$. For the divergence of the velocity field, this dominance does not happen until $\sigma_\ell^2 \approx 1$. The results show a weak dependence on the power spectral index, $n$, induced by logarithmic terms coming from the non-locality of the perturbative solutions. This spectral dependence is absent at tree-level (for unsmoothed fields), because tree diagrams correspond to averaging out the tidal field which contains the non-local contribution. In fact, the dynamics at the tree-level has been shown to be equivalent to the spherical collapse model, and can be characterized by the "monopole" (angular average) of the perturbation theory kernels [7,17,18].

We also calculated loop corrections in different non-linear approximation schemes: the Zel'dovich approximation (ZA), linear potential approximation (LPA) and the frozen-flow approximation (FFA). Of these three theories, ZA shows the closest loop corrections to the ED, followed by LPA, and then FFA, in qualitative agreement with the known results at the tree-level. Calculations in this set of approximations are much simpler, due to the absence of non-local contributions which necessarily appear when solving the Poisson equation beyond linear perturbation theory. This advantage of course dissapears when one considers the calculation of higher-point cumulants, since now the statistic itself becomes non-local. Of all the non-linear approximations, ZA is the only one which gives loop corrections convergent in the infrared; we showed that this fact is related to the invariance of the equations of motion under a Galilean transformation.

These results show that loop corrections are generally not negligible in the weakly non-linear regime $\sigma \leq 1$ for unsmoothed fields. However, to see whether NLCPT can describe the $\sigma$ dependence of one-point cumulants, the effects of smoothing must be included in order to compare the predictions with numerical simulations. Results obtained recently for one-loop corrections to the variance with gaussian smoothing show very good agreement with N-Body simulations [32]. This is certainly encouraging news, and gives support to the perturbative approach to gravitational instability. In a subsequent paper, we will present loop corrections for smoothed cumulants and multi-point correlation functions.

## ACKNOWLEDGMENTS

We would like to thank F. Bernardeau, S. Colombi, J. Fry, R. Juszkiewicz, and L. Kofman for useful conversations. This research was supported in part by the DOE and by NASA grant NAG5-2788 at Fermilab.

## APPENDIX A: ANGULAR INTEGRALS

In this appendix we calculate the necessary angular integrals for loop corrections to 1-point cumulants. For corrections of order $\mathcal{O}(\sigma_\ell^6)$ to cumulants in the non-linear approximations we need the integral:

$$\int d\Omega_1 \int d\Omega_2 \int d\Omega_3 \ (\mathbf{q}_1 \cdot \mathbf{q}_2)^a (\mathbf{q}_2 \cdot \mathbf{q}_3)^b (\mathbf{q}_3 \cdot \mathbf{q}_1)^c, \tag{A1}$$

where $a$, $b$ and $c$ are non-negative integers. This integral, and its generalizations to higher number of wave-vectors can be calculated by using the result that:

$$\int d\Omega_q \ q_{i_1} \ldots q_{i_p} = \begin{cases} q^p A_{i_1 \ldots i_p} & \text{if } p \text{ is even}, \\ 0 & \text{if } p \text{ is odd}, \end{cases} \tag{A2}$$

where:

$$A_{i_1 \ldots i_p} \equiv \frac{4\pi}{(p+1)!} \sum_\pi \delta_{\pi(i_1)\pi(i_2)} \ldots \delta_{\pi(i_{p-1})\pi(i_p)}, \tag{A3}$$

and $\delta_{ij}$ denotes the Kronecker symbol. These results follow from antisymmetry under inversion of a component of $\mathbf{q}$, symmetry under permutations of the $i$'s, and rotational invariance. The coefficient in Eq. (A3) is obtained from the evaluation of a particular component of Eq. (A2). Eqs. (A2) and (A3) are the only integrations needed in the case of the non-linear approximations considered in the main text. Note that the result of the angular integrations gives a function in which the $q$'s are decoupled, so the radial integrations reduce to a product of one dimensional integrals.

Loop corrections in ED require the calculation of more complex angular integrals, due to the presence of angular variables in denominators coming from the Poisson Green's function. This introduces the complication that the integration does not decouple into one-dimensional integrals, a reflection of the non-locality introduced by the exact dynamics of gravity. (This results in the spectral index dependence of loop corrections.) For corrections up to order $\mathcal{O}(\sigma_\ell^6)$ we need:

$$I_1 \equiv \int d\Omega_1 \int d\Omega_2 \int d\Omega_3 \ \frac{(\mathbf{q}_1 \cdot \mathbf{q}_2)^a (\mathbf{q}_2 \cdot \mathbf{q}_3)^b (\mathbf{q}_3 \cdot \mathbf{q}_1)^c}{(\mathbf{q}_1 + \mathbf{q}_2)^2}, \tag{A4}$$

We choose the polar axis in the direction of $\mathbf{q}_1$ and the $x$-axis so that $\mathbf{q}_2$ lies in the $x$-$z$ plane. Then:

$$\mathbf{q}_1 \equiv q_1(0, 0, 1), \tag{A5a}$$
$$\mathbf{q}_2 \equiv q_2(\sin\theta, 0, \cos\theta), \tag{A5b}$$
$$\mathbf{q}_3 \equiv q_3(\sin\omega\cos\phi, \sin\omega\sin\phi, \cos\omega), \tag{A5c}$$

and therefore:

$$I_1 = 8\pi^2 q_1^{c+a} q_2^{a+b} q_3^{b+c} \int_{-1}^1 dx \int_{-1}^1 dy \int_0^{2\pi} d\phi \ \frac{x^a \left(\sqrt{1-x^2}\sqrt{1-y^2}\cos\phi + xy\right)^b y^c}{q_1^2 + q_2^2 + 2q_1 q_2 x}, \tag{A6}$$

where $x \equiv \cos\theta$ and $y \equiv \cos\omega$. Since $b$ is non-negative, we can use the binomial expansion to get:

$$I_1 = 8\pi^2 q_1^{c+a} q_2^{a+b} q_3^{b+c} \sum_{i=0}^b \binom{b}{i} \int_{-1}^1 dx \ \frac{x^{a+b-i}(1-x^2)^{i/2}}{q_1^2 + q_2^2 + 2q_1 q_2 x} \int_{-1}^1 dy \ y^{c+b-i}(1-y^2)^{i/2} \int_0^{2\pi} d\phi \ (\cos\phi)^i. \tag{A7}$$

Now note that for $i = 2k$:

$$\int_0^{2\pi} d\phi \ (\cos\phi)^{2k} = 2\pi \frac{(2k-1)!!}{(2k)!!}, \tag{A8}$$

the corresponding integral being zero when $i$ is odd. This result allows further binomial expansion in (A7) to get:

$$I_1 = 16\pi^3 q_1^{c+a} q_2^{a+b} q_3^{b+c} \sum_{k=0}^{[b/2]} \binom{b}{2k} \frac{(2k-1)!!}{(2k)!!} \sum_{i=0}^{k} \binom{k}{i} \sum_{j=0}^{k} \binom{k}{j} (-1)^{i+j} \left( \int_{-1}^{1} dx \ \frac{x^{a+b-2i}}{q_1^2 + q_2^2 + 2q_1 q_2 x} \right) \left( \int_{-1}^{1} dy \ y^{c+b-2j} \right), \tag{A9}$$

where $[b/2]$ denotes the integer part of $b/2$. The integral over $y$ is elementary, and using Eq. (B3) (see Appendix B) we arrive at the desired final expression:

$$I_1 = \frac{(4\pi)^3}{2} q_1^{c+a-1} q_2^{a+b-1} q_3^{b+c} \sum_{k=0}^{[b/2]} \binom{b}{2k} \frac{(2k-1)!!}{(2k)!!} \sum_{i=0}^{k} \binom{k}{i} \sum_{j=0}^{k} \binom{k}{j} \frac{(-1)^{i+j}}{(b+c-2j+1)} \left[ \sum_{r=0}^{[(a+b-2i-1)/2]} \frac{1}{2r+1} \right.$$
$$\left. \times \left( -\frac{q_1^2 + q_2^2}{2q_1 q_2} \right)^{a+b-2i-2r-1} + \left( -\frac{q_1^2 + q_2^2}{2q_1 q_2} \right)^{a+b-2i} \ln \frac{|q_1 + q_2|}{|q_1 - q_2|} \right], \tag{A10}$$

if $(b+c)$ is even, and zero otherwise. Another angular integral of interest is:

$$I_2 \equiv \int d\Omega_1 \int d\Omega_2 \int d\Omega_3 \ \frac{(\mathbf{q}_1 \cdot \mathbf{q}_2)^a (\mathbf{q}_2 \cdot \mathbf{q}_3)^b (\mathbf{q}_3 \cdot \mathbf{q}_1)^c}{(\mathbf{q}_1 + \mathbf{q}_2)^2 (\mathbf{q}_1 + \mathbf{q}_3)^2}. \tag{A11}$$

Following the same steps that lead to Eq. (A10) we obtain:

$$I_2 = \frac{(4\pi)^3}{4} q_1^{c+a-2} q_2^{a+b-1} q_3^{b+c-1} \sum_{k=0}^{[b/2]} \binom{b}{2k} \frac{(2k-1)!!}{(2k)!!} \sum_{i=0}^{k} \binom{k}{i} \sum_{j=0}^{k} \binom{k}{j} (-1)^{i+j} \left[ \sum_{r=0}^{[(a+b-2i-1)/2]} \left( -\frac{q_1^2 + q_2^2}{2q_1 q_2} \right)^{a+b-2i-2r-1} \right.$$
$$\times \frac{1}{2r+1} + \left( -\frac{q_1^2 + q_2^2}{2q_1 q_2} \right)^{a+b-2i} \ln \frac{|q_1 + q_2|}{|q_1 - q_2|} \left] \right[ \sum_{s=0}^{[(b+c-2j-1)/2]} \left( -\frac{q_1^2 + q_3^2}{2q_1 q_3} \right)^{b+c-2j-2s-1} \frac{1}{2s+1} + \left( -\frac{q_1^2 + q_3^2}{2q_1 q_3} \right)^{b+c-2j}$$
$$\times \ln \frac{|q_1 + q_3|}{|q_1 - q_3|} \right], \tag{A12}$$

We also need angular integrals involving coupling of three wave-vectors in the denominator:

$$I_3 \equiv \int d\Omega_1 \int d\Omega_2 \int d\Omega_3 \ \frac{(\mathbf{q}_1 \cdot \mathbf{q}_2)^a (\mathbf{q}_2 \cdot \mathbf{q}_3)^b (\mathbf{q}_3 \cdot \mathbf{q}_1)^c}{(\mathbf{q}_1 + \mathbf{q}_2 + \mathbf{q}_3)^2}. \tag{A13}$$

In this case it is more convenient to choose the $x$-$z$ plane as the plane spanned by $\mathbf{q}_1$ and $\mathbf{q}_2$, and take the $z$-axis to be along $\mathbf{Q} \equiv \mathbf{q}_1 + \mathbf{q}_2$. With these conventions we have:

$$\mathbf{q}_1 \equiv q_1(\sin\theta_1, 0, \cos\theta_1), \tag{A14a}$$
$$\mathbf{q}_2 \equiv q_2(-\sin\theta_2, 0, \cos\theta_2), \tag{A14b}$$
$$\mathbf{q}_3 \equiv q_3(\sin\theta_3 \cos\phi, \sin\theta_3 \sin\phi, \cos\theta_3). \tag{A14c}$$

All the necessary dot products can be expressed in terms of $\phi$, $z \equiv \cos\theta_3$ and $x \equiv \cos\theta$ where $\theta$ is the angle between $\mathbf{q}_1$ and $\mathbf{q}_2$:

$$\mathbf{q}_1 \cdot \mathbf{q}_2 \equiv q_1 q_2 x, \tag{A15a}$$
$$\mathbf{q}_2 \cdot \mathbf{q}_3 \equiv q_2 q_3 \left[ -q_1 \sqrt{1-x^2} \sqrt{1-z^2} \cos\phi + (q_2 + q_1 x) z \right]/Q, \tag{A15b}$$
$$\mathbf{q}_3 \cdot \mathbf{q}_1 \equiv q_3 q_1 \left[ q_2 \sqrt{1-x^2} \sqrt{1-z^2} \cos\phi + (q_1 + q_2 x) z \right]/Q, \tag{A15c}$$

where $Q^2 = q_1^2 + q_2^2 + 2q_1q_2x$. We therefore have:

$$I_3 = 8\pi^2 q_1^{c+a} q_2^{a+b} q_3^{b+c} \int_{-1}^{1} dx \int_{-1}^{1} dz \int_0^{2\pi} d\phi \; \frac{x^a \left[ -q_1\sqrt{1-x^2}\sqrt{1-z^2}\cos\phi + (q_2+q_1x)z \right]^b}{(Q^2 + q_3^2 + 2Qq_3z)Q^{b+c}}$$
$$\times \left[ q_2\sqrt{1-x^2}\sqrt{1-z^2}\cos\phi + (q_1+q_2x)z \right]^c. \tag{A16}$$

Using binomial expansions, Eq. (A8) and the following identity for any function $F$:

$$\sum_{i=0}^{b} \sum_{\substack{j=0 \\ i+j=2k}}^{c} F(i,j) = \sum_{i=0}^{b} \sum_{k=[(i+1)/2]}^{[(i+c)/2]} F(i, 2k-i), \tag{A17}$$

we get after some lengthy algebra:

$$I_3 = 16\pi^3 q_1^{c+a} q_2^{a+b} q_3^{b+c} \sum_{i=0}^{b} \sum_{k=[(i+1)/2]}^{[(i+c)/2]} \binom{b}{i}\binom{c}{2k-i}(-1)^i \frac{(2k-1)!!}{(2k)!!} \sum_{t=0}^{c-2k+i} \sum_{u=0}^{b-i} \sum_{v=0}^{k} (-1)^v \binom{k}{v}\binom{b-i}{u}\binom{c-2k+i}{t}$$
$$\times q_1^{c+u-2k-t+2i} q_2^{b-u+t+2k-2i} \int_{-1}^{1} dx \frac{x^{a+t+u+2v}}{Q^{b+c}} \sum_{r=0}^{k} (-1)^r \binom{k}{r} \int_{-1}^{1} dz \; \frac{z^{b+c-2k+2r}}{(Q^2+q_3^2+2Qq_3z)}. \tag{A18}$$

Now, using Eq. (B3) we have:

$$I_3 = 16\pi^3 q_1^{c+a} q_2^{a+b} q_3^{b+c} \sum_{i=0}^{b} \sum_{k=[(i+1)/2]}^{[(i+c)/2]} \sum_{t=0}^{c-2k+i} \sum_{u=0}^{b-i} \sum_{v=0}^{k} \sum_{r=0}^{k} \binom{b}{i}\binom{c}{2k-i}\binom{k}{v}\binom{b-i}{u}\binom{c-2k+i}{t}\binom{k}{r}\frac{(2k-1)!!}{(2k)!!}$$
$$\times (-1)^{i+v+r} q_1^{c+u-2k-t+2i} q_2^{b-u+t+2k-2i} \left[ \sum_{s=0}^{[(b+c-2k+2r-1)/2]} \left(\frac{-1}{2q_3}\right)^{b+c+2r-2k-2s-1} \frac{1}{2s+1} \sum_{w=0}^{b+c-2k+2r-2s-1} \right.$$
$$\times \binom{b+c-2k+2r-2s-1}{w} q_3^{2w-1} \int_{-1}^{1} dx \frac{x^{a+t+u+2v}}{Q^{2(k-r+s+w+1)}} + \left(\frac{-1}{2q_3}\right)^{b+c+2r-2k} \sum_{w=0}^{b+c-2k+2r} \binom{b+c-2k+2r}{w} q_3^{2w-1}$$
$$\times \left. \int_{-1}^{1} \frac{dx}{Q} \frac{x^{a+t+u+2v}}{Q^{2(k-r+w)}} \ln\frac{|Q+q_3|}{|Q-q_3|} \right]. \tag{A19}$$

Finally, noting that:

$$\int_{-1}^{1} \frac{dx}{Q} \frac{x^m}{Q^{2c}} \ln\frac{|Q+q_3|}{|Q-q_3|} = \frac{2}{(2q_1q_2)^{m+1}} \sum_{h=0}^{m} \binom{m}{h}(-1)^{m-h}(q_1^2+q_2^2)^{m-h} \int_{|q_1-q_2|}^{|q_1+q_2|} dz \; z^{2(h-c)} \ln\frac{|z+q_3|}{|z-q_3|}, \tag{A20}$$

we arrive at the final expression:

$$I_3 = \frac{(4\pi)^3}{4} q_1^{c+a} q_2^{a+b} q_3^{b+c} \sum_{i=0}^{b} \sum_{k=[(i+1)/2]}^{[(i+c)/2]} \sum_{t=0}^{c-2k+i} \sum_{u=0}^{b-i} \sum_{v=0}^{k} \sum_{r=0}^{k} \binom{b}{i}\binom{c}{2k-i}\binom{k}{v}\binom{b-i}{u}\binom{c-2k+i}{t}\binom{k}{r}\frac{(2k-1)!!}{(2k)!!}$$
$$\times (-1)^{i+v+r} q_1^{c+u-2k-t+2i} q_2^{b-u+t+2k-2i} \left[ \sum_{s=0}^{[(b+c-2k+2r-1)/2]} \left(\frac{-1}{2q_3}\right)^{b+c+2r-2k-2s-1} \frac{1}{2s+1} \sum_{w=0}^{b+c-2k+2r-2s-1} \right.$$
$$\times \binom{b+c-2k+2r-2s-1}{w} \frac{q_3^{2w-1}}{(2q_1q_2)^{k-r+s+w+1}} M\left(a+t+u+2v, k-r+s+w+1, \frac{q_1^2+q_2^2}{2q_1q_2} ; -1, 1\right)$$
$$+ \left(\frac{-1}{2q_3}\right)^{b+c+2r-2k} \sum_{w=0}^{b+c-2k+2r} \binom{b+c-2k+2r}{w} \frac{2 \; q_3^{2w-1}}{(2q_1q_2)^{a+t+u+2v+1}} \sum_{h=0}^{a+t+u+2v} \binom{a+t+u+2v}{h}$$
$$\times \left. (-1)^{a+t+u+2v-h}(q_1^2+q_2^2)^{a+t+u+2v-h} \Delta J(2h-2k+2r-2w, q_3 ; |q_1-q_2|, |q_1+q_2|) \right], \tag{A21}$$

where we have used the definitions:

$$M(m, c, \alpha\ ; a, b) \equiv \int_a^b dx\ \frac{x^m}{(x+\alpha)^c}, \tag{A22}$$

with $M(m, 1, \alpha\ ; a, b) \equiv M(m, 1, \alpha\ ; b) - M(m, 1, \alpha\ ; a)$ (see Eq. (B3)), and:

$$J(m, \alpha\ ; a, b) \equiv \int_a^b dx\ x^m \ln|x+\alpha|, \tag{A23}$$

with $\Delta J(m, \alpha\ ; a, b) \equiv J(m, \alpha\ ; a, b) - J(m, -\alpha\ ; a, b)$. These integrals will be calculated in the next appendix since they are also needed for radial integrations.

From Eq. (A21) one can get the result for the angular integral:

$$I_4 \equiv \int d\Omega_1 \int d\Omega_2 \int d\Omega_3\ \frac{(\mathbf{q}_1 \cdot \mathbf{q}_2)^a (\mathbf{q}_2 \cdot \mathbf{q}_3)^b (\mathbf{q}_3 \cdot \mathbf{q}_1)^c}{(\mathbf{q}_1 + \mathbf{q}_2 + \mathbf{q}_3)^2 (\mathbf{q}_1 + \mathbf{q}_2)^2}, \tag{A24}$$

since the only difference with Eq. (A13) is the additional factor of $1/Q^2$. We get:

$$\begin{aligned}
I_4 = &\frac{(4\pi)^3}{4} q_1^{c+a} q_2^{a+b} q_3^{b+c} \sum_{i=0}^{b} \sum_{k=[(i+1)/2]}^{[(i+c)/2]} \sum_{t=0}^{c-2k+i} \sum_{u=0}^{b-i} \sum_{v=0}^{k} \sum_{r=0}^{k} \binom{b}{i}\binom{c}{2k-i}\binom{k}{v}\binom{b-i}{u}\binom{c-2k+i}{t}\binom{k}{r} \frac{(2k-1)!!}{(2k)!!} \\
&\times (-1)^{i+v+r} q_1^{c+u-2k-t+2i} q_2^{b-u+t+2k-2i} \left[ \sum_{s=0}^{[(b+c-2k+2r-1)/2]} \left(\frac{-1}{2q_3}\right)^{b+c+2r-2k-2s-1} \frac{1}{2s+1} \sum_{w=0}^{b+c-2k+2r-2s-1} \right. \\
&\times \binom{b+c-2k+2r-2s-1}{w} \frac{q_3^{2w-1}}{(2q_1 q_2)^{k-r+s+w+2}} M\left(a+t+u+2v, k-r+s+w+2, \frac{q_1^2+q_2^2}{2q_1 q_2}\ ; -1, 1\right) \\
&+ \left(\frac{-1}{2q_3}\right)^{b+c+2r-2k} \sum_{w=0}^{b+c-2k+2r} \binom{b+c-2k+2r}{w} \frac{2\ q_3^{2w-1}}{(2q_1 q_2)^{a+t+u+2v+1}} \sum_{h=0}^{a+t+u+2v} \binom{a+t+u+2v}{h} \\
&\left. \times (-1)^{a+t+u+2v-h} (q_1^2+q_2^2)^{a+t+u+2v-h} \Delta J(2h-2k+2r-2w-2, q_3\ ; |q_1-q_2|, |q_1+q_2|) \right].
\end{aligned} \tag{A25}$$

## APPENDIX B: RADIAL INTEGRALS

We now proceed to calculate the radial integrals involved in loop corrections to cumulants in the ED case. As the results from the angular integrations show, most of the radial integrations needed involve logarithms. The number of radial integrals involved is quite large, especially because in general radial integrals do not decouple into products of independent one-dimensional integrations. However, all these integrals can be calculated in terms of the ones given in this appendix using integration by parts and partial fraction decompositions. Many of the integrals of interest are generated by:

$$M(m, c, \alpha\ ; x) \equiv \int dx\ \frac{x^m}{(x+\alpha)^c} = \frac{(-1)^{c-1}}{(c-1)!} \frac{\partial^{c-1}}{\partial \alpha^{c-1}} M(m, 1, \alpha\ ; x). \tag{B1}$$

The integral for $c = 1$ can be solved by noting the following identities for $m \geq 0$ and $\alpha \neq 0$:

$$\frac{x^m}{x+\alpha} = \frac{(-\alpha)^m}{x+\alpha} + \sum_{i=1}^{m} x^{i-1}(-\alpha)^{m-i}, \tag{B2a}$$

$$\frac{1}{x^m(x+\alpha)} = \frac{1}{(-\alpha)^m(x+\alpha)} - \sum_{i=1}^{m} \frac{1}{x^i(-\alpha)^{m+1-i}}. \tag{B2b}$$

Therefore we have:

$$M(m, 1, \alpha\ ; x) = \sum_{j=0}^{m} \frac{(-\alpha)^{m-j} x^j}{j} + (-\alpha)^m \ln|x + \alpha| \qquad , m \geq 0 ,\tag{B3a}$$

$$M(m, 1, \alpha\ ; x) = \sum_{j=1}^{-m-1} \frac{(-\alpha)^{m+j}}{j\, x^j} + (-\alpha)^m \Big( \ln|x + \alpha| - \ln|x| \Big) \qquad , m \leq -1.\tag{B3b}$$

By taking derivatives with respect to $\alpha$ (denoted by $\partial_\alpha$) according to Eq. (B1) one can generate $M(m, c, \alpha\ ; x)$ with $c \neq 1$. We get:

$$M(m, c, \alpha\ ; x) = \sum_{j=1}^{m} \frac{x^{m-j+1}(-1)^{c+j}}{(m-j+1)(c-1)!} \partial_\alpha^{c-1}(\alpha^{j-1}) + \frac{(-1)^{c+m-1}}{(c-1)!} \partial_\alpha^{c-1}(\alpha)^m \ln|x+\alpha| + \sum_{j=1}^{c-1} \binom{c-1}{j} \frac{(j-1)!}{(c-1)!}$$

$$\times \frac{(-1)^{c+m+j}}{x+\alpha} \partial_\alpha^{c-1-j}(\alpha^m) \qquad , m \geq 0 ,\tag{B4a}$$

$$M(m, c, \alpha\ ; x) = \sum_{j=1}^{-m-1} \frac{x^{m+j}(-1)^{c+j}}{(m+j)(c-1)!} \partial_\alpha^{c-1}(\alpha^{-j}) + \frac{(-1)^{c+m-1}}{(c-1)!} \partial_\alpha^{c-1}(\alpha)^m \Big( \ln|x+\alpha| - \ln|x| \Big) + \sum_{j=1}^{c-1} \binom{c-1}{j}$$

$$\times \frac{(j-1)!}{(c-1)!} \frac{(-1)^{c+m+j}}{x+\alpha} \partial_\alpha^{c-1-j}(\alpha^m) \qquad , m \leq -1.\tag{B4b}$$

Another important integral is:

$$J(m, \alpha\ ; x) \equiv \int dx\ x^m \ln|x + \alpha|,\tag{B5}$$

which can be solved by shifting the integration variable $x \longrightarrow x + \alpha$ and then using integration by parts for $m \geq 0$ and $m = -2$. The $m < -2$ integrals are generated by derivatives of $J(m, \alpha\ ; x - \alpha)$ with respect to $\alpha$. The $m = -1$ case deserves special treatment and needs the introduction of a new function, the dilogarithm. Shifting variables back we get:

$$J(m, \alpha\ ; x) = \frac{x^{m+1} - (-\alpha)^{m+1}}{m+1} \ln|x + \alpha| - \sum_{j=0}^{m} \binom{m}{j} (-\alpha)^{m-j} \frac{(x+\alpha)^{j+1}}{(j+1)^2} \qquad , m \geq 0 ,\tag{B6a}$$

$$J(-1, \alpha\ ; x) = \ln|\alpha|\ \ln|x| - \mathrm{Re}\big[\mathrm{Li}_2(-x/\alpha)\big],\tag{B6b}$$

$$J(m, \alpha\ ; x) = \frac{x^{m+1} - (-\alpha)^{m+1}}{m+1} \ln|x + \alpha| + \frac{(-\alpha)^{m+1}}{m+1} \ln|x| - \frac{(-\alpha)^{m+1}}{m+1} \sum_{j=1}^{-m-2} \frac{(-\alpha)^j}{j\, x^j} \qquad , m \leq -2,\tag{B6c}$$

where we have used $\mathrm{Re}[\ln(x)] = \ln|x|$, and $\mathrm{Li}_2$ denotes the dilogarithm, defined by [69]:

$$\mathrm{Li}_2(x) \equiv -\int_0^x dz\ \frac{\ln(1-z)}{z}.\tag{B7}$$

This definition is extended to higher order polylogarithms in the following way [69]:

$$\mathrm{Li}_n(x) \equiv \int_0^x dz\ \frac{\mathrm{Li}_{n-1}(z)}{z},\tag{B8}$$

where $\mathrm{Li}_1(x) \equiv -\ln(1-z)/z$. Polylogarithms have the series expansion for small argument ($x \leq 1$):

$$\mathrm{Li}_n(x) = \sum_{i=1}^{\infty} \frac{x^i}{n^i},\tag{B9}$$

which gives the connection to the Riemann zeta function, $\mathrm{Li}_n(1) = \zeta(n)$. Polylogarithms of real argument $x$ are real as long as $x \leq 1$, and for $x > 1$ we have $\mathrm{Im}[\mathrm{Li}_n(x)] = -\pi \ln^{n-1}(x)/(n-1)!$. To carry out series expansions of polylogarithms in powers of $\epsilon/k_c$ we also need the so-called functional equations [69]:

$$\text{Li}_2(x) = -\text{Li}_2(1/x) - \frac{1}{2}\ln^2(x) + \frac{\pi^2}{3} - i\pi \ln x \qquad , x \geq 1, \tag{B10}$$

$$\text{Li}_2(-x) = -\text{Li}_2(-1/x) - \frac{1}{2}\ln^2(x) - \frac{\pi^2}{6}, \tag{B11}$$

$$\text{Li}_3(x) = \text{Li}_3(1/x) - \frac{1}{6}\ln^3(x) + \frac{\pi^2}{3}\ln(x) - \frac{i}{2}\pi \ln^2 x \qquad , x \geq 1, \tag{B12}$$

$$\text{Li}_3(-x) = \text{Li}_3(-1/x) - \frac{1}{6}\ln^3(x) - \frac{\pi^2}{6}\ln(x), \tag{B13}$$

$$\text{Li}_4(x) = -\text{Li}_4(1/x) - \frac{1}{24}\ln^4(x) + \frac{\pi^2}{6}\ln^2(x) + \frac{\pi^4}{45} - \frac{i}{6}\pi \ln^3 x \qquad , x \geq 1, \tag{B14}$$

$$\text{Li}_4(-x) = -\text{Li}_4(-1/x) - \frac{1}{24}\ln^4(x) + \frac{\pi^2}{6}\ln^2(x) - \frac{7\pi^4}{360}, \tag{B15}$$

which show the behavior of polylogarithms for large positive and negative arguments. The integrals involving polylogarithms combined with powers and/or logarithms that appear in the calculation of 1-loop corrections to $\sigma^2$ and $S_3$ can be obtained in terms of $M$, $J$ and the following last "master" integral given by:

$$T(m, \alpha, \beta; x) \equiv \int dx \, x^m \, \ln|x+\alpha| \, \ln|x+\beta|, \tag{B16}$$

which can be easily solved in terms of $J$ integrals by integrating by parts when $m \neq -1$. In this last case, unless $\alpha$ or $\beta$ are zero, the integration by parts does not work and one has to calculate it explicitly. By factoring out $\alpha$ and $\beta$ from inside the logarithms one finds:

$$T(-1, \alpha, \beta; x) = \ln|\alpha| \, \ln|\beta| \, \ln|x| - \text{Re}\big[\text{Li}_2(-\frac{x}{\alpha})\big]\ln|\beta| - \text{Re}\big[\text{Li}_2(-\frac{x}{\beta})\big]\ln|\alpha| + \int^{-x/\alpha} \frac{dy}{y} \, \ln|1-y| \, \ln|1-cy|, \tag{B17}$$

where $c \equiv \alpha/\beta$. This last integral is solved by using that [69]:

$$\int \frac{dx}{x} \, \ln(1-x) \, \ln(1-cx) = \text{Li}_3\left[\frac{1-cx}{1-x}\right] + \text{Li}_3(1/c) + \text{Li}_3(1) - \text{Li}_3(1-x) - \text{Li}_3\left[\frac{1-cx}{c(1-x)}\right] + \frac{1}{2}\ln(c)\ln^2(1-x)$$
$$+ \ln(1-cx)\Big[\text{Li}_2(1/c) - \text{Li}_2(x)\Big] + \ln(1-x)\Big[\text{Li}_2(1-cx) - \text{Li}_2(1/c) + \frac{\pi^2}{6}\Big]. \tag{B18}$$

Finally, we must mention that special cases arise when the constants $\alpha$ or $\beta$ become zero or such that there is a singularity in the region of integration. Most of the time these situations can be solved by taking the appropriate limits (by using the functional equations for the polylogarithms, for example). There are cases, however, where individual integrals are singular, but these divergences can be regulated by introducing another small parameter, say $\delta$, which gets canceled at the end of the calculation by taking $\delta \ll \epsilon \ll k_c$. In practice, the symbolic integration routine developed checks for singularities in the integration region and introduces infinitesimal cutoffs accordingly.